\documentclass{article}
\usepackage{graphicx}
\usepackage{float}
\usepackage{topcapt}

\usepackage{lipsum}
\usepackage{epigraph}

\usepackage{amsmath}
\usepackage{amssymb}
\usepackage{url}
\usepackage{eurosym}

\usepackage[T2A]{fontenc}
\usepackage[utf8]{inputenc}
\usepackage[russian,english,british]{babel}

\usepackage{longtable}
\usepackage{fancyvrb}
\usepackage{tocbibind}

\begin{document}

\title{The Wandering of Corn}
\author{Valerii Salov}
\date{}
\maketitle

\begin{abstract}
Time and Sales of corn futures traded electronically on the CME Group Globex are studied. Theories of continuous prices turn upside down reality of intra-day trading. Prices and their increments are discrete and obey lattice probability distributions. A function for systematic evolution of futures trading volume is proposed. Dependence between sample skewness and kurtosis of waiting times does not support hypothesis of Weibull distribution. Kumaraswamy distribution is more suitable for waiting times. Relationships between trading volume and maximum profit strategies are presented. Frequencies of absolute b-increments are approximated by a Hurwitz Zeta distribution. Relative b-increments are non-Gaussian too. Dependence between b- and a-increments allows to interpret the sample variances of b-increments as a stochastic process. Mean sample variance of b-increments vs. a-increments is presented. The L1 distance and Log-likelihood statistics for independence between a- and b-increments are controversial. Corn price jumps remind of chain branching reactions. Bi-logarithmic plots of the empirical frequencies of extreme b-increments vs. ranks are presented. Corresponding distributions resemble snakes forked tongues. The maximum profit strategy is discussed as a measure of non-equilibrium.
\end{abstract}

\section{Introduction}

\setlength{\epigraphwidth}{0.77\textwidth}

\epigraph{It is quite probable that with the development of modern computing techniques, it will become understood that in many cases it is reasonable to study real phenomena without making use of intermediate step of their stylization in the form of infinite and continuous mathematics, passing directly to discrete models.}
{\textsc{Andrey Nikolaevich Kolmogorov, \cite{kolmogorov1983}}}

Empirical frequencies of \textit{price increments} $\Delta P_i=P_i - P_{i - 1}$ and \textit{log-returns} $\ln(\frac{P_i}{P_{i - 1}}) = \ln(P_i) - \ln(P_{i - 1})$ are often approximated by continuous \textit{probability density functions}, PDF. Natural $i$ changes from $2$ to $N$. $i$ and $i-1$ is "current" and  "previous". The less $|\Delta P_i|$, the more accurate $\ln(\frac{P_i}{P_{i - 1}}) \approx \frac{\Delta P_i}{P_{i - 1}}$. The latter fraction is a \textit{relative price increment} or \textit{return} infinite or undefined for $P_{i - 1} = 0$. For \textit{real} logarithm $\frac{P_i}{P_{i - 1}} > 0$. $P=0$ indicates \textit{bankruptcy}. $P < 0$ resembles a \textit{garbage removal cost}. For \textit{futures}, $P_i > 0$ $\forall i$. It is both the sum and product
\begin{equation}
\label{EqPriceAsSum}
P_i=P_1 + \sum_{k=2}^{k=i} \Delta P_k, \; i = 1, \dots, N, \; \sum_{k=2}^{k=1} \Delta P_k = 0,
\end{equation}
\begin{equation}
\label{EqPriceAsProduct}
P_i=P_1 \times \prod_{k=2}^{k=i} e^{\ln(\frac{P_k}{P_{k-1}})}, \; i = 1, \dots, N, \; \prod_{k=2}^{k=1} e^{\ln(\frac{P_k}{P_{k-1}})} = 1.
\end{equation}

Bachelier \cite{bachelier1900} assumed that $\Delta P_i$ are \textit{random Gaussian variables} with variances $\sigma_i^2$ proportional to \textit{time increments} $\Delta t_i = t_i - t_{i - 1}$. Being \textit{independent identically distributed}, i.i.d., for constant $\Delta t_i$, they make the sum in Equation \ref{EqPriceAsSum} and $P_i$ Gaussian variables with variance $(i - 1) \times \sigma^2$ \cite{gnedenko1949}, \cite{gnedenko1988}. In contrast with a futures price, such $P_i$ can become negative, even, for large $P_1$.

Remery \cite{remery1946}, Laurent \cite{laurent1957}, \cite{laurent1959}, Osborne \cite{osborne1959}, \cite{osborne1959_2}, Samuelson \cite[see Foreword about 'geometric' Brownian motion]{davis2006} suggested that log-returns $\ln(\frac{P_i}{P_{i - 1}})$ are Gaussian. Then, from Equation \ref{EqPriceAsProduct}
\begin{equation}
\label{EqLogPriceAsSum}
\ln(P_i)=\ln(P_1) + \sum_{k=2}^{i} \ln(\frac{P_k}{P_{k - 1}}), \; i = 1, \dots, N,
\end{equation}
and $\ln(P_i)$ is also Gaussian, if log-returns are i.i.d. The latter is easier to assume for constant $\Delta t_k$. While log-returns and  $\ln(P_i)$ can be negative or zero, the expression under the logarithm is always positive.

An intra-day futures trader knows that \textit{waiting times} or \textit{durations}, time intervals between arriving neighboring \textit{price ticks}, are irregular. Mandelbrot and Taylor \cite{mandelbrot1967}, Clark \cite{clark1970}, \cite{clark1973} emphasized importance of randomness of time durations for prices. Bochner developed the theory of \textit{subordinated processes} \cite{bochner1955}. Rubin introduced \textit{regular point processes}, and studied them theoretically with different \textit{intensity functions} \cite{rubin1972}. Modern applications are Madan and Seneta \cite{madan1990}, Carr and Maden \cite{carr1997}, Goodhart and O'Hara \cite{goodhart1997}, Engle and Russel \cite{engle1998}, \cite{engle2000}, McCulloch \cite{mcculloch2005}. \textit{To judge on priority, read Kolmogorov's} \cite{kolmogorov1940}.

Theories of continuous prices and rates overfill modern finance. Gaussian i.i.d increments and log-returns were only the beginning. They yield continuous but never differentiable Brownian motions. Deterministic time dependencies of mean and variance of Gaussian distributions were next. They are replaced with moments following random processes frequently also Gaussian. Adding correlations between levels of continuous stochastic processes improves fitting demanded by pricing \textit{derivatives} based on \textit{absence of arbitrage} and \textit{martingale measures}. The former is a rational way to a unique option value and a complement to an otherwise insufficient \textit{underlying} price model. In contrast, trading futures does not need this assumption. \textit{Postulating absence of arbitrage to please theoretical pricing neglects a research supporting practical trading}.

Apologists of continuous prices and absence of arbitrage should be indebted to Kolmogorov \cite{kolmogorov1931} and Doob \cite{doob1990} for computational framework and martingales. Emphasizing novelty of his own contribution, Kolmogorov writes: \textit{"Author systematically considers the simplest cases of stochastically determined processes and in the first order - processes continuous in time ..."} (VS's translation of the revised version, page 5, where Kolmogorov also comments on the contribution of Fokker and Planck). He both estimates priority of systematic study of scheme with continuous in time changes of probability and criticizes a lack of mathematical rigorous of Bachelier's approach. This was said in 1931 but epigraph to this article is from 1983 with roots in 1970th. This time distance is comparable with another, where the father of the probability theory axioms moves from its foundation to definitions of \textit{randomness} based on the \textit{theory of algorithms} \cite{kolmogorov1987}. A historian of mathematics and philosopher will understand the evolution of views of the great mathematician of the 20th century. Meanwhile, following to this dramatic shift in paradigms of "continuous" and "discrete", let us examine futures corn prices.

\section{ZCH16 Friday January 22, 2016}

\begin{figure}[!h]
  \centering
  \includegraphics[width=130mm]{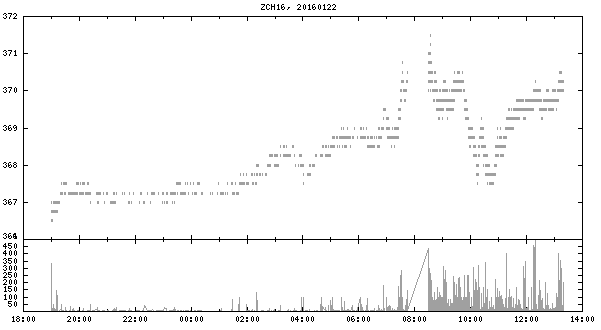}
  \caption[FigESH16_20160122]
   {ZCH16 Time \& Sales, \url{http://www.cmegroup.com/}, transaction prices (top) and sizes (bottom) during the overnight and day time ranges of the trading session closed on Friday January 22, 2016. Plots are done using custom C++  and Python programs and gnuplot \url{http://www.gnuplot.info/}.}
  \label{FigESH16_20160122}
\end{figure}

ZC is the ticker of the corn futures contract traded on the \textit{CME Group Globex electronic platform}. H16 indicates expiration month, March, and year 2016. While each trading session is unique, the one on Friday January 22, 2016, due to absence of extraordinary events, is \textit{typical}. Time \& Sales data were collected from \url{http://www.cmegroup.com/}. This is where contract specifications can be found.  ZCH16 was traded in overnight [01/21/2016 19:00:00, 01/22/2016 07:45:00], and day [01/22/2016 08:30:00, 01/22/2016 13:20:00] time ranges, Figure \ref{FigESH16_20160122}, where time is the Central Standard Time, CST.
\paragraph{The minimal absolute non zero price increment} for ZCH16 is $\delta_{ZC}=0.25$ cents per bushel. The contract is for 5,000 bushels of corn and the value of this change is \$12.50. The prices on Figure \ref{FigESH16_20160122} are in cents per bushel. We clearly see 18 equidistant discrete levels $366.50 + 0.25 i, i = 0, \; \dots, \; 17$ in the overnight and 17 levels $367.50 + 0.25i, i = 0, \; \dots, \; 16$ in the day range.

Each tick is a triplet of time, price, and size or volume - the number of bought and sold contracts, for instance, \{2016-01-22 07:44:59, 370.25, 10\}. Price ticks associate with \textit{transactions}, \textit{indicative}, \textit{cancel}, and other market conditions. In this paper C++ programs process records with nonzero size. The overnight and day ranges contain 2,164 and 11,309 such ticks. \textit{Some traders sleep at night}.

\paragraph{A-, b-, and c-increments.} In \cite{salov2013}, for classification, the author names $\Delta t_i$ and $\Delta P_i$ between neighboring ticks the \textit{a-} and \textit{b-increments} and the price increments between the first price in a next and last price in a previous trading session or time range - \textit{c-increments}. Time intervals between ranges and sessions are typically much longer than a-increments for \textit{liquid} contracts and almost constant. They did not get special name in the classification. There are 2,163 a- and b-increments in the overnight range, one c-increment 370.50 (08:30:00) - 370.25 (07:44:59) = 0.25 = $\delta_{ZC}$ between two ranges, and 11,308 a- and b-increments in the day range. The time interval corresponding to c-increment was 45 minutes and one second. Currently, times in Time \& Sales data are reported with accuracy of one second.

\paragraph{Non-Gaussian b-increments.} Discreteness of prices and their increments simplifies plotting their frequency histograms because of a natural bin width determined by levels forming a \textit{lattice} for the \textit{empirical probability mass functions}, EPMF, Figure \ref{FigESH16_Distrib_20160122}. The overnight and day b-increments histograms are similar and resemble a \textit{bell curve}. Being expressed in integer numbers of $\delta_{ZC}$, b-increments yield overnight and day statistics:
\begin{figure}[!h]
  \centering
  \includegraphics[width=130mm]{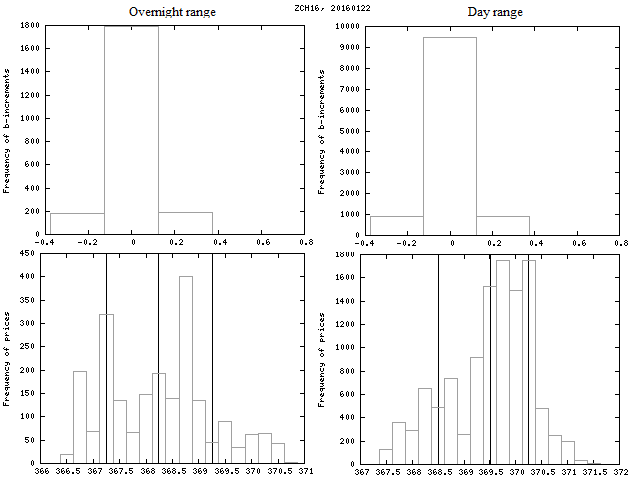}
  \caption[FigESH16_Distrib_20160122]
   {ZCH16 Time \& Sales, \url{http://www.cmegroup.com/}, empirical histograms of transaction b-increments (top) and prices (bottom) during the overnight and day time ranges of the trading session on Friday January 22, 2016. Plots are done using custom C++  and Python programs and gnuplot \url{http://www.gnuplot.info/}.}
  \label{FigESH16_Distrib_20160122}
\end{figure}
\begin{verbatim}
Overnight
Mean                = 0.00647249
Samples size        = 2163
Maximum value       = 2
Maximum value count = 1
Minimum value       = -1
Minimum value count = 179
Variance            = 0.172946
Std. deviation      = 0.415868
Skewness            = 0.0820005
Excess kurtosis     = 2.96807
0 (-2, -1] 179
1 (-1, 0] 1792
2 (0, 1] 191
3 (1, 2] 1

Day
Mean                = -0.000176866
Samples size        = 11308
Maximum value       = 2
Maximum value count = 1
Minimum value       = -1
Minimum value count = 923
Variance            = 0.163262
Std. deviation      = 0.404057
Skewness            = 0.00667709
Excess kurtosis     = 3.16607
0 (-2, -1] 923
1 (-1, 0] 9465
2 (0, 1] 919
3 (1, 2] 1
\end{verbatim}
For Gaussian PDF the excess kurtosis is zero. The found values are 3.0 and 3.2. Skewness is close to zero as it should be. The means are close to zero. In the overnight range the maximum deviation from the mean expressed in standard deviations is $\frac{2 - 0.00647249}{0.415868} \approx 4.8$. With the found mean and variance, Gaussian probability to get an equal or greater positive deviation is $8.2\times 10^{-7}$. We get one in 2,163: $\frac{1}{2,163} \approx 4.6 \times 10^{-4}$. This is 561 times more frequent and risky than it was assumed by Bachelier. For the day range the maximum positive deviation from the mean is $\frac{2 - (-0.000176866)}{0.404057} \approx 5.0$. Gaussian probability of the equal or greater positive deviation is $3.7\times 10^{-7}$. We get one in 11,308: $\frac{1}{11,308} \approx  8.8\times 10^{-5}$. The \textit{risk underestimation ratio} is $\frac{8.8\times 10^{-5}}{3.7\times 10^{-7}} \approx 238$.

The overnight intervals of b-increments, their counts $n$, and Gaussian probabilities $p$ for mean and variance 0.00647249 and 0.172946 are: $(-\infty, -0.5]$, $n_1=179$, $p_1=0.1116$; $(-0.5, 0.5]$, $n_2=1,792$, $p_2=0.7707$; $(0.5, 1.5]$, $n_3=191$, $p_3=0.1175$; $(1.5, \infty)$, $n_4=1$, $p_4=0.0001645$. For $N=2,163$ $\chi^2=\sum_{i=1}^{i=4}\frac{(n_i-p_iN)^2}{p_iN} \approx 42.4 > \chi^2(p=0.005, \textrm{degrees of freedom} = 3) \approx 12.8$. The \textit{Pearson's goodness of fit test} rejects hypothesis of the Gaussian b-increments.

For the day range and $Gaussian(-0.000176866, 0.163262)$: $(-\infty, -0.5]$, $n_1=923$, $p_1=0.1080$; $(-0.5, 0.5]$, $n_2=9,465$, $p_2=0.7841$; $(0.5, 1.5]$, $n_3=919$, $p_3=0.1078$; $(1.5, \infty)$, $n_4=1$, $p_4=0.0001025$. For $N=11,308$ $\chi^2 \approx 187.2$. Agreement with the Gaussian distribution of b-increments hypothesis is worse.

Inadequateness of the Gaussian hypothesis has been emphasized from 1950th, soon after Bachelier's ideas \cite{bachelier1900}, well known to mathematicians \cite{kolmogorov1931}, invaded economics \cite{davis2006}. Sir Kendall \cite{kendall1953}, Mandelbrot \cite{mandelbrot1963}, Fama \cite{fama1965} presented evidences of what is widely named today \textit{fat tails}. \textit{Luckily}, this did not stop in 1973 Black, Scholes \cite{black1973}, and Merton \cite{merton1973} to derive their \textit{option value formulas} (Merton accounts \textit{continuous dividend yield}) relied on the \textit{mathematical Brownian motion} of $\frac{d\ln(P)}{P}$ and lognormal prices. Both are similar to Bachelier's version of 1900, where price increments are normal. All are based on a continuous Brownian motion or \textit{Wiener process} \cite{neftci1996}. \textit{Futures contracts are not followers of these underlying processes either \cite{salov2011}, \cite{salov2013}. In the latter papers, author also criticizes application of continuous distributions such as Gaussian.}

\paragraph{Continuous vs. discrete distributions.} Probability that a random variable $\xi$ takes a value less than an arbitrary real number $x$ is named a \textit{function of distribution} of probabilities of the random variable $\xi$: $F(x)=\mathbb{P}\{\xi < x\}$, \cite[page 28]{gnedenko1949}, \cite[page 117]{gnedenko1988}, \cite{korn1968}. There is an alternative definition, where $F(x)=\mathbb{P}\{\xi \le x\}$ \cite[page 52]{karr1993}. It is also known under the name \textit{cumulative distribution function}, CDF. For a \textit{continuous} $\xi$, probability that its value is equal to $x$ is zero $\mathbb{P}\{\xi=x\}=0$. \textit{Probability measure} 0 assigned by a continuous distribution to a point makes less critical which CDF ($<$ or $\le$) is applied. However, it is needed to be specific using one for discrete distributions. Computations in previous paragraphs apply the Microsoft Excel function NORMDIST. Given $x$, mean $\alpha_1$, and standard deviation $\sigma = \sqrt{\sigma^2}=\sqrt{\mu_2}$, where $\mu_2$ is the \textit{second central moment} - variance, it returns probability $\mathbb{P}\{\xi \le x\}$ expressed by the integral of the Gaussian PDF $f(x) = \frac{1}{\sqrt{2\pi \mu_2}}e^{\frac{(x-\alpha_1)^2}{2\mu_2}}$: $F(x)=\frac{1}{\sqrt{2\pi \mu_2}}\int_{-\infty}^x e^{\frac{(y-\alpha_1)^2}{2\mu_2}}dy$. $\forall x, f(x) > 0$. In contrast with $F(x)$ and probabilities the density can be greater than one. Applying Gaussian distribution with $\alpha_1 = 0.00647249$ and $\sqrt{\mu_2} = 0.415868$, we get that probability of the price increment to be greater than 0.01 and less than 0.99 is equal to NORMDIST(0.99, 0.00647249, 0.415868, TRUE) - NORMDIST(0.01, 0.00647249, 0.415868, TRUE) $\approx 0.488$, where TRUE and FALSE means CDF and PDF. With $P_1=366.75$ we get $\mathbb{P}\{366.75 + 0.01 * 0.25 =  366.7525 < P_2 < 366.75 + 0.99 * 0.25 = 366.9975\} = 0.488$. Gaussian increments yield $\mathbb{P}\{P_2 = 366.50\} = 0$, $\mathbb{P}\{P_2 = 366.75\} = 0$, $\mathbb{P}\{P_2 = 367.00\} = 0$, $\mathbb{P}\{P_2 = 367.25\} = 0$. \textit{Theory presents reality upside down. Time \& Sales frequencies are } $\mathbb{P}\{366.7525 < P_2 < 366.9975\} = 0$ \textit{but not} 0.488 \textit{and} $\mathbb{P}\{P_2 = 366.50\} = \frac{179}{2,163}$, $\mathbb{P}\{P_2 = 366.75\} = \frac{1,792}{2,163}$, $\mathbb{P}\{P_2 = 367.00\} = \frac{191}{2,163}$, $\mathbb{P}\{P_2 = 367.25\} = \frac{1}{2,2163}$ \textit{but not zero! The reality of intra-day trading is not only fat-tails but discreteness of prices and their increments}.

\section{Discreteness from atoms to trading}

\paragraph{Bohr \cite[pp. 8, 9]{bohr1913}}: \textit{"The amount of energy emitted by the passing of the system from a state corresponding to $\tau=\tau_1$ to one corresponding to $\tau=\tau_2$, is consequently $W_{\tau_2}-W_{\tau_1}=\frac{2\pi^2 m e^4}{h^2}\left(\frac{1}{\tau_2}-\frac{1}{\tau_1}\right)$. If $\dots$ the amount of energy emitted is equal to $h\nu$, where $\nu$ is the frequency of the radiation, we get $\dots$ $\nu=\frac{2\pi^2 m e^4}{h^3}\left(\frac{1}{\tau_2}-\frac{1}{\tau_1}\right)$."} $\tau = 1, 2, \; \dots \; \infty$ enumerates states in atom of hydrogen. Bohr applies the \textit{Gaussian Centimeter-gram-second}, CGS, system of units, where other constants with modern values found in \textit{Wikipedia} are the electron mass $m \approx 9.10938356 \times 10^{-28}g$, \textit{gram}, and charge $e \approx 4.80320427 \times 10^{-10}statC$, \textit{statcoulomb}, or $Fr$, \textit{Franklin}, the Planck constant $h \approx 6.62607004 \times 10^{-27}\frac{cm^2 g}{s}$, where \textit{cm} and \textit{s} are \textit{centimeter} and \textit{second}, $\pi \approx 3.14159265358979323846$, and $\frac{2\pi^2 m e^4}{h^3} \approx 3.2898409 \times 10^{15} \frac{1}{s}$ or $Hz$, \textit{hertz}.

Bohr's formulas are not for the \textit{International System of Units}, SI, with $m \approx 9.10938356 \times 10^{-31} kg$, \textit{kilogram}, $e \approx 1.60217662 \times 10^{-19}C$, \textit{coulombs}, $h \approx 6.62607004 \times 10^{-34}\frac{m^2 kg}{s}$, where \textit{m} is \textit{meter}, and the \textit{dielectric constant of vacuum} $\epsilon_0 \approx 8.854187817 \times 10^{-12} \frac{F}{m}$, where $F$ is \textit{farad}. The factor for $\nu$ is $\frac{m e^4}{8 \epsilon_0^2 h^3}$. Second is common for CGS and SI. The factor gets the same value in $Hz$. Frequency can be converted to \textit{wavelength} $\lambda = \frac{c}{\nu}$, $c \approx 299,792,458 \frac{m}{s}$ is the \textit{speed of light in vacuum}. Enjoy computing wavelengths of emitted light for transitions between energy levels: $\lambda_{3\rightarrow2} \approx \frac{299,792,458}{ 3.2898409 \times 10^{15} \left(\frac{1}{2^2}-\frac{1}{3^2}\right)} \approx 656.1nm$, where \textit{nm} is \textit{nanometer} $=10^{-9}m$, $\lambda_{4\rightarrow 2} \approx 486.1nm$, $\lambda_{5\rightarrow 2} \approx 434.1nm$, $\lambda_{6\rightarrow 2} \approx 410.2nm$, $\lambda_{7\rightarrow 2} \approx 397.0nm$, $\lambda_{8\rightarrow 2} \approx 388.9nm$.
\begin{figure}
  \centering
  \includegraphics[width=130mm]{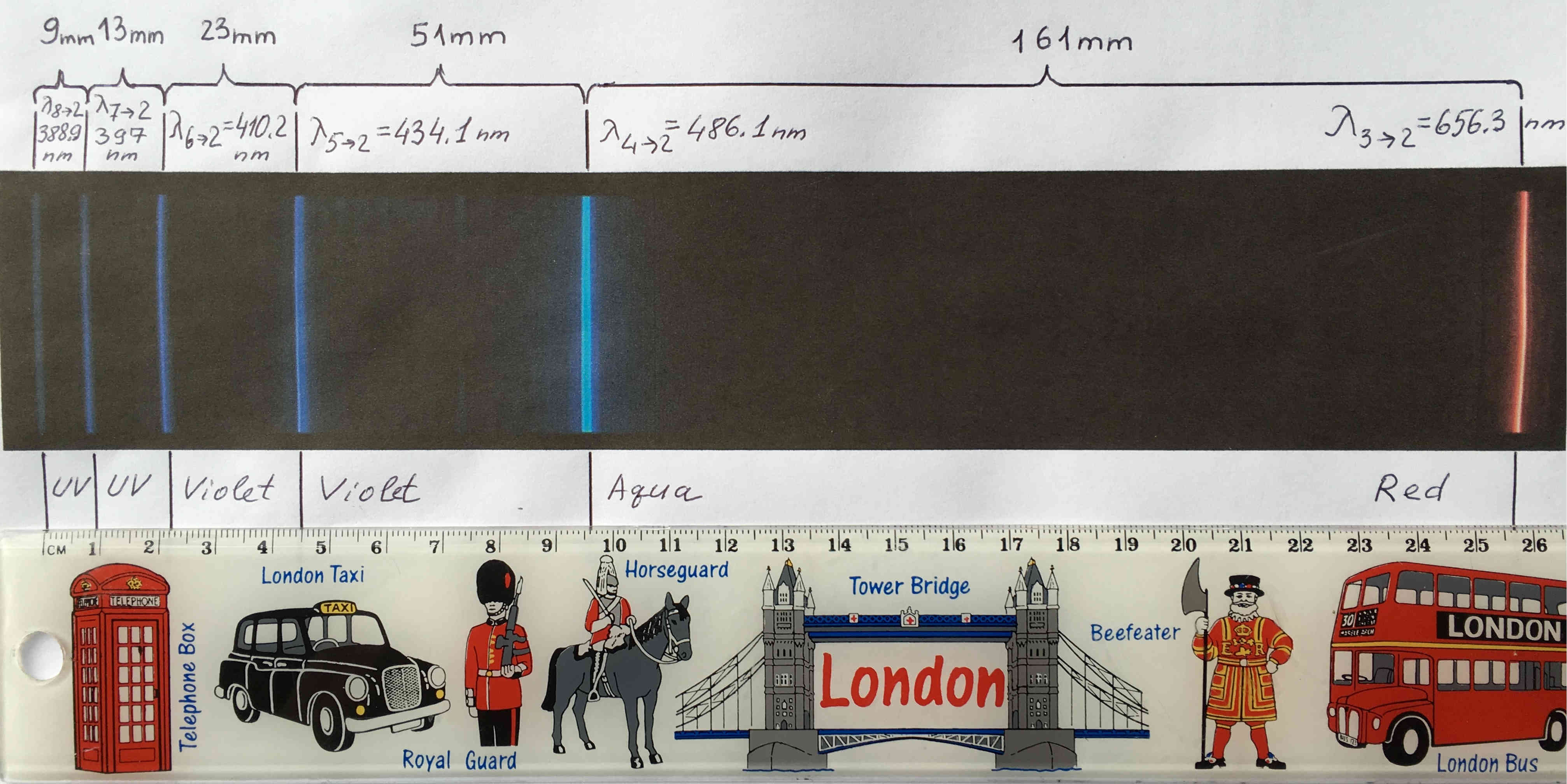}
  \caption[FigHSpectrum]
   {A few hydrogen emission spectrum lines in the Balmer series. The original image is from \url{https://en.wikipedia.org/wiki/Balmer_series}. Collage with souvenir ruler, wavelengths, lengths, and color descriptors is made by the author.}
  \label{FigHSpectrum}
\end{figure}
Distances and wavelengths differences on Figure \ref{FigHSpectrum} are proportional
\begin{displaymath}
\frac{\lambda_{3\rightarrow 2} - \lambda_{5\rightarrow 2}}{\lambda_{3\rightarrow 2} - \lambda_{4\rightarrow 2}} = \frac{\frac{1}{\frac{1}{2^2}-\frac{1}{3^2}}-\frac{1}{\frac{1}{2^2}-\frac{1}{5^2}}}{\frac{1}{\frac{1}{2^2}-\frac{1}{3^2}}-\frac{1}{\frac{1}{2^2}-\frac{1}{4^2}}}=\frac{64}{49}\approx 1.306 \approx \frac{161mm + 51mm}{161mm} \approx 1.312.
\end{displaymath}

\paragraph{\textit{Maximum Profit Strategies}, MPS.} One can associate MPS with any time series of prices and transaction costs \cite{salov2007}, \cite{salov2008}, \cite{salov2011}, \cite{salov2011b}, \cite{salov2012}, \cite{salov2013}. A strategy can be expressed by a chain of integer numbers of contracts: \textit{buy} - positive, \textit{sell} - negative, and/or \textit{do nothing} - zero. Transactions are executed at corresponding prices and costs.
\begin{figure}
  \centering
  \includegraphics[width=150mm]{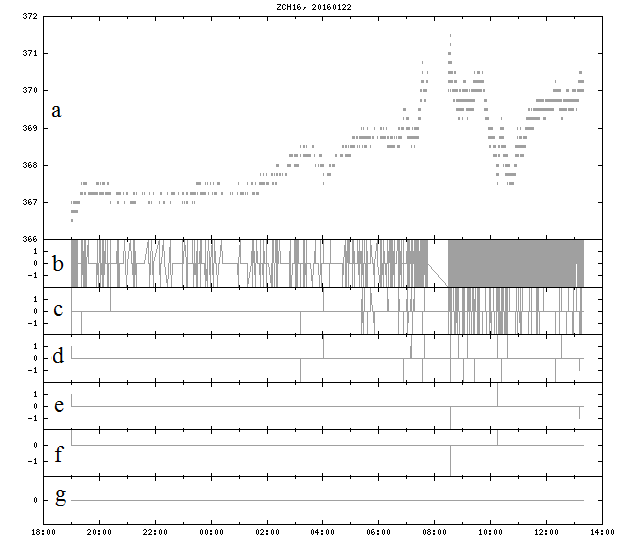}
  \caption[FigMpsSpectra]
   {ESH16 Time \& Sales, \url{http://www.cmegroup.com/}, a - transaction prices during the overnight and day time ranges of the trading session closed on Friday January 22, 2016, and the maximum profit strategies for the transaction costs per contract per transaction: b - \$4.99, c - \$12.49, d - \$24.99, e - \$49.99, f - \$74.99, and g - \$199.99 (degenerate case). Plots are done using custom C++  and Python programs and gnuplot \url{http://www.gnuplot.info/}.}
  \label{FigMpsSpectra}
\end{figure}
They can be drawn as vertical segments from do nothing zero level: up - buy, and down - sell. For a time interval the number of \textit{best} transactions depends on the cost per contract per transaction, Figure \ref{FigMpsSpectra}. Starting from high costs all round-trip trades, pairs of offsetting transactions, lose and the MPS is a \textit{degenerate do nothing strategy}, always available. Introduction and details of MPS are in \cite{salov2007} and \cite{salov2013}. Images of MPS resemble Balmer series and emphasize obvious discreteness of trading. It follows from discreteness of prices, costs, transaction times, and local price maximums and minimums. MPS and \textit{optimal trading elements}, OTE, \cite{salov2011}, \cite{salov2011b}, \cite{salov2012}, \cite{salov2013} are objective market properties.

\paragraph{Quantum, Quanta, Kvant, Quant.} Nobel Lectures in physics \cite{planck1920}, \cite{einstein1921}, \cite{bohr1922}, \cite{millikan1923}, \cite{broglie1929}, \cite{heisenberg1932}, \cite{schrodinger1933}, \cite{dirac1933} apply "quantum" and plural "quanta" to denote the minimum amount of a physical entity. Planck writes \textit{"elementary quantum of action"}; \textit{"\dots Einstein \dots pointed out that the introduction of the energy quanta, determined by the quantum of action \dots"}; \textit{"a light quantum or photon"} \cite{planck1920}. Bohr \cite[p. 13]{bohr1922} says \textit{"so-called energy-quanta"}. Einstein mentions \textit{"quantum problems"} \cite[p. 485]{einstein1921}. de Broglie references \textit{"the strange quantum concept introduced by Planck in 1900"} \cite[p. 244]{broglie1929} and explains \textit{"\dots it had to be assumed on the contrary that it emits equal and finite quantities, quanta. The energy of each quantum \dots is equal to $h\nu$"}  \cite[p. 245]{broglie1929}. Heisenberg presents \textit{"quantum mechanics"} and cites \textit{"Einstein's hypothesis of light quanta"} \cite[p. 290]{heisenberg1932}. Schrodinger discusses \textit{"the so-called quantum conditions and quantum postulates"} \cite[p. 309]{schrodinger1933}. Dirac analyses \textit{"quantum equations"} \cite[p. 322]{dirac1933}.

In contrast with "quantum" and "quanta" Millikan coins in the title of the lecture \textit{"the light-quant"} \cite[p. 54]{millikan1923}. Typo? Hardly. On page 64 we read \textit{"the impact between a light-quant and a free electron"}. Accordant "KVANT" but not "quantum" was borrowed by Russian language. The cover page of the first 1970 issue of the popular in Russia journal "Kvant" contained the formula $E=h\nu$. When the author first time in 1990th had heard \textit{quant} with respect to modeling and programming financial software, he did not think about \textit{quantitative analysis} but \textit{kvant}, \textit{quantum mechanics}, \textit{quantum chemistry}. Derman has popularized "quant" \cite{derman2004}. He draws parallels with quantum mechanics, mentions Planck, Einstein, Schrodinger, and extends the meaning:  \textit{Quants and their cohorts practice "financial engineering" - an awkward neologism coined to describe the jumble of activities that would better be termed quantitative finance}.

A definition found on Web: \textit{"a highly paid computer specialist with a degree in a quantitative science, employed by a financial house to predict the future price movements of securities, commodities, currencies, etc"}. This sounds too mercantile. Following to this logic a prestigious award of \textit{Risk} magazine "Quant of the Year" should be granted to a quant with the highest annual income.

Keeping in mind the "kvant" interpretation, when the author sees "quant" in a resume, he wants to ask "What is your frequency"? A smile or bewilderment accompanying the answer builds statistics how the word "quant" is understood.

Figure \ref{FigMpsSpectra} illustrates the "quantum properties" of the intra-day trading. In contrast with yearly, monthly, daily, hourly, etc. price bars, the a-, b-, c-increments are \textit{indecomposable} further - \textit{elementary}. They are atoms constituting Time \& Sales. In contrast with indistinguishable atoms these have random properties. The minimal non zero price fluctuation of a U.S. Treasury Bond future contract is equivalent to \$31.25. \textit{This decent lunch can be neither ignored nor divided. It is a quantum or kvant}.

\section{Prices vs. increments sample distributions}

From Equation \ref{EqPriceAsSum} $P_i$ is  the sum of $P_1$ and $i-1$ random b-increments. If the latter are i.i.d. random variables with variance $\sigma^2$, then $P_i$'s variance is $(i-1)\sigma^2$. Each $P_i$ in a session is from \textit{own distribution}. \textit{The number of random i.i.d. summands differs for each $P_i$ and the latter cannot be i.i.d. random variables}.

The price histograms can be plotted, Figure \ref{FigESH16_Distrib_20160122} (bottom), and are the basis for the \textit{Market Profile} \cite{salov2011}. The vertical lines illustrate\textit{value area}. The central is the mean price. The left and right surrounding lines correspond to 15 percents of area counted from the left and right sides of a histogram.

If $P_i$ from one session would be i.i.d. random variables, then b-increments could not, in general, have this property. This rises \textit{chicken and egg question} \cite{salov2011}, \cite[pp. 34 - 36]{salov2013}: what is primary the $N$ sequential \textit{observed} transaction prices $P_1,$ $P_2$, $\dots$, $P_i$, $\dots$, $P_N$ arriving with Time \& Sales or their \textit{computed} $N-1$ increments $\Delta P_2$, $\dots$, $\Delta P_i$, $\dots$, $\Delta P_{N}$?

\textit{Sample moments} are \textit{symmetric functions} of a sample: reordering values does not change statistics. Reordering $P_i$ keeps the moments intact but changes $\Delta P_i$ and their statistics. Reordering $\Delta P_i$ with $P_1$ intact does not affect their moments but modifies $P_i$ and their statistics. The sample mean of prices is
\begin{equation}
\label{EqSamplePDeltaP}
\begin{split}
a_1^P=\frac{\sum_{i=1}^{i=N}P_i}{N}=\frac{\sum_{i=1}^{i=N}(P_1 + \sum_{k=2}^{k=i}\Delta P_k)}{N}=P_1+\frac{\sum_{i=1}^{i=N}\sum_{k=2}^{k=i}\Delta P_k}{N}=\\
P_1+\frac{\sum_{i=2}^{i=N}(N-(i-1))\Delta P_i}{N}=P_1+(N-1)a_1^{\Delta P}-\frac{\sum_{i=2}^{i=N}(i-1)\Delta P_i}{N}=\\
P_1+\frac{N^2-1}{N}a_1^{\Delta P}-\frac{\sum_{i=2}^{i=N}i\Delta P_i}{N},
\end{split}
\end{equation}
where $a_1^{\Delta P}=\frac{\sum_{i=2}^{i=N}\Delta P_i}{N-1}=\frac{P_N-P_1}{N-1}$ is the sample mean of price increments. The order of $\Delta P_i$ in the sum matters because of the weights $\frac{i}{N}$. If $\Delta P_i$ are i.i.d. with the probability measure $\phi_{\Delta P}$ and mean $\alpha_1^{\Delta P}$ $\forall i \in [2, N]$, then the mathematical expectations are $E_{\phi_{\Delta P}}(a_1^{\Delta P}) = \alpha_1^{\Delta P} = E_{\phi_{\Delta P}}(\Delta P_i)$. The sample mean is an unbiased estimate of the population mean. This yields
\begin{equation}
\label{EqESamplePEDeltaP}
\begin{split}
E_{\phi_{\Delta P}}(a_1^P)=P_1+\frac{N^2-1}{N}\alpha_1^{\Delta P}-\frac{\alpha_1^{\Delta P}}{N} \sum_{i=2}^{i=N}i=\\
P_1+\frac{\alpha_1^{\Delta P}}{N}\left(N^2-1 - \frac{N(N+1)-2}{2}\right)=P_1+\frac{N-1}{2}\alpha_1^{\Delta P}.
\end{split}
\end{equation}
Let us notice that $\sum_{i=2}^{i=N}i\Delta P_i=2P_2-2P_1+3P_3-3P_2+\dots + NP_N-NP_{N-1}=-P_1-Na_1^P+(N+1)P_N$. Therefore, if, in contrast, $P_i$ are i.i.d. with the probability measure $\phi_P$, then the mathematical expectations are $E_{\phi_P}(a_1^P)=\alpha_1^P=E_{\phi_P}(P_i)$ $\forall i \in [1, N]$ and $E_{\phi_P}(a_1^{\Delta P})=0$.

\section{Price limits}

Corn futures price limits are known in advance. The limits and expanded limits change in time: "Corn Futures Contracts Specs" \url{http://www.cmegroup.com/trading/agricultural/grain-and-oilseed/corn_contract_specifications.html} and "CBOT Rulebook, Chapter 10 Corn Futures" \url{http://www.cmegroup.com/rulebook/CBOT/II/10/10.pdf}. On Friday April 1, 2016 the May 2016 corn futures ZCK16 settlement price was $P_S=354.00$ cents per bushel. The price limit for Monday April 4, 2016 was set to $L_{ZC}=\$0.25$ per bushel. The \textit{up} and \textit{down limit prices} for a next session are computed adding and subtracting $L_{ZC}$ from the previous settlement price. Therefore, the limit prices for April 4, 2016 were $P_U=379.00$ and $P_D=329.00$. If a corn contract is traded after the second business day of an expiration month, then the limits are not applied. If the price reaches the limit, then for a next session the expanded limit is set to \$0.40 per bushel.

The number of equidistant price levels in the limited lattice distribution is equal to $\frac{P_U-P_D}{\delta_{ZC}}+1=\frac{P_S + L_{ZC} - (P_S - L_{ZC})}{\delta _{ZC}}+1=\frac{2L_{ZC}}{\delta_{ZC}}+1=\frac{2\times \$0.25}{\$0.0025}+1=201$. Next session prices are $P_k = P_S - L_{ZC} + k \times \delta_{ZC}$, where $k=0, \dots, \frac{2L_{ZC}}{\delta_{ZC}}$.

Due to price limits, conditional probabilities $\mathbb{P}\{\Delta P_{i+1}>0|P_i=P_U\} = \mathbb{P}\{\Delta P_{i+1}<0|P_i=P_D\}=0$. \textit{This implies that $\Delta P_i$ are not i.i.d.}

\section{Action, transaction, trade, size, volume}

\textit{Actions} expressed by numbers of contracts are sequential elements of strategies. Buy (positive) and sell (negative) actions are \textit{transactions}. Do nothing (zero) action is not a transaction. If a chain of actions with zero sum contains transactions, then buys and sells can be combined in \textit{round trip trades}, each with zero \textit{net action}. A strategy with zero net action does not change the number of contracts in a \textit{trading position}. Transactions and round-trip trades relate to a single account. The absolute value of an action is \textit{size} or \textit{volume}.

The numbers of bought and sold contracts are equal for any Time \& Sales tick. \textit{Trade ticks}, with non-zero size, are trades combining opposite transactions. A \textit{limit order} in an electronic \textit{trading book} can be matched with an offsetting order sent from the same account. For the account it would be a loss of commissions and exchange fees per contract per round-trip trade times the number of contracts. More often ticks combine transactions from different accounts.

The speed and size of arriving ticks characterize \textit{liquidity} \cite{salov2013}. News, seasons, weather, overnight and day ranges, and contract evolution affect them. Statistics of waiting times, a-increments, change from session to session.

\begin{figure}
  \centering
  \includegraphics[width=110mm]{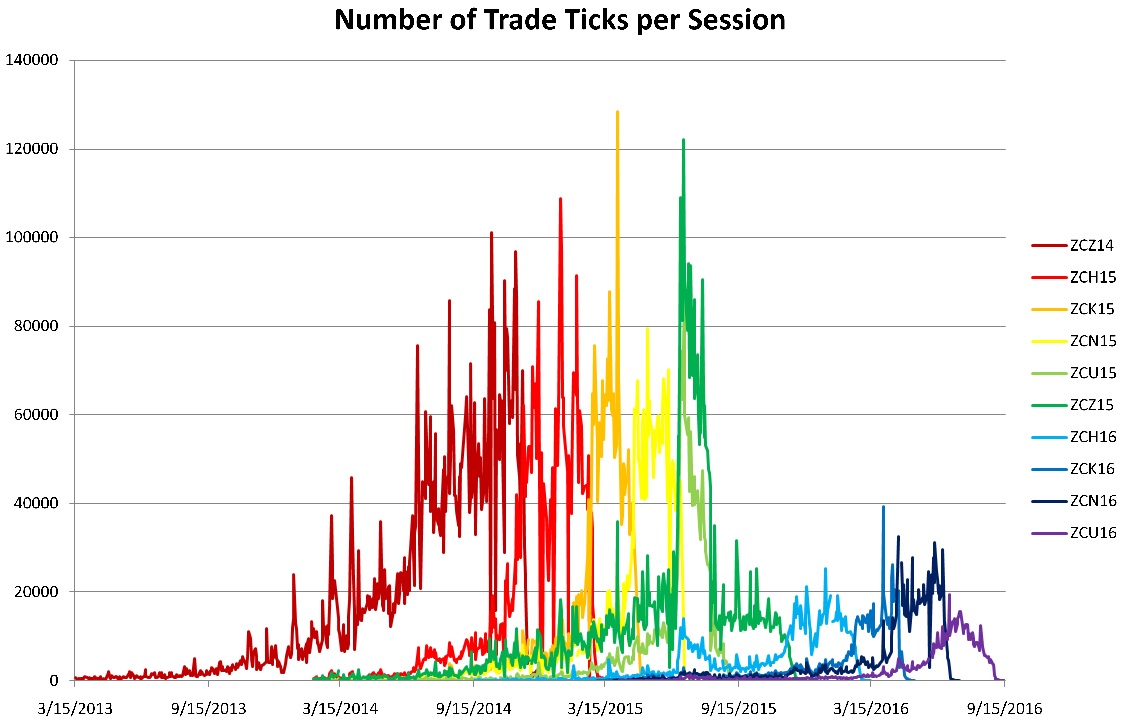}
  \caption[FigN]
   {Time \& Sales \url{http://www.cmegroup.com/} for 3507 sessions of ZCZ14, ZCH15, ZCK15, ZCN15, ZCU15, ZCZ15, ZCH16, ZCK16, ZCN16, and ZCU16 traded between Friday March 15, 2013 and Wednesday September 14, 2016. Microsoft Excel Chart.}
  \label{FigN}
\end{figure}

\begin{figure}
  \centering
  \includegraphics[width=110mm]{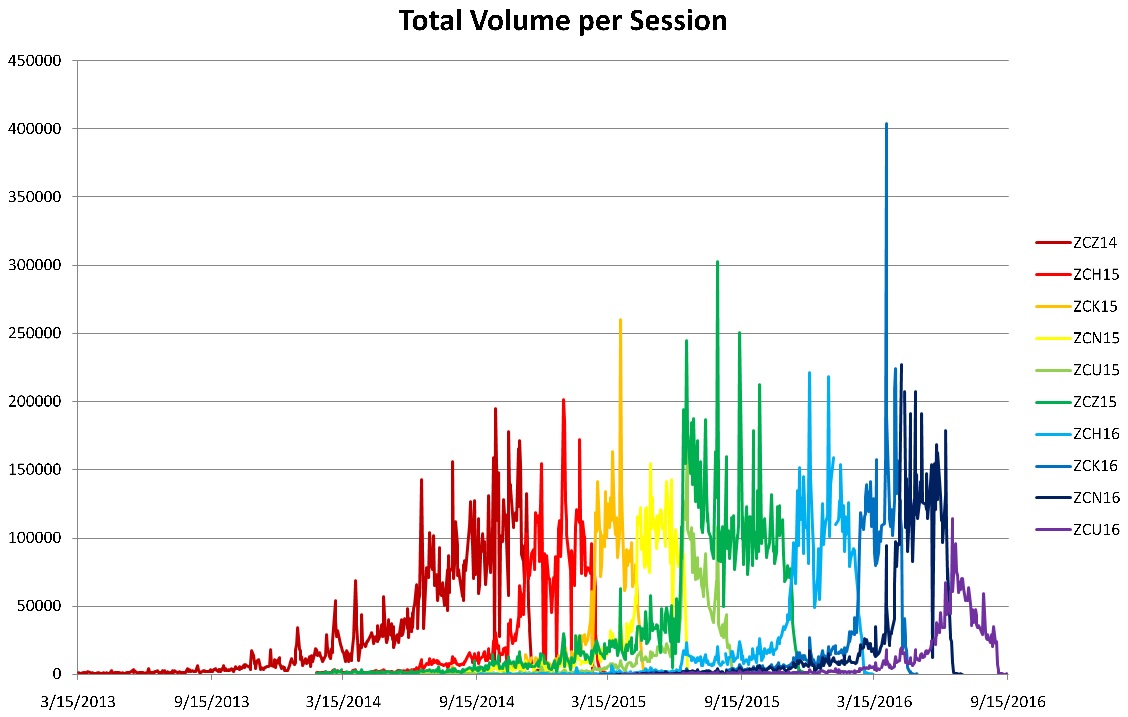}
  \caption[FigV]
   {Time \& Sales \url{http://www.cmegroup.com/} for 3507 sessions of ZCZ14, ZCH15, ZCK15, ZCN15, ZCU15, ZCZ15, ZCH16, ZCK16, ZCN16 and ZCU16 traded between Friday March 15, 2013 and Wednesday September 14, 2016. Microsoft Excel Chart.}
  \label{FigV}
\end{figure}

\begin{figure}
  \centering
  \includegraphics[width=100mm]{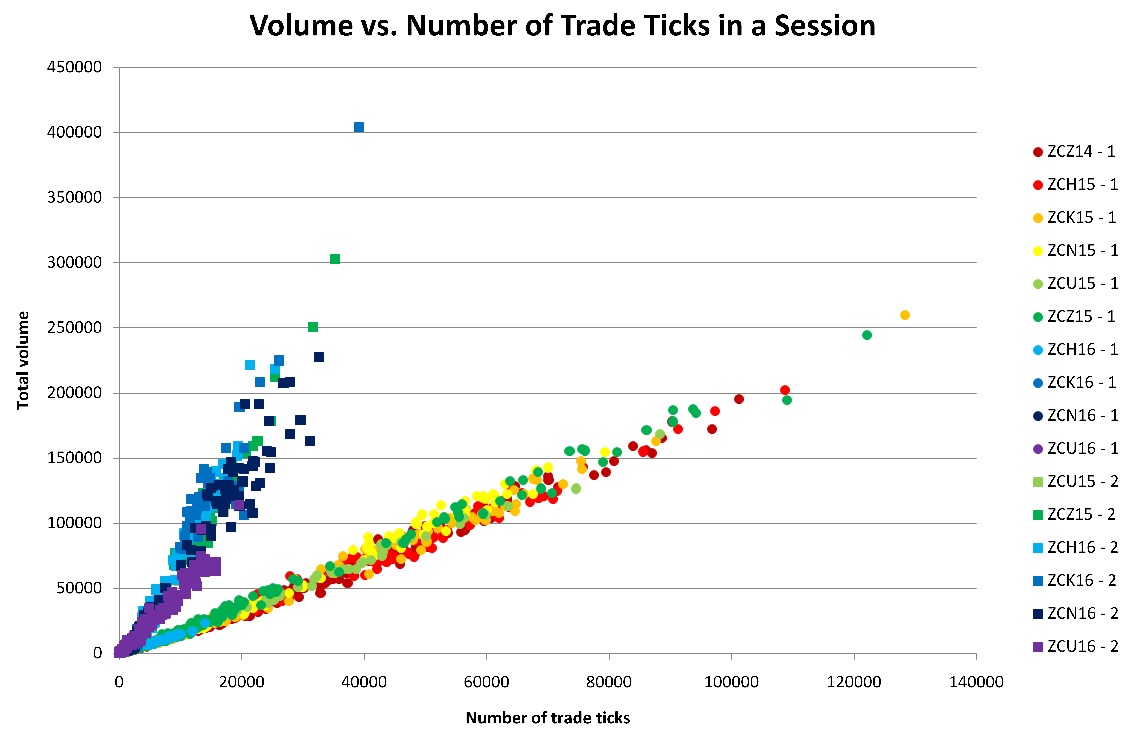}
  \caption[FigVN]
   {Time \& Sales \url{http://www.cmegroup.com/} for 3507 sessions of ZCZ14, ZCH15, ZCK15, ZCN15, ZCU15, ZCZ15, ZCH16, ZCK16, ZCN16, and ZCU16 traded between Friday March 15, 2013 and Wednesday September 14, 2016. Cluster 1 with smaller slope is for sessions prior Friday August 7, 2015. Microsoft Excel Chart.}
  \label{FigVN}
\end{figure}

\paragraph{Number of trade ticks $N$ and total volume $V$ from session to session.} The smallest size of a trade tick is one and $0 \le N \le V$. $N$ and $V$ reflect trading activity changing both \textit{systematically} and \textit{randomly}, Figures \ref{FigN}, \ref{FigV}. Maximums are systematically reached, when a corn contract becomes \textit{nearby}. However, during short time intervals $N$ and $V$ look random. Their distributions change in time. Mixing all values in one sample yields a big standard deviation exceeding the mean. For instance, in 431 sessions collected between Friday March 15, 2013 and Friday December 12, 2014 sample means for ZCZ14 $V$ and $N$ are 29359 and 17271, while sample standard deviations  are 39345 and 21436.
\paragraph{Total volume vs. number of trade ticks.} If $N=0$, then $V=0$. $V$ and $N$ were computed for 3507 sessions of ZCZ14 December, ZCH15 March, ZCK15 May, ZCN15 July, ZCU15 September, ZCZ15, ZCH16, ZCK16, ZCN16, and ZCU16. 2559 points prior and 948 since Friday August 7, 2015 form two clusters, Figure \ref{FigVN}.
The Microsoft Excel Data Analysis Regression with zero intercept yields $V=(1.828 \pm 0.006)N$, correlation coefficient $R=0.996$, $n=2559$ (prior) and $V=(7.12 \pm 0.09)N$, $R=0.980$, $n=948$ (since). In both cases the confidence probability for slope estimates is 95\%. The reported zero $P$\textit{-value} and $F$\textit{-significance} emphasize accuracy of slopes and  strength of regressions. The ratio of slopes after summing relative errors is $\frac{7.12}{1.828}\approx 3.89 \pm 0.06$.
\begin{figure}
  \centering
  \includegraphics[width=120mm]{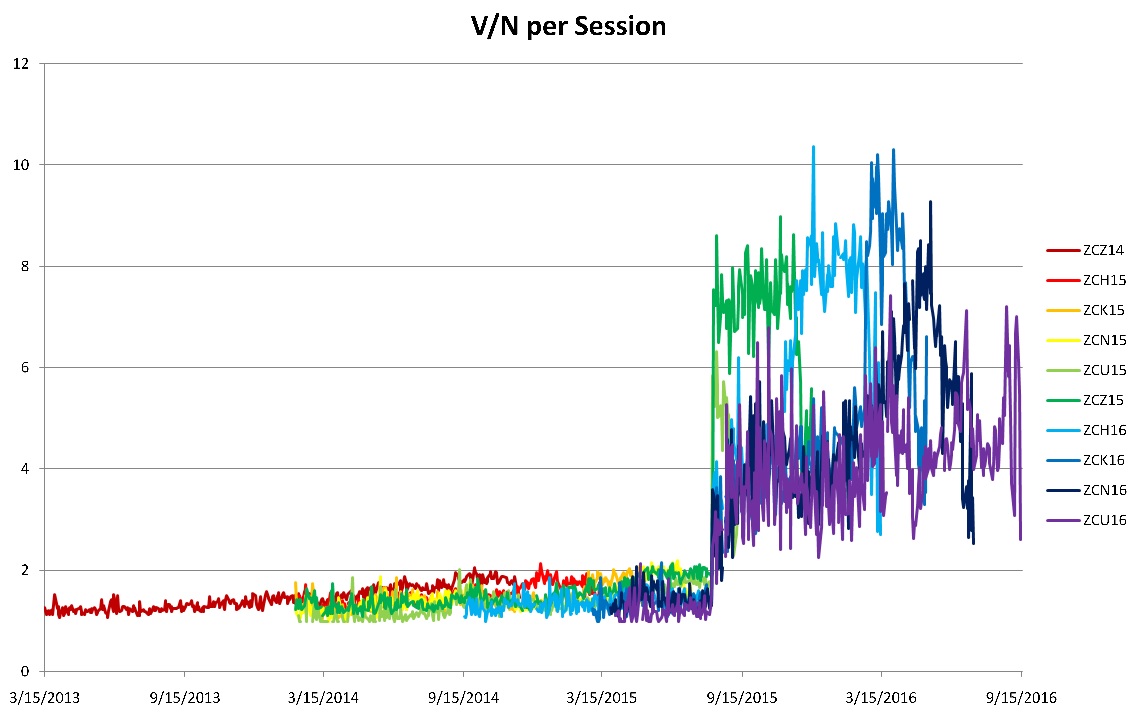}
  \caption[FigR]
   {Time \& Sales \url{http://www.cmegroup.com/} for 3507 sessions of ZCZ14, ZCH15, ZCK15, ZCN15, ZCU15, ZCZ15, ZCH16, ZCK16, ZCN16, and ZCU16 traded between Friday March 15, 2013 and Wednesday September 14, 2016. Microsoft Excel Chart.}
  \label{FigR}
\end{figure}

Figure \ref{FigR} demonstrates that ratio $\frac{V}{N}$ since August 7, 2015 was concentrating at the levels 4 and 8. This is not obvious from Figure \ref{FigVN}, where 948 points in the cluster with greater slope get greater variance due to mixing points with small and big slopes. During evolution of the 10 contracts the visible average levels are 1.2, 1.8, 4, 8.

\paragraph{Sharpe increase of $\frac{V}{N}$ ratios} on August 7, 2015 synchronous for corn contracts appears as decreasing the number of trade ticks with the total volume remaining at the same levels. For liquid contracts of that time ZCU15/ZCZ15 on August 4, 5, and 6, 2015 $N_{ZCU15}/N_{ZCZ15}$ were 26183/47692, 25313/46890, 24249/43508 and $V_{ZCU15}/V_{ZCZ15}$ were 44029/91468, 40558/87930, 38205/84789 or as mean values with $95\%$ confidence intervals for three points $(25.2 \pm 4.2) \times 10^3/(46.0 \pm 9.6) \times 10^3$ and $(41 \pm 13) \times 10^3/(88 \pm 14) \times 10^3$. The total volumes on August 7, 2015 were 40558/84860 - within the confidence intervals, while the numbers of trade ticks were 6845/11248 or 3.7 - 4.1 times less than mean $N$. \textit{This looks as a change in the underlying trading systems and reporting rather than decreasing trades' activity measured by $N$. We shall prefer simulating the more stable total volume in time dependencies crossing August 7, 2015.}

\section{Systematic evolution of volume}

\begin{figure}[!h]
  \centering
  \includegraphics[width=120mm]{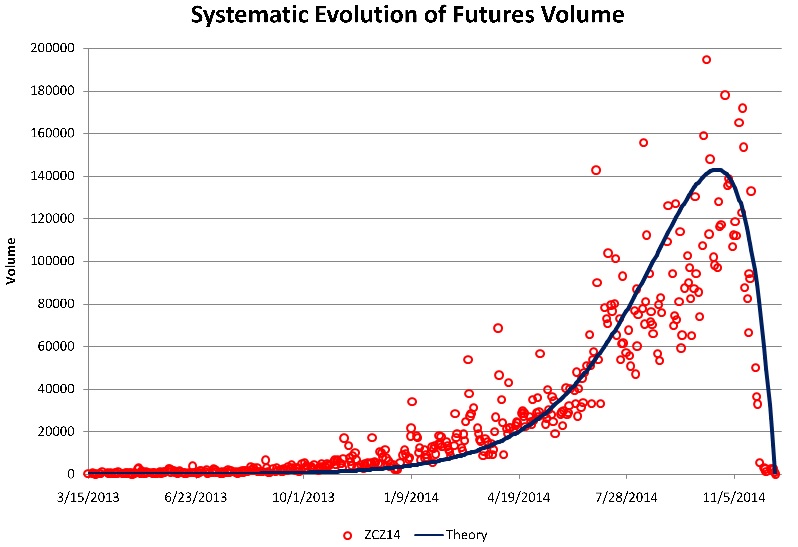}
  \caption[FigZCZ14Volume]
   {Total volume of ZCZ14, Friday March 15, 2013 - Friday December 12, 2014. $T$ = Saturday December 13, 2014 (the day after termination of trading) - $T_0$ = Wednesday December 13, 2012 (the day before listing) = 730 calendar days, $0 \le \tau $ = date - 12/13/2012 $ \le 730$, $V=4 \times 10^{-5}(730-\tau)\tau e^{0.017\tau}.$}
  \label{FigZCZ14Volume}
\end{figure}

Systematic evolution of total daily volume $V$ and \textit{open interest} for a contract is known to traders \cite[pp. 42 - 55, Figures 5-1, 5-3, 5-4, 5-5]{shaleen1991}: a slow long growth is followed by an accelerating quick maximum and fast drop to zero at expiration, Figures \ref{FigV}, \ref{FigZCZ14Volume}. In contrast with stocks and foreign exchange, FX, futures expire. Contracts specifications describe \textit{termination of trading} for ZC: \textit{"the business day prior to the 15th calendar day of the contract month"}. Unexpected extraordinary events can influence on this rule.

Contract birth is less certain. For trading, a contract must be listed on an exchange such as the \textit{Chicago Board of Trade}, CBOT, - a \textit{Designated Contract Market}, DCM. It is listed, for instance, by \textit{self-certification} determined by the \textit{U.S. Commodity Futures Trading Commission}, CFTC, and the \textit{U.S. Securities and Exchange Commission}, SEC. However, DCM decides, which contracts are listed. For \textit{Eurodollar Futures} GE, 44 contracts are specified. For ZC the months H, K, N, U, Z are specified leaving uncertain a next birthday. On Tuesday May 31, 2016 ZCZ19 had the longest maturity but ZCU19 was not yet listed. Two years of life is a reasonable expectation for ZC.

On the day $T_0$, before listing a contract, and $T$, after termination of trading, $V=0$. The author has noticed that curves $V(t)=A(T-t)^B(t-T_0)^Ce^{D(t-T_0)}$ resemble systematic evolution of $V$ for $A > 0$, $B > 0$, $C > 0$, $D > 0$, $T_0 \le t \le T$, Figure \ref{FigZCZ14Volume}. $\tau = t - T_0$, $L=T-T_0$, and $L-\tau$ are contract age, lifespan, and time until expiration, where $0 \le \tau \le L$. Then,

\begin{equation}
\label{EqVolumeSession}
V(\tau)=A(L-\tau)^B\tau^Ce^{D\tau}, \; V(0)=V(L)=0,
\end{equation}
\begin{equation}
\label{EqDerivVolumeSession}
\frac{dV}{d\tau}=A(D(L-\tau)\tau + C(L-\tau) - B\tau)(L-\tau)^{B-1}\tau^{C-1}e^{D\tau}.
\end{equation}
A few values are $0 < C < 1$, $\lim_{\tau\rightarrow 0}\frac{dV}{d\tau}(\tau)=\infty$; $C=1$, $\frac{dV}{d\tau}(0)=AL^B$; $1 < C$, $\frac{dV}{d\tau}(0)=0$; $0 < B < 1$, $\lim_{\tau\rightarrow L}\frac{dV}{d\tau}(\tau)=-\infty$; $B=1$, $\frac{dV}{d\tau}(L)=-AL^Ce^{DL}$; $1 < B$, $\frac{dV}{d\tau}(L)=0$. For $0 < \tau < L$, $\frac{dV}{dt}=0$, if $D(L-\tau)\tau + C(L-\tau) - B\tau = 0$ with the root $\tau_{max}$ and $V_{max}$
\begin{equation}
\label{EqMaxVolumeSession}
\begin{split}
\tau_{max}=\frac{DL-C-B+\sqrt{(DL-C-B)^2+4DCL}}{2D},\\
V_{max}=\frac{A}{(2D)^{B+C}}(DL+M)^B(DL-M)^Ce^{\frac{DL-M}{2}},\\
M=C+B-\sqrt{(DL-C-B)^2+4DLC}.
\end{split}
\end{equation}
Multiplying and dividing the right side of Equation \ref{EqDerivVolumeSession} by $(L-\tau)\tau$ for $0 < \tau < L$ and accounting Equation \ref{EqVolumeSession} yields the differential equation
\begin{equation}
\label{EqDiffVolumeSession}
\begin{split}
\frac{dV}{d\tau}=\left(D+\frac{C}{\tau}-\frac{B}{L-\tau}\right)V(\tau)=G(\tau)V,\\
\frac{1}{V} \frac{dV}{d\tau} = G(\tau) = \left(D+\frac{C}{\tau}-\frac{B}{L-\tau}\right),
\end{split}
\end{equation}
where $G(\tau)$ is the \textit{relative growth rate} of $V$. This is a \textit{linear differential equation of the first order} $\frac{dy}{dx}+P(x)y=Q(x)$ \cite[pp. 92 - 98]{tenenbaum1963}. It is \textit{homogeneous} in the sense that $Q(x)=0$ and \textit{non-autonomous} because the right side $f(\tau, V)=G(\tau)V$ explicitly depends on age $\tau$ \cite[Chapter 3, \$27]{arnold2000}. The variables $\tau$ and $V$ are \textit{separable} $\frac{dV}{V}=G(\tau)d\tau$ \cite[p. 52]{tenenbaum1963}.

By the \textit{general Leibniz rule} for $n=0, 1, 2, \dots$, $\frac{d^{n+1}V}{d\tau^{n+1}}=(G(\tau)V(\tau))^{[n]} = \sum_{i=0}^{i=n}\frac{n!}{i!(n-i)!}G^{[n-i]}(\tau)V^{[i]}(\tau)$, where $G^{[0]}(\tau)=G(\tau)$, $V^{[0]}(\tau)=V(\tau)$. For $k=1, 2, \dots$, $G^{[k]}(\tau)=\frac{d^kG}{d\tau^k}$, $V^{[k]}(\tau)=\frac{d^kV}{d\tau^k}$. By \textit{mathematical induction}
\begin{equation}
\label{EqDiffG}
G^{[k]}(\tau)=\frac{d^kG}{d\tau^k}=k!\left((-1)^k\frac{C}{\tau^{k+1}} - \frac{B}{(L-\tau)^{k+1}}\right), \; k = 1, 2, \dots
\end{equation}
Indeed, setting $k=1$ in Equation \ref{EqDiffG} and differentiating expression in brackets of Equation \ref{EqDiffVolumeSession} yields identical $\frac{dG}{d\tau}=-\frac{C}{\tau^2}-\frac{B}{(L-\tau)^2}$. Let Equation \ref{EqDiffG} is valid for any $k$. For $(k+1)$: $(k+1)!\left((-1)^{(k+1)}\frac{C}{\tau^{(k+1)+1}} - \frac{B}{(L-\tau)^{(k+1)+1}}\right)$. Differentiating the right side of Equation \ref{EqDiffG} yields equivalent $k!(k+1)\left((-1)^{k+1}\frac{C}{\tau^{k+2}} - \frac{B}{(L-\tau)^{k+2}}\right)$. Further generalization assumes recursive application of the Leibniz rule to $V^{[i]}=\frac{d^iV}{d\tau^i}=(G(\tau)V(\tau))^{[i-1]}$. This and Equation \ref{EqDiffVolumeSession} ensure that for $i=1,2,\dots$ we get a differential equation of the order $i+1$ with the separable variables $\tau$ and $V$. The second, third, and fourth order derivatives are
\begin{equation}
\label{EqVDerivatives}
\begin{split}
\frac{d^2V}{d\tau^2}=\left(\frac{dG}{d\tau} + G^2\right)V,\\
\frac{d^3V}{d\tau^3}=\left(\frac{d^2G}{d\tau^2} + 3\frac{dG}{d\tau}G + G^3\right)V,\\
\frac{d^4V}{d\tau^4}=\left(\frac{d^3G}{d\tau^3} + 4\frac{d^2G}{d\tau^2}G + 6\frac{dG}{d\tau}G^2 + 3\left(\frac{dG}{d\tau}\right)^2 + G^4\right)V,
\end{split}
\end{equation}
where $G$ and derivatives of $G$ are given by Equations \ref{EqDiffVolumeSession} and \ref{EqDiffG}. A \textit{necessary condition} of the \textit{inflection point(s)} of $V$ is $\frac{d^2V}{d\tau^2}=0$ and, therefore, for $0 < \tau < L$ $\frac{dG}{d\tau} = -G^2$ or $C(L-\tau)^2+B\tau^2=(D\tau(L-\tau)+C(L-\tau)-B\tau)^2$. The latter can be solved analytically, as an \textit{algebraic equation of the fourth order}, numerically, or graphically. The graphic method is easier to apply after taking square roots of the positive left and right sides.
\paragraph{Cumulative Volume.} \textit{Population growth} expressed by a \textit{logistic curve} \cite{verhulst1838}, solutions of the \textit{Lotka-Volterra} autonomous differential equations \cite{lotka1920}, \cite{volterra1926}, describing \textit{predator-prey} interactions, solutions of \textit{more realistic and universal predator-prey Kolmogorov autonomous differential equations} \cite{kolmogorov1936}, \cite{kolmogorov1972} are \textit{integral curves}. In contrast, popular daily volumes of futures and stocks and open interests of futures are differential curves. They are \textit{step functions} $\frac{\Delta V_c}{\Delta \tau}(\tau)$, where $\Delta \tau$ is one day. What is denoted by $V$ is already the first derivative of the cumulative volume $V_c(\tau) = \int_0^{\tau}{A(L-x)^Bx^Ce^{Dx}dx}$, where $0 \le \tau \le L$.

Brokerage companies, exchanges, \textit{National Futures Association}, NFA, clearing houses are interested in great values of this integral due to collecting transaction fees from each trade and contract. The "HB0106 Revenue-Financial Transaction Tax Act" considered by 99th General Assembly State of Illinois can increase the number of interested sides. There is a concern that the Act can negatively influence on markets \cite{duffy2016}.

If $B$ is natural, then $(L-x)^B=\sum_{i=0}^{i=B}\frac{B!}{i!(B-i)!}L^{B-i}x^i(-1)^i$ is a polynomial with finite number of summands and
\begin{equation}
\label{EqCumulativeV}
V_c(\tau)=A\sum_{i=0}^{i=B}\frac{B!}{i!(B-i)!}(-1)^i L^{B-i}\int_0^{\tau}x^{C+i}e^{Dx}dx, \; 0 \le \tau \le L.
\end{equation}
In contrast, if $C$ is natural, then the substitution $y = L - x$ yields 
\begin{equation}
\label{EqCumulativeVy}
V_c(\tau)=Ae^{DL}\sum_{i=0}^{i=C}\frac{C!}{i!(C-i)!}(-1)^i L^{C-i}\int_{L-\tau}^Ly^{B+i}e^{-Dy}dy, \; 0 \le \tau \le L.
\end{equation}
Integrals in Equations \ref{EqCumulativeV} and \ref{EqCumulativeVy} can be expressed via the \textit{upper incomplete gamma function} $\Gamma(s,x)=\int_x^{\infty}t^{s-1}e^{-t}dt$. As long as $p \ne -2$, $p \ne -1$, $q \ne 0$,
\begin{equation}
\label{EqCommonIntegral}
\begin{split}
\int x^pe^{qx}dx = \frac{q^{-p-1}\Gamma(p+1,-qx)}{(-1)^p}+ Constant,\\
\int x^pe^{-qx}dx = -q^{-p-1}\Gamma(p+1,qx) + Constant,\\
\end{split}
\end{equation}
Then,
\begin{equation}
\label{EqCumulativeVC123}
\begin{split}
V_c(\tau, C=1) = Ae^{DL}D^{-B-2}\{LD[\Gamma(B+1,D(L-\tau))-\Gamma(B+1,DL)]-\\
-[\Gamma(B+2,D(L-\tau))-\Gamma(B+2,DL)]\};\\
V_c(\tau, C=2) = Ae^{DL}D^{-B-3}\{L^2D^2[\Gamma(B+1,D(L-\tau))-\Gamma(B+1,DL)]-\\
2LD[\Gamma(B+2,D(L-\tau))-\Gamma(B+2,DL)]+\\
+[\Gamma(B+3,D(L-\tau))-\Gamma(B+3,DL)]\};\\
V_c(\tau, C=3) = Ae^{DL}D^{-B-4}\{L^3D^3[\Gamma(B+1,D(L-\tau))-\Gamma(B+1,DL)]-\\
3L^2D^2[\Gamma(B+2,D(L-\tau))-\Gamma(B+2,DL)]+\\
+3LD[\Gamma(B+3,D(L-\tau))-\Gamma(B+3,DL)]-\\
-[\Gamma(B+4,D(L-\tau))-\Gamma(B+4,DL)]\};
\end{split}
\end{equation}
$\Gamma(s,x)$ can be computed in Microsoft Excel as EXP(GAMMALN($s,x$)) * (1 - GAMMADIST($x,s,1,$TRUE)).
\begin{figure}[!h]
  \centering
  \includegraphics[width=120mm]{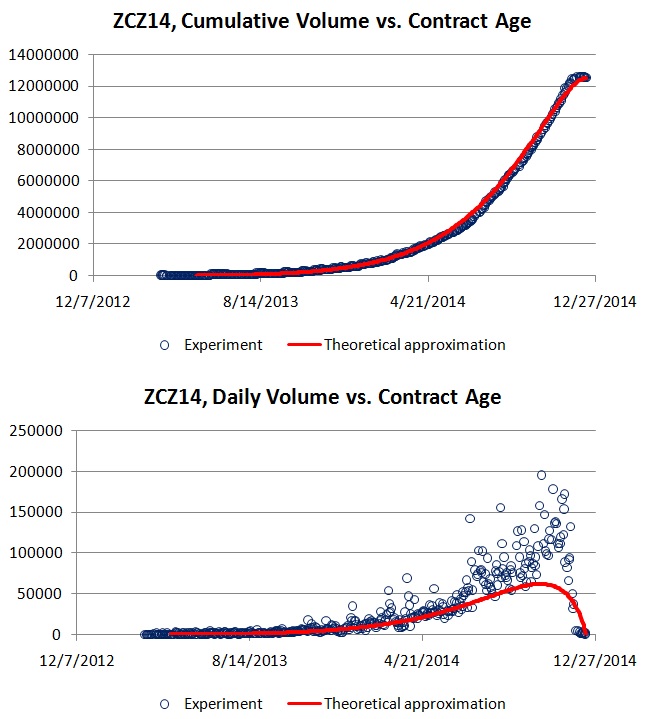}
  \caption[FigZCZ14VcV]
   {Total volume of ZCZ14, Friday March 15, 2013 - Friday December 12, 2014. $T$ = Saturday December 13, 2014 (the day after termination of trading) - $T_0$ = Wednesday December 13, 2012 (the day before listing) = 730 calendar days, $0 \le \tau $ = date - 12/13/2012 $ \le 730$, $V(\tau)=3.71 \times 10^{-3}(730-\tau)^{0.783037883}\tau e^{0.010327916\tau}$ (bottom), and $V_c(\tau) = \int_{92}^{\tau}V(t)dt, \; 92 = 03/15/2013 - T_0$ (top) computed using Equation \ref{EqCumulativeVC123} for $C = 1$.}
  \label{FigZCZ14VcV}
\end{figure}

Summation smooths random shocks of total daily volume $V$, Figure \ref{FigZCZ14VcV} (top). Fitting experimental values of $V_c$ by Equation \ref{EqCumulativeVC123} with $C=1$ is good. $A, B, D$ estimates reported on Figure \ref{FigZCZ14VcV} are obtained minimizing the maximum absolute difference between computed and observed values of $V_c$ - \textit{Chebyshev criterion}. Pafnuty Lvovich Chebyshev was not only a mathematician but a mechanic and cared that coordinates of moving and interacting parts of mechanisms predicted theoretically would not deviate from experimental values too much to cause damages \cite{chebyshev}. The $A, B, D$ fitting the integral curve $V_c(\tau)$ lower the differential curve $V(\tau)$, compare Figures \ref{FigZCZ14Volume} and \ref{FigZCZ14VcV} (bottom). This is because several daily volumes are missed and Equations \ref{EqCumulativeVC123} "integrate" non-contributing weekends and holidays. If Equation \ref{EqVolumeSession} fits $V$ and its integration is replaced with summation of the curve values at integer $\tau$ skipping weekends and holidays, then \textit{"the wolves have eaten much and the sheep have not been touched"}: both integral $V_c(\tau)$ and differential $V(\tau)$ curves can share the same $A$, $B$, $C$, $D$.

In general, the three factors $(L-\tau)^B \tau^Ce^{D\tau}$ prevent reduction of the $V_c(\tau)$ to the \textit{finite} number of incomplete gamma or \textit{incomplete beta functions} $\mathrm{B}(x; a, b) = \int_0^x t^{a-1}(1-t)^{b-1}dt$, $0 \le x \le 1$. The $e^{Dx}=\sum_{i=0}^{i=\infty}\frac{(Dx)^i}{i!}$ and $x=Lt$ yield
\begin{equation}
\label{EqVcInf}
\begin{split}
V_c(\tau) = \int_0^{\tau}A(L-x)^Bx^C\sum_{i=0}^{i=\infty}\frac{(Dx)^i}{i!}dx =\\
= A \int_0^{\tau} \sum_{i=0}^{i=\infty} \frac{D^i}{i!} (L-x)^Bx^{C+i}dx =\\
= AL^{B+C+1}\sum_{i=0}^{i=\infty} \frac{(DL)^i}{i!}\mathrm{B}(\frac{\tau}{L}; C+i+1, B+1).
\end{split}
\end{equation}
Since for $0 \le x \le L$, $0 \le D$, $0 \le B$, $0 \le C$, $|\frac{D^i}{i!} (L-x)^Bx^{C+i}| \le \frac{D^i}{i!} L^{B+C+i}$, and $\sum_{i=0}^{i=\infty} \frac{D^i}{i!} L^{B+C+i} = L^{B+C} \sum_{i=0}^{i=\infty} \frac{(DL)^i}{i!} = L^{B+C} e^{DL}$, we conclude that the middle series under the integral in Equations \ref{EqVcInf} \textit{converges uniformly} \cite[pp. 427 - 429, Weierstrass test]{fihtengoltz1970} and the integral of the sum can be replaced with the sum of the integrals \cite[pp. 437 - 438, Theorem 6]{fihtengoltz1970} expressed by $\mathrm{B}(x;a,b)$. The latter can be computed in Microsoft Excel: $\mathrm{BETADIST}(x, a, b) * \mathrm{EXP}(\mathrm{GAMMALN}(a) + \mathrm{GAMMALN}(b) - \mathrm{GAMMALN}(a + b))$.

\begin{figure}[!h]
  \centering
  \includegraphics[width=70mm]{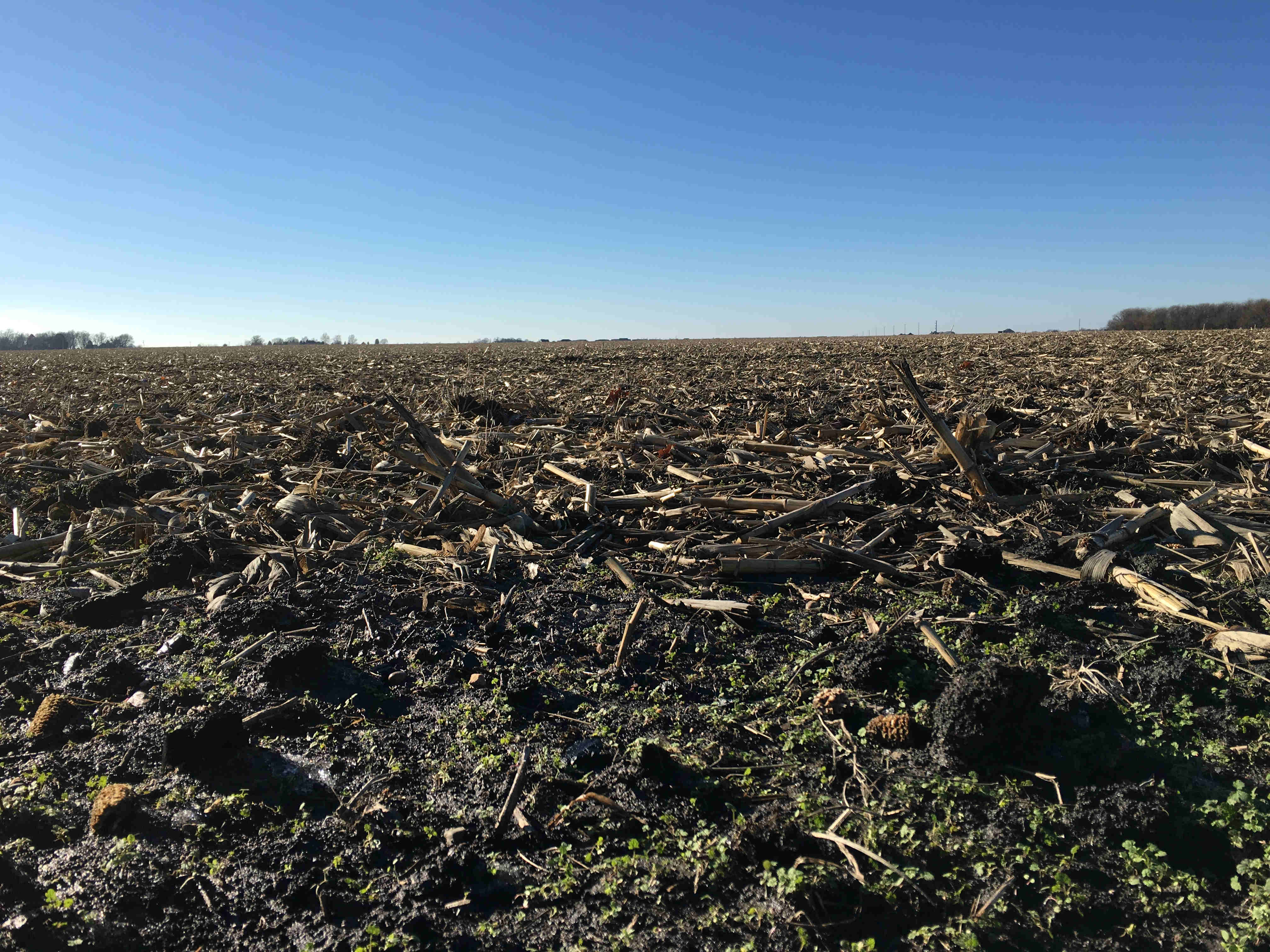}
  \caption[FigEmptyField]
   {Saturday January 2, 2016, Savoy, Illinois. A field after 2015 corn crop.}
  \label{FigEmptyField}
\end{figure}

\paragraph{The Life Strength Function} is a suitable name for Equation \ref{EqVolumeSession}. The number of publications per year by a scientist or journalist, daily biomass of corn on a field during growth and harvest, Figures \ref{FigEmptyField}, \ref{FigCornField}, \ref{FigNewCornField}, \ref{FigNewCorn}, daily attractiveness of a woman and man, daily health of an organism might look similar. \textit{The latter two are waiting  a measure.}

In contrast with a hypothetic health function of age, for a futures contract the expiration date is known in advance. Only extraordinary events can change it. Asymptotic, $\tau \rightarrow \infty$,  dependencies of properties cannot satisfy cases, where \textit{the dates of birth and death are known in advance}. Let us review two examples: time dependencies of the body weight $W$ \cite{bertalanffy1957} and intermediate compound concentration [B] in two sequential chemical reactions of the first order \cite{semiokhin1995}.

Bertalanffy: \textit{"Animal growth can be considered a result of a counteraction of synthesis and destruction, of the anabolism and catabolism of the building materials of the body"}. Differential equation \cite[p. 223, Equation (5)]{bertalanffy1957} $\frac{dW}{dt}=\eta W^m - k W^n$ for $n = 1$ is solved by \cite[p. 224, Equation (6)]{bertalanffy1957} $W = \{\frac{\eta}{k} - [\frac{\eta}{k} - W_0^{(1-m)}]e^{-(1-m)kt}\}^{\frac{1}{1-m}}$ with $W_0 = $ weight at time $t = 0$; \textit{"$\eta$ and $k$ are constants of anabolism and catabolism respectively, and the exponents $m$ and $n$ indicate that the latter are proportional to some power of the weight $W$"}. To map $\frac{dW}{dt}$ to $V(\tau) =\frac{dV_c}{d\tau}$, we set $W_0=0$ yielding $W=\{\frac{\eta}{k}[1-e^{-(1-m)kt}]\}^{\frac{1}{1-m}}$, $\lim_{t \rightarrow \infty} W = \{\frac{\eta}{k}\}^{\frac{1}{1-m}}$, $\frac{dW}{dt}=\{\frac{\eta}{k}[1-e^{-(1-m)kt}]\}^{\frac{m}{1-m}}\eta e^{-(1-m)kt}$, $\frac{dW}{dt}(t = 0) = 0$, $\lim_{t \rightarrow \infty} \frac{dW}{dt} = 0$, and

\begin{math}
\frac{d^2W}{dt^2} =\dfrac{\eta k \left (\frac{\eta(1-e^{-k(1-m)t})}{k}\right )^{\frac{m}{1-m}} e^{-k(1-m)t} \left ((m-1)e^{k(1-m)t}+1 \right )}{e^{k(1-m)t}-1}.
\end{math}

$\frac{d^2W}{dt^2}=0$ at $t_{max} = \frac{\ln(\frac{1}{1-m})}{k(1-m)}$ and $\frac{dW}{dt}(t_{max}) = \{\frac{\eta m}{k}\}^{\frac{m}{1-m}}\eta (1-m)$.  Both $\frac{dW}{dt}$ and $V(\tau)$ are equal to zero at $t = 0$ and $\tau = 0$ and have a maximum. \textit{However, the $\frac{dW}{dt}$ asymptotically approaches zero with $t \rightarrow \infty$, while the daily trading volume $V(\tau)$ is equal to zero exactly after the contract expiration.}

For two sequential chemical reactions A $\xrightarrow[]{k_1}$ B $\xrightarrow[]{k_2}$ C, the concentration [B] = $\frac{k_1 a}{k_2 - k_1}(e^{-k_1t} - e^{-k_2t})$, where $k_1$ and $k_2$ are the constants of chemical reactions and at $t = 0$: [A] = $a$, [B] = 0, [C] = 0 \cite[p. 29, Equation (3.27)]{semiokhin1995}. The [B] has maximum at $\frac{d[\textrm{B}]}{dt} = 0$ and $t_{max} = \frac{\ln(\frac{k_2}{k_1})}{k_2 - k_1}$ \cite[p. 30, Equation (3.29)]{semiokhin1995}. Except the initial condition, [B] approaches zero only asymptotically with $t \rightarrow \infty$. \textit{Again, this is in contrast with $V(\tau)$ equal to zero exactly at $\tau = 0$ and $\tau = L$.}

The recent monograph reviews the following models of absolute and relative growth rates, AGR and RGR, \cite[Chapter 3]{panik2014}: 1) linear, 2) logarithmic reciprocal, 3) logistic, 4) Compertz, 5) Weibull, 6) negative exponential, 7) von Bertalanffy, 8) log-logistic, 9) Brody, 10) Janoschek, 11) Lundqvist-Korf, 12) Schumacher, 13) Hossfeld, 14) Stannard, 15) Schnute, 16) Morgan-Mercer-Flodin, 17) McDill-Amateis, 18) Levacovic I, 19) Levacovic III, 20) Yoshida I, 21) Sloboda, 22) monomolecular, 23) Chapman-Richards, 24) generalized Michaelis-Menten; models of yield-density curves \cite[Chapter 7]{panik2014}: 25) Shinozaki and Kira, 26) Holliday, 27) Farazadaghi and Harris, 28) Bleasdale and Nelder, 29) Bleasdale simplified equation, 30) variations using allometric equation, 31) expolinear growth equation of Goudriaan and Monteith, 32) beta growth function \cite{yin2003}, 33) asymmetric growth equations. From 1) - 33) only 32) is defined on a finite time interval. Let us show that Yin et al. equation \cite[pp. 362 - 363, Equation (7)]{yin2003} is a particular case of Equation \ref{EqVolumeSession}:
\begin{equation}
\label{EqYin}
\dfrac{dw}{dt} = c_m\left [ \left (\dfrac{t_e - t}{t_e - t_m} \right ) \left (\dfrac{t - t_b}{t_m - t_b} \right )^{\dfrac{t_m - t_b}{t_e - t_m}} \right ]^\delta, \; t_b \le t \le t_e, \; t_b < t_m < t_e.
\end{equation}
Indeed, if $L - \tau = t_e - t$, $\tau = t - t_b$, $A = c_m \left [(t_e - t_m)(t_m - t_b)^{\frac{t_m - t_b}{t_e - t_m}} \right ]^{-\delta}$, $B = \delta$, $C = \frac{t_m - t_b}{t_e - t_m}\delta$, $D = 0$, then the right side of Equation \ref{EqVolumeSession} is transformed to the right side of equation above. The $e^{D\tau}$ in Equation \ref{EqVolumeSession} becomes an additional factor of asymmetry depending on the sign and value of $D$. Applying the L'Hopital's rule and $C$ and $B$ values, we get from Equation \ref{EqMaxVolumeSession} $\lim_{D \rightarrow 0} \tau_{max} = \frac{CL}{C + B} = t_m - t_b$.

\textit{Equation \ref{EqVolumeSession} is a flexible function suitable for simulation of systematic evolution of futures daily trading volume.}

\begin{figure}[!h]
  \centering
  \includegraphics[width=70mm]{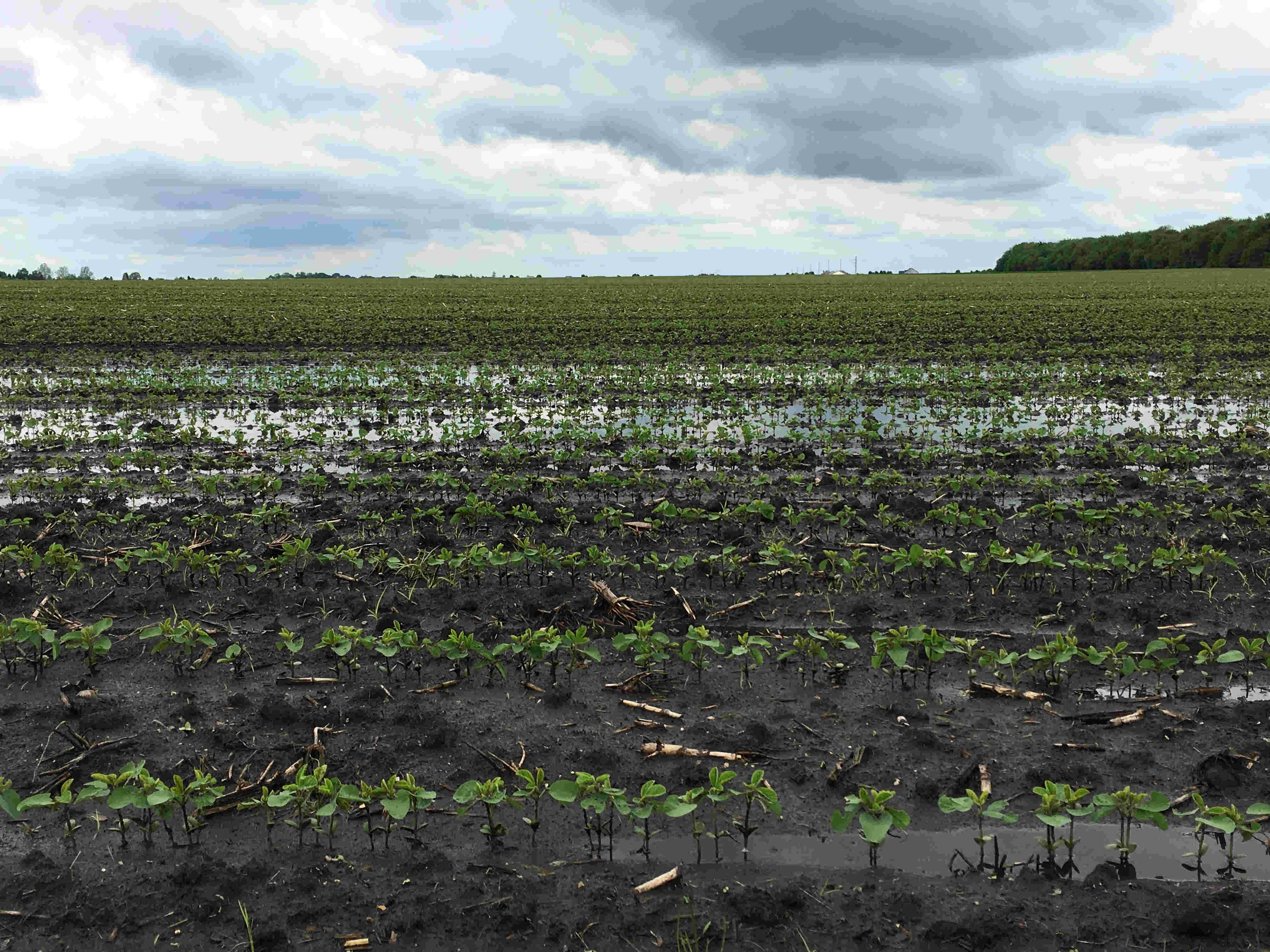}
  \caption[FigSoybeanField]
   {Sunday June 5, 2016, Savoy, Illinois. Risk Factors: a) decision to grow soybeans instead of corn, b) temperature 70F, c) rains, d) minor flooding. }
  \label{FigSoybeanField}
\end{figure}
\begin{figure}[!h]
  \centering
  \includegraphics[width=70mm]{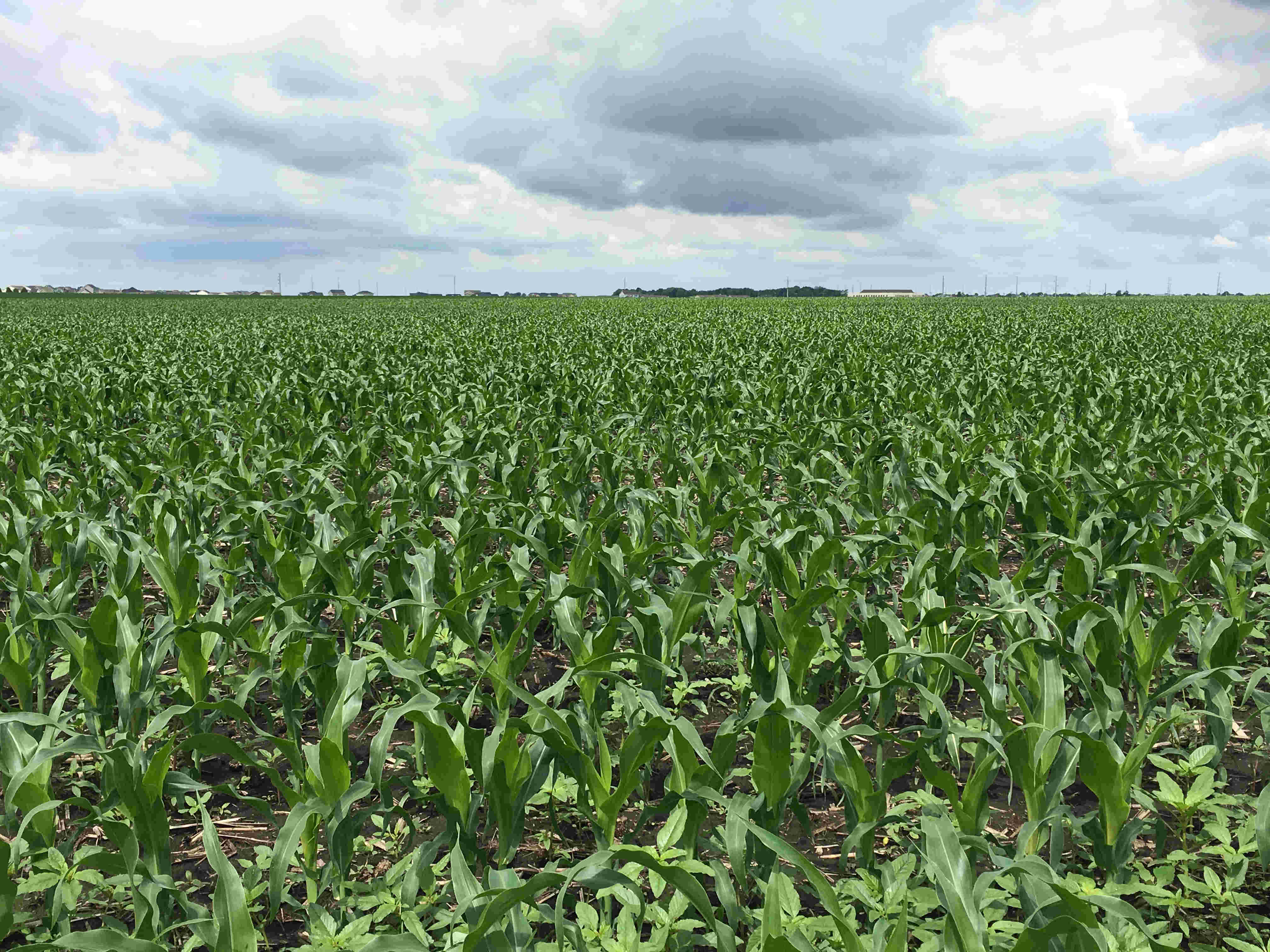}
  \caption[FigCornField]
   {Sunday June 5, 2016, Savoy, Illinois. A corn field.}
  \label{FigCornField}
\end{figure}
\begin{figure}[!h]
  \centering
  \includegraphics[width=70mm]{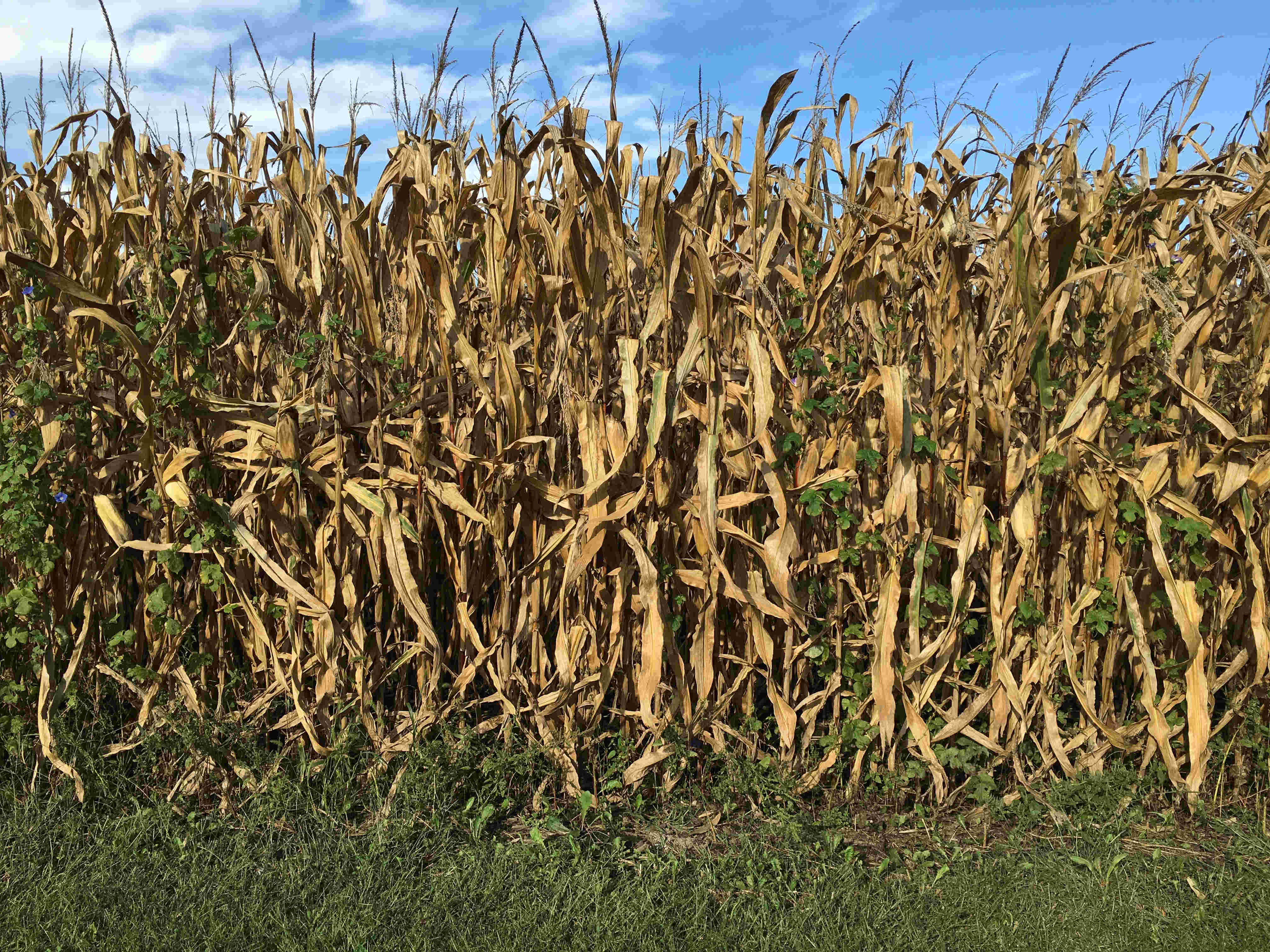}
  \caption[FigNewCornField]
   {Sunday September 18, 2016, Savoy, Illinois. New corn is ready.}
  \label{FigNewCornField}
\end{figure}
\begin{figure}[!h]
  \centering
  \includegraphics[width=70mm]{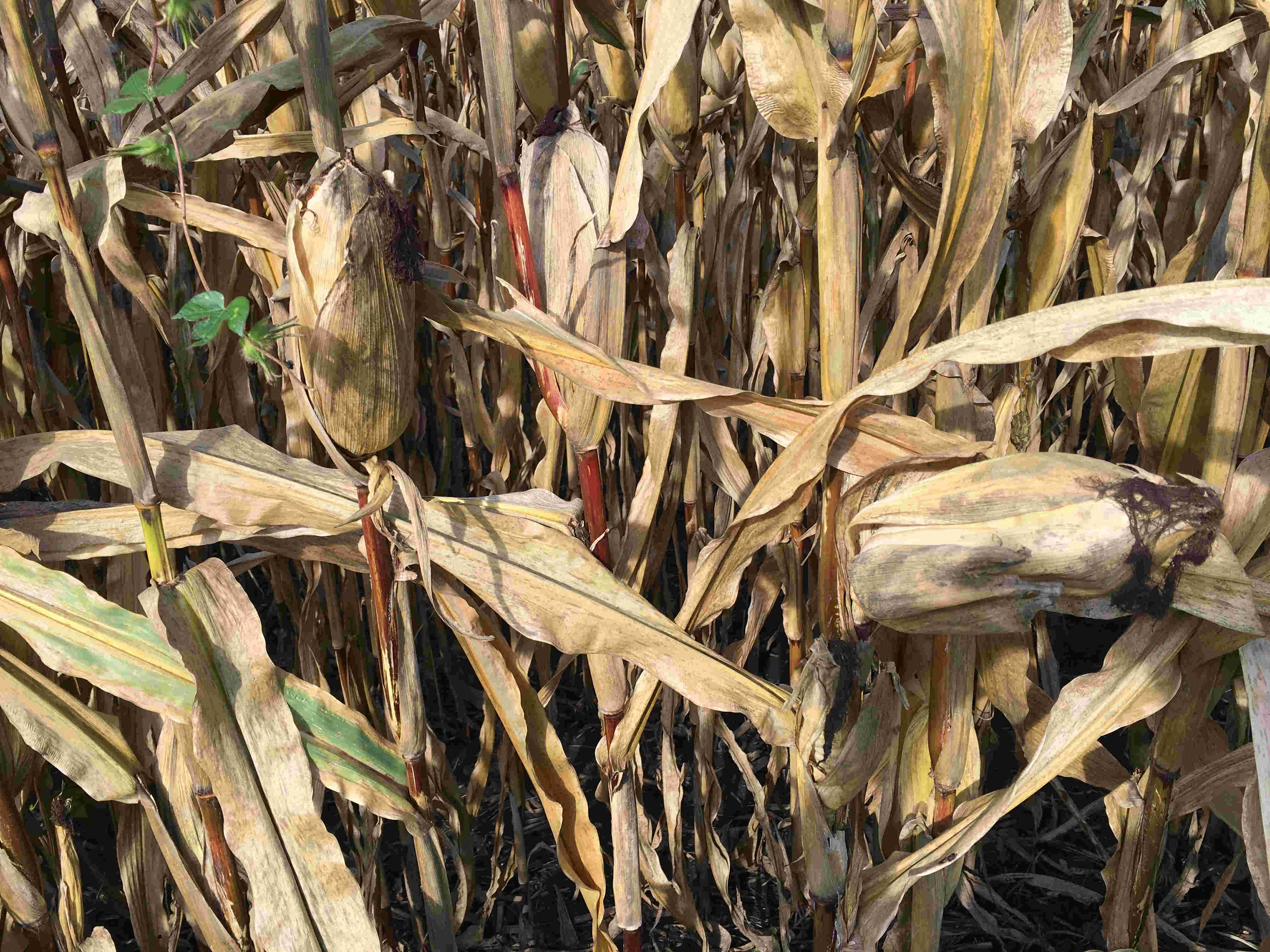}
  \caption[FigNewCorn]
   {Sunday September 18, 2016, Savoy, Illinois. New corn is ready.}
  \label{FigNewCorn}
\end{figure}

\section{Randomness of Waiting Times and Volume}

\begin{figure}[!h]
  \centering
  \includegraphics[width=105mm]{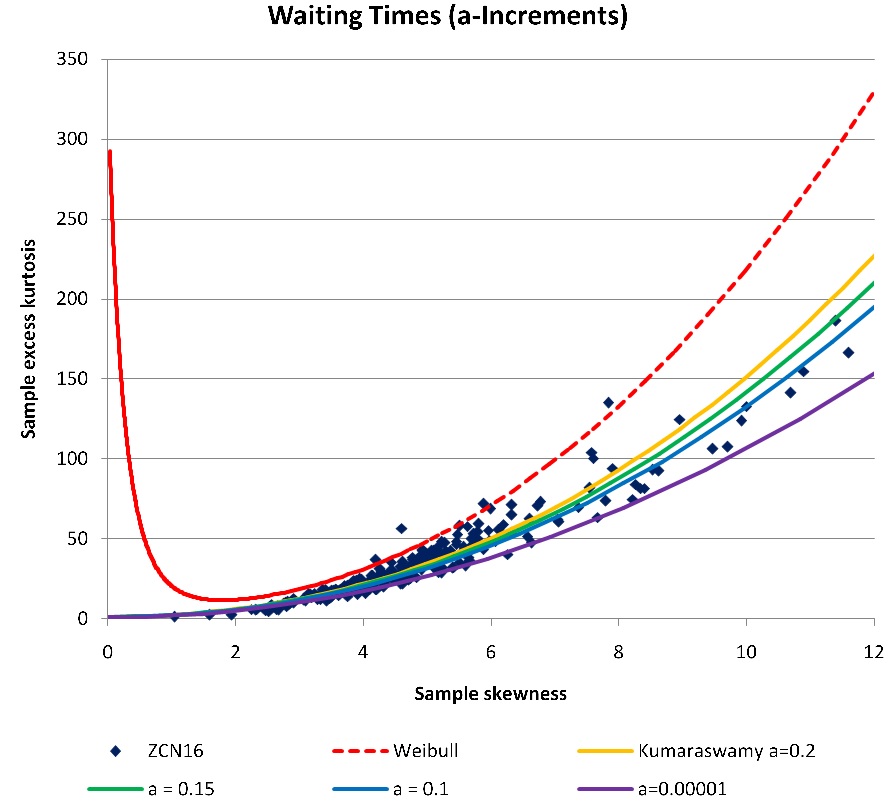}
  \caption[FigExKurtosisSkewness]
   {ZCN16,  311 sessions in [2015-03-26, 2016-07-01], [08:30:00, 13:15:00]: Sample excess kurtosis vs. skewness (dots). The only possible Weibull curve and four Kumaraswamy family curves are presented.}
  \label{FigExKurtosisSkewness}
\end{figure}

Let $N_{(s,r)}$ is the number of trading ticks in the $r$th range of the $s$th trading session, $n=N_{(s,r)}-1$. The sample estimates of the mean $a_{1(s,r)}^{\Delta t}$, variance $\mu_{2(s,r)}^{\Delta t}$, skewness $\frac{\mu_{3(s,r)}^{\Delta t}}{(\mu_{2(s,r)}^{\Delta t})^{\frac{3}{2}}}$, and excess kurtosis $\frac{\mu_{4(s,r)}^{\Delta t}}{(\mu_{2(s,r)}^{\Delta t})^2} - 3$ are computed for a-increments $t_{i(s,r)} - t_{i-1(s,r)}$ using the formulas (see also \cite{korn1968}) 
\begin{equation}
\label{EqSampleStatistics}
\begin{split}
a_{k(s,r)}^{\Delta t}=\dfrac{\sum_{i=2}^{i=N_{(s,r)}} \left (t_{i(s,r)} - t_{i-1(s,r)} \right )^k}{n},\\
m_{k(s,r)}^{\Delta t} = \dfrac{\sum_{i=2}^{i=N_{(s,r)}} \left (t_{i(s,r)} - t_{i-1(s,r)} - a_{1(s,r)}^{\Delta t} \right )^k}{n},\\
\mu_{2(s,r)}^{\Delta t} = \dfrac{n}{n-1}m_{2(s,r)}^{\Delta t},\\
\mu_{3(s,r)}^{\Delta t} = \dfrac{n^2}{(n-1)(n-2)}m_{3(s,r)}^{\Delta t},\\
\mu_{4(s,r)}^{\Delta t} = \dfrac{n(n^2-2n + 3)m_{4(s,r)}^{\Delta t}-3n(2n-3)(m_{2(s,r)}^{\Delta t})^2}{(n-1)(n-2)(n-3)}.
\end{split}
\end{equation}
For 311 trading sessions [2015-03-26, 2016-07-01] and second range [08:30:00, 13:15:00] of ZCN16, the sample excess kurtosis is plotted against skewness for a-increments as dots, Figure \ref{FigExKurtosisSkewness}.
Earlier \cite[pp. 22 - 27, Figures 8, 9]{salov2013} and, to the best of author's knowledge, first time, similar systematic deviation from theoretical Weibull distribution curve \cite{rousu1973}, \cite[Equations 24, 25]{salov2013} has been reported. In \cite[pp. 27 - 32]{salov2013}, the Kumaraswamy distribution has demonstrated better fitting properties \cite{kumaraswamy1980}
\begin{equation}
\label{EqKumaraswamy}
\begin{split}
CDF(z) = F(z) = F_0 + (1 - F_0)\left (1 - \left (1 - \left (\frac{z - z_{min}}{z_{max} - z_{min}} \right )^a \right )^b \right ),\\
PDF(z) = \frac{ab(1 - F_0)}{z_{max} - z_{min}} \left (\frac{z - z_{min}}{z_{max} - z_{min}} \right )^{a-1}      \left (1 - \left (\frac{z - z_{min}}{z_{max} - z_{min}} \right )^a \right )^{b-1}
\end{split}
\end{equation}
with $z_{min} \le z \le z_{max}$. Using Equations 35 - 38 from \cite[p. 29]{salov2013} for $z_{min}=0$, and $F_0=0$, four parametric dependencies between skewness and excess kurtosis are plotted on Figure \ref{FigExKurtosisSkewness} for $a=0.2$, $b \in [0.34, 5.5]$; $a=0.15$ $b \in [0.3, 3.8]$; $a=0.1$, $b \in [0.3, 2.8]$; $a=0.00001$, $b \in [0.058, 0.435]$. In contrast with the Weibull curve, they have extra degree of freedom forming a family for different $a$, where choosing a better curve is possible.

\begin{figure}[!h]
  \centering
  \includegraphics[width=105mm]{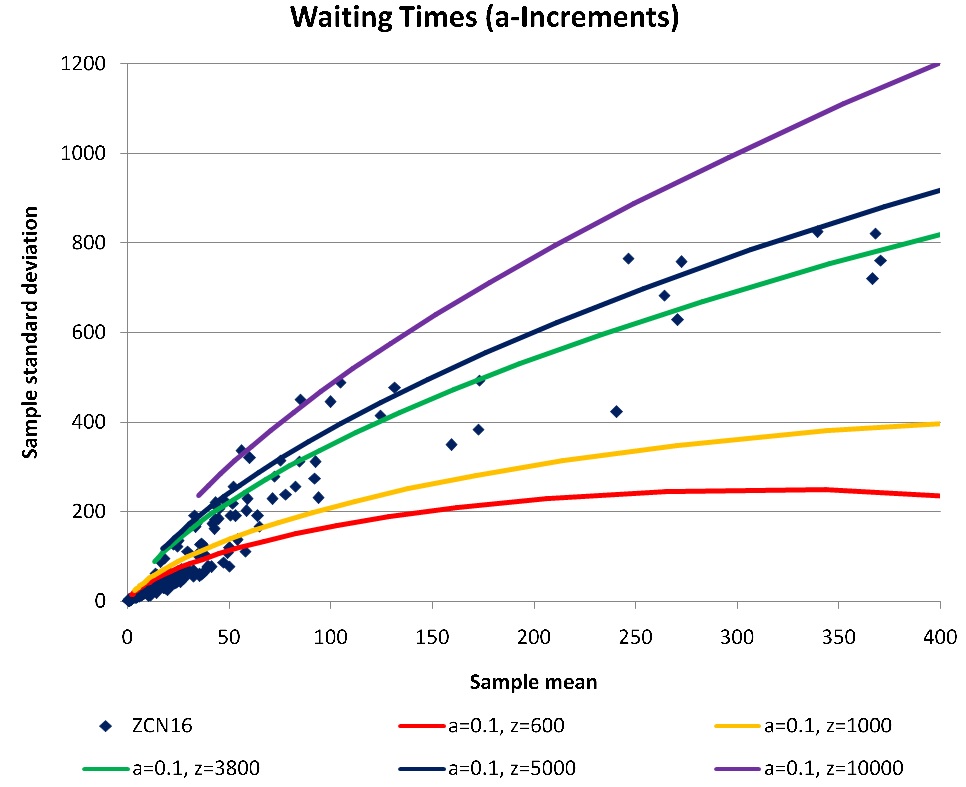}
  \caption[FigStdDevMean]
   {ZCN16,  311 sessions in [2015-03-26, 2016-07-01], [08:30:00, 13:15:00]: Sample standard deviation vs. mean of waiting times, a-increments, in seconds (dots). The five Kumaraswamy family curves are presented.}
  \label{FigStdDevMean}
\end{figure}

Equation 36 from \cite[p. 29]{salov2013} $\sqrt{\mu_2}=\frac{\sqrt{\mathrm{B}(1+\frac{2}{a},b)-b\mathrm{B}(1+\frac{1}{a},b)^2}}{\mathrm{B}(1+\frac{1}{a},b)\sqrt{b}}\alpha_1$ is a dependence between the standard deviation $\sqrt{\mu_2}=z_{max}\sqrt{b\left(\mathrm{B}(1+\frac{2}{a},b)-b\mathrm{B}(1+\frac{1}{a},b)^2\right)}$ and mean $\alpha_1=z_{max}b\mathrm{B}(1+\frac{1}{a},b)$ of the Kumaraswamy distribution. Figure \ref{FigStdDevMean} plots sample standard deviations vs. sample means for the same 311 sessions of ZCN16 together with five parametric curves with $a=0.1$  and varying $z_{max}$.

Figures \ref{FigExKurtosisSkewness} and \ref{FigStdDevMean} suggest that fitting Kumaraswamy distributions to the sets of waiting time sample statistics \{mean, standard deviation, skewness, excess kurtosis\} can be done in steps: 1) select optimal $a$ by fitting parametric dependencies of the skewness and excess kurtosis, and 2) select optimal $z_{max}$ for the chosen $a$ fitting parametric dependencies of the mean and standard deviation. Clearly, that a single combination of $a$, $b$, and $z_{max}$ for $z_{min} = 0$ and $F_0 = 0$ cannot satisfy the daily sessions sets: distributions of waiting times change. But they change in a manner yielding depicted parametric dependencies.

\textit{The remarkable properties of the four Kumaraswamy moments are not those expressed by solid curves on Figures \ref{FigExKurtosisSkewness} and \ref{FigStdDevMean}. This is only interesting mathematics involving complete beta or gamma functions \cite[pp. 27 - 32, Equations 28 - 33, 35 - 38]{salov2013}. The truly remarkable circumstances are the curves close to the experimental estimates of the moments. What could be arbitrary combinations of the four quantities is elegantly organized by the Market.}

\section{Volume and MPS0}
\begin{figure}[!h]
  \centering
  \includegraphics[width=120mm]{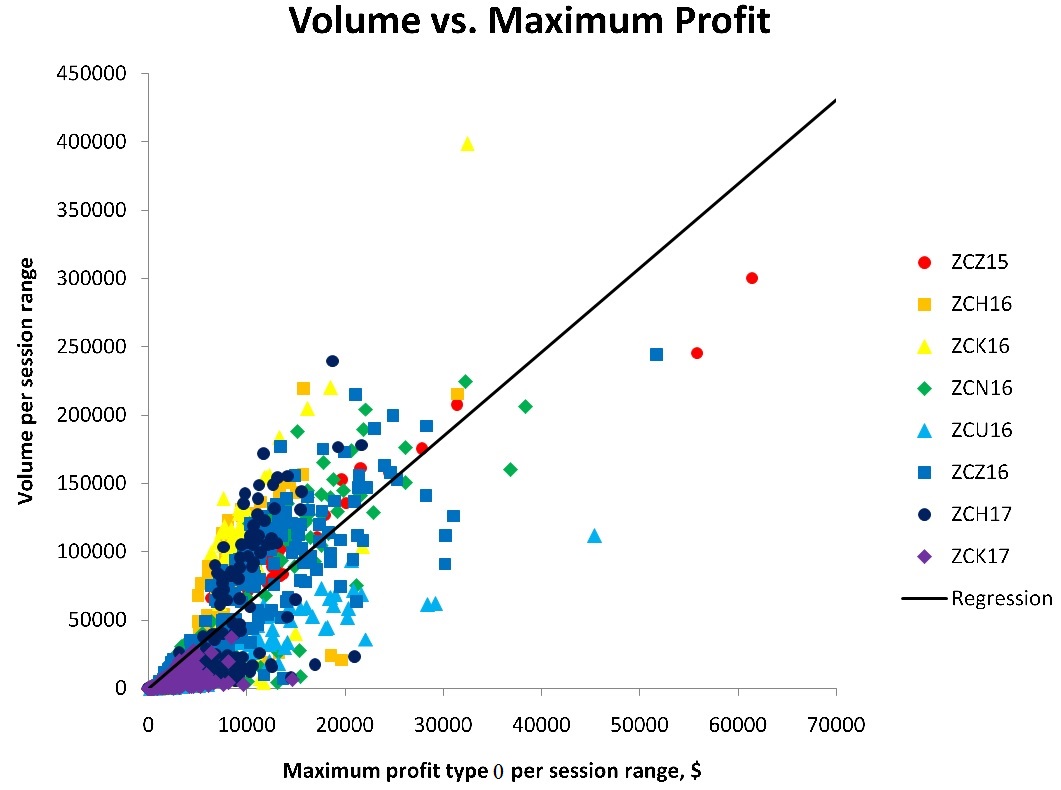}
  \caption[FigVolumeMP0ZC]
   {Corn. Volume vs. MP0, [2015-08-07, 2017-02-09], [08:30:00, 13:15:00], 1922 sessions: ZCZ15 88, ZCH16 144, ZCK16 189, ZCN16 230, ZCU16 271, ZCZ16 333, ZCH17 353, ZCK17 314. Initial margin \$544.50, maintenance margin \$495.00, transaction cost \$4.68 (round trip \$9.36). Regression with one parameter (depicted): slope = $6.16 \pm 0.07$, intercept = 0 (forced), coefficient of linear correlation $r = 0.89$, confidence interval 0.95\%. Regression with two parameters: slope = $6.6 \pm 0.1$, intercept = $-5601 \pm 788$, $r = 0.84$, confidence interval 0.95\%. Using Microsoft Excel, Data Analysis, Regression.} 
  \label{FigVolumeMP0ZC}
\end{figure}

If the MPS0 enters a market, then later it reverses long to short and vice versa positions of fixed size (1 contract) until it exits \cite{salov2007}, \cite[p. 84]{salov2013}. If the number of trading ticks is zero, then the volume and \textit{maximum profit}, MP, are zero too. If the number of trading ticks and volume are positive but \textit{all} absolute dollar price fluctuations are not greater than transaction costs, then the best strategy is \textit{do nothing} and the MP is still zero. The \cite{salov2013} proposes that traders are attracted to markets by frequent and big potential profit opportunities observed \textit{till now}. \textit{The MPS is a measure of these frequencies and magnitudes. One can expect that the greater MP is, the greater trading volumes is}. To the best of author's knowledge, Figure \ref{FigVolumeMP0ZC} \textit{is the first confirmation}. While the variances of MP and volume do not put 1922 dots on one regression line, the coefficient of linear correlation 0.89 points to two closely related factors.

\begin{figure}[!h]
  \centering
  \includegraphics[width=70mm]{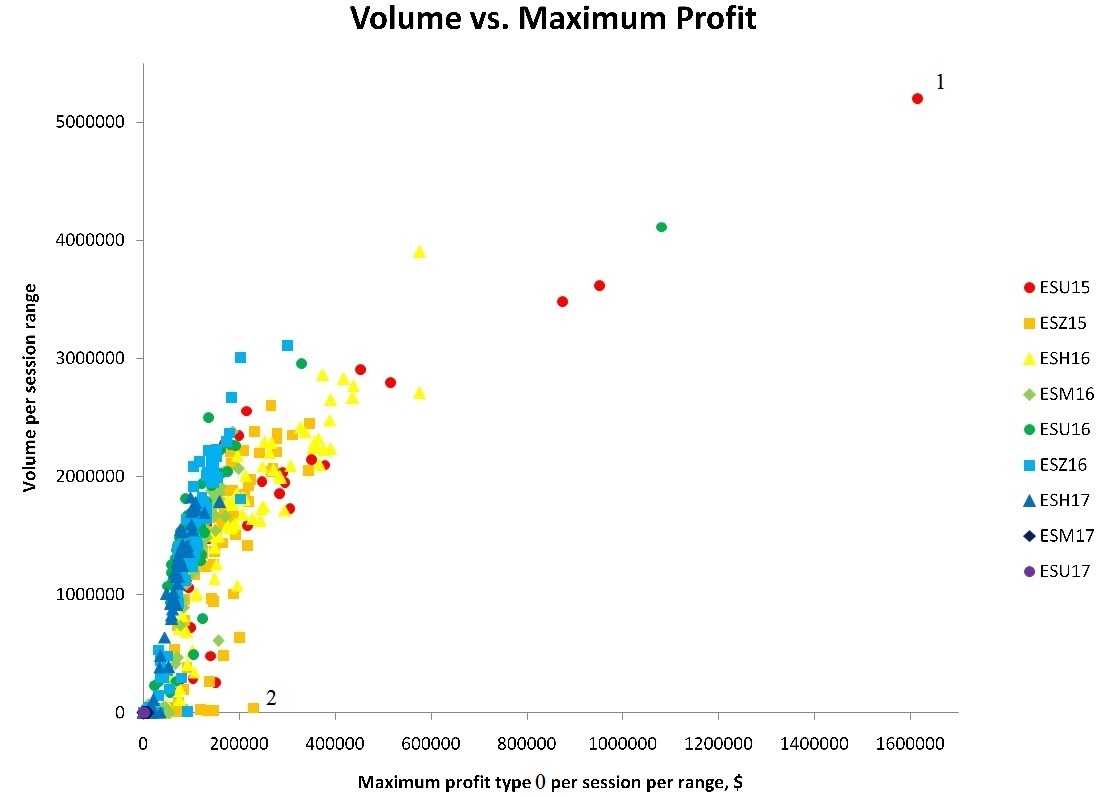}
  \caption[FigVolumeMP0ES]
   {E-mini S\&P 500. Volume vs. MP0, [2015-08-07, 2017-02-10], [08:30:00, 15:15:00], 1215 sessions: ESU15 30, ESZ15 93, ESH16 150, ESM16 208, ESU16 208, ESZ16 233, ESH17 167, ESM17 99, ESU17 27. Initial margin \$1306.25, maintenance margin \$1187.50, transaction cost \$4.68. Monday August 24, 2015, China Yuan Devaluation and Stock decline: ESU15 (\$1615255.38, 5200474) 1, ESZ15 (\$230007.86, 39788) 2.} 
  \label{FigVolumeMP0ES}
\end{figure}

\begin{figure}[!h]
  \centering
  \includegraphics[width=70mm]{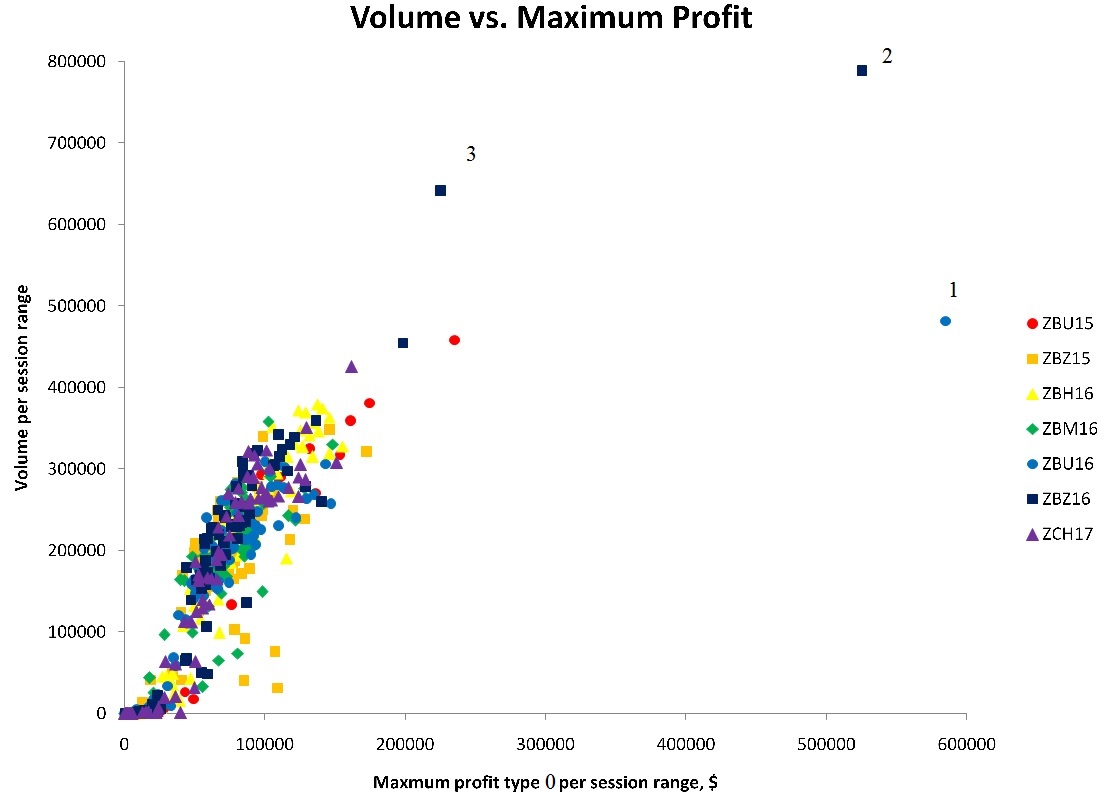}
  \caption[FigVolumeMP0ZB]
   {U.S. Treasury Bond. Volume vs. MP0, [2015-08-07, 2017-02-10], [07:20:00, 16:00:00], 660 sessions: ZBU15 31, ZBZ15 83, ZBH16 113, ZBM16 113, ZBU16 120, ZBZ16 120, ZBH17 80. Initial margin \$2420.00, maintenance margin \$2200.00, transaction cost \$4.68. Friday June 24, 2016, Brexit: ZBU16 (\$585164.22, 481598) 1. Wednesday November 9, 2016, U.S. Election: ZBZ16 (\$525576.52, 788648) 2, Thursday November 10, 2016 (\$225193.95, 641929) 3.} 
  \label{FigVolumeMP0ZB}
\end{figure}

\begin{figure}[!h]
  \centering
  \includegraphics[width=70mm]{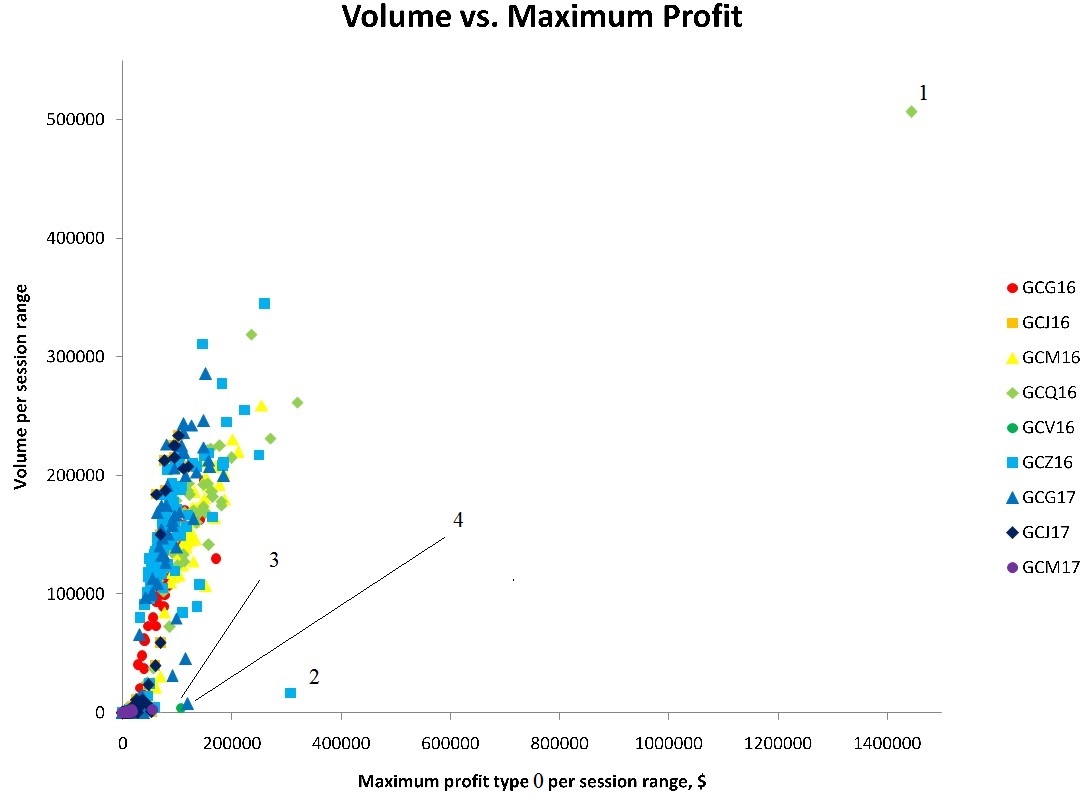}
  \caption[FigVolumeMP0GC]
   {Gold. Volume vs. MP0, [2015-08-07, 2017-02-10], [previous day 17:00:00, 16:00:00], 1859 sessions: GCU15 135, GCJ15 179, GCM16 218, GCQ16 212, GCV16 205, GCZ16 326, GCG17 232, GCJ17 179, GCM17 173. Initial margin \$6534.00, maintenance margin \$5940.00, transaction cost \$4.68. Friday June 24, 2016, Brexit: GCQ16 (\$1443922.88, 506809) 1, GCZ16 (\$307210.32, 16601) 2, GCU17 (\$106834.64, 3928) 3. Wednesday November 9, 2016, U.S. Election: GCG17 (\$119281.28, 7670) 4.} 
  \label{FigVolumeMP0GC}
\end{figure}

\begin{figure}[!h]
  \centering
  \includegraphics[width=70mm]{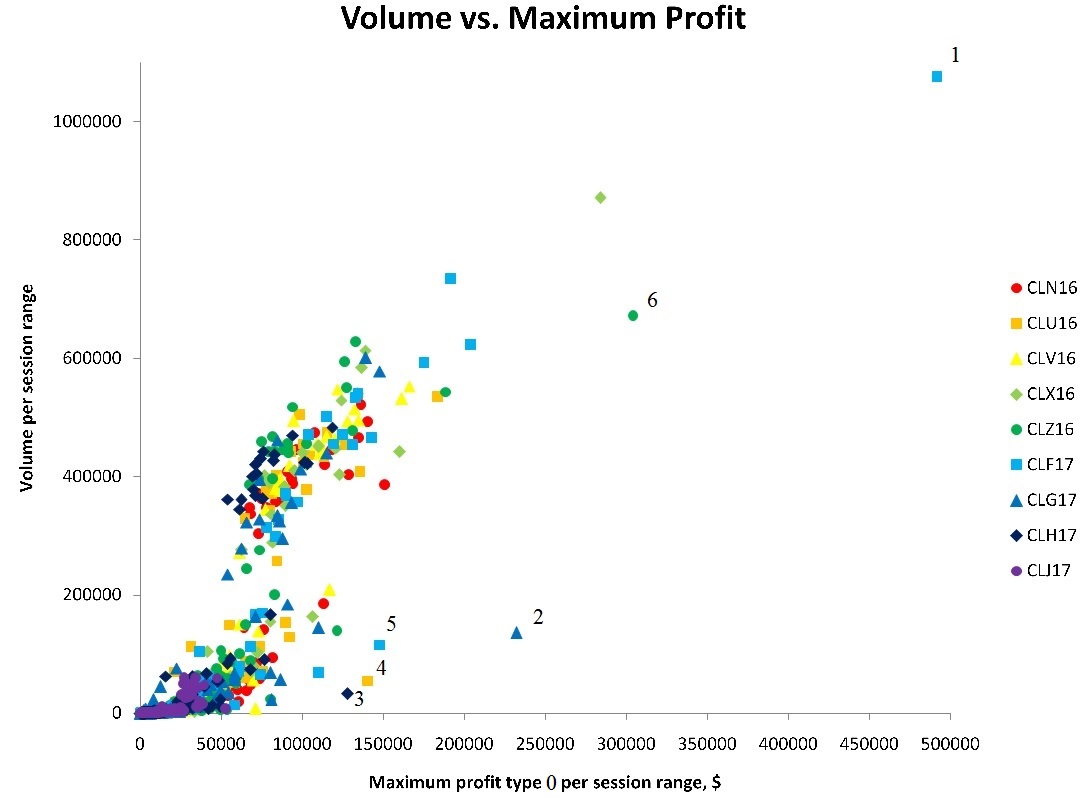}
  \caption[FigVolumeMP0CL]
   {Crude Oil. Volume vs. MP0, [2015-08-07, 2017-02-10], [previous day 17:00:00, 16:00:00], 2106 sessions: CLN16 199, CLU16 244, CLV16 205, CLX16 221, CLZ16 316, CLF17 238, CLG17 231, CLH17 247, CLJ17 205. Initial margin \$1936.00, maintenance margin \$1760.00, transaction cost \$4.68. Wednesday November 30, 2016, Vienna OPEC Meeting: CLF17 (\$491728.88, 1076001) 1, CLG17 (\$231982.64, 135678) 2, CLH17 (\$128141.60, 32323) 3. Friday June 24, 2016, Brexit: CLF17 (\$140618.08, 54048) 4. Wednesday November 9, 2016, U.S. Election: CLF17 (\$147884.40, 115097) 5, CLZ16 (\$304275.68, 672784) 6.} 
  \label{FigVolumeMP0CL}
\end{figure}

Trading volume and MP reach greater values for E-Mini S\&P500 than Corn contracts, Figure \ref{FigVolumeMP0ES}. Similar plots for U.S. Treasury Bonds, Gold, Crude Oil are on Figures \ref{FigVolumeMP0ZB}, \ref{FigVolumeMP0GC}, \ref{FigVolumeMP0CL}.


Often, prices for one commodity and different expiration months and years correlate but the volume concentrates on a nearby contract. Under such conditions big price fluctuations and micro trends substantially contribute to MP but volume increases mainly for nearby contracts creating outlying points: Figure \ref{FigVolumeMP0ES} 1, 2; Figure \ref{FigVolumeMP0GC} 1 - 3; Figure \ref{FigVolumeMP0CL} 1 - 3. China Yuan Deflation, Brexit, U.S. Presidential Election 2016, and Vienna OPEC Meeting dramatically influenced on markets and were responsible for these outliers.  \textit{All 7762 points in a single coordinate system form clusters, where on average greater MPs associate with greater trading volumes.}

\begin{figure}[!h]
  \centering
  \includegraphics[width=120mm]{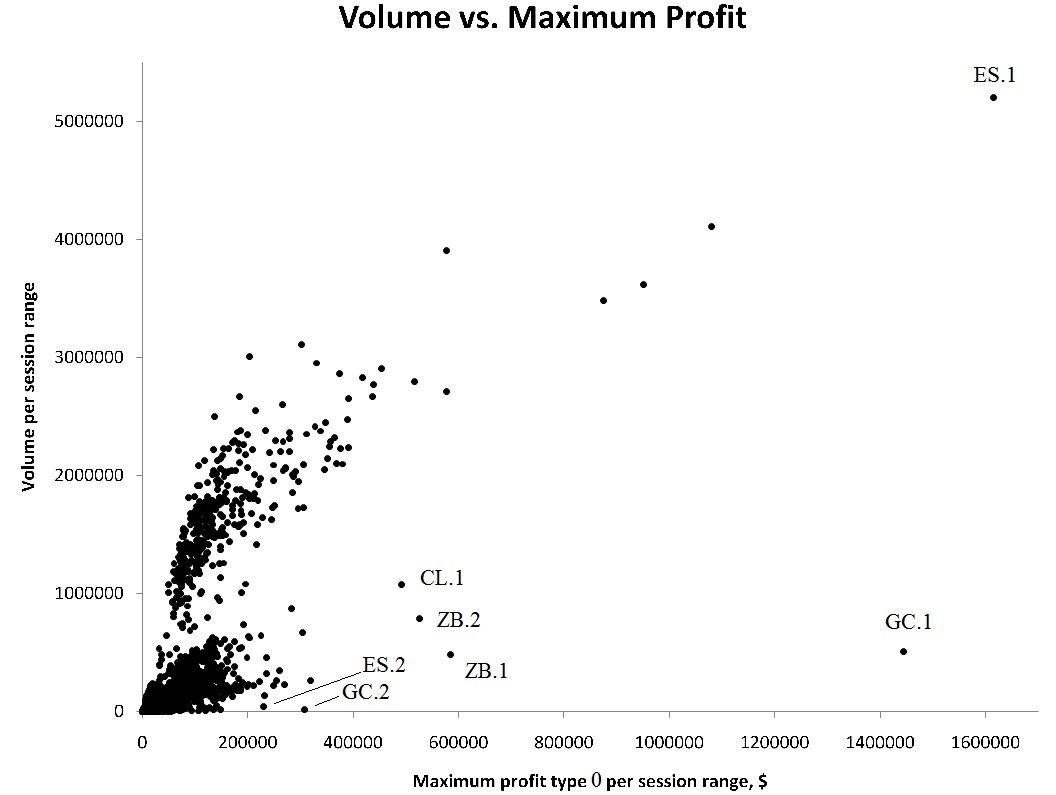}
  \caption[FigVolumeMP0]
   {Volume vs. MP0, [2015-08-07, 2017-02-10], 7762 sessions. ES.1 denotes Point 1 on ES Figure. Similarly, other labels are applied.} 
  \label{FigVolumeMP0}
\end{figure}

\section{Randomness of Price Increments}

Futures corn price increments, b-increments, are $k=..., -2, -1, 0, 1, 2, ...$ multiplies of $\delta_{ZC} = 0.25$ cents per bushel and map to integers $k$. This simplifies selection of bins of empirical frequency distributions. Sample moments were computed for $k$, Table \ref{b-increments}. Combining in one sample b-increments for all contracts does not change statistically mean, standard deviation, skewness, and excess kurtosis. Comparing with the standard deviations, the absolute means are tiny. Subtracting the means from the maximum and minimum increments, taking absolute values of the differences, and dividing them by the standard deviation yields values measured in dozens of standard deviations hinting that b-increments are not from a Gaussian distribution. They are also discrete. Similar results are presented in \cite{salov2013}.
\begin{center}
\begin{longtable}{|l|l|r|r|r|r|r|r|r|r|r|r|}
\caption[Sample Statistics of b-Increments]{Sample Statistics of b-Increments in $\delta_{ZC}$. Each minimum and maximum b-increment has occurred one time. The session range is 08:30:00 - 13:15:00 CST.} \label{b-increments} \\

 \hline
 \multicolumn{1}{|c|}{Ticker} &
 \multicolumn{1}{|c|}{Interval} &
 \multicolumn{1}{c|}{Sess.} &
 \multicolumn{1}{c|}{Size} &
 \multicolumn{1}{c|}{Mean} &
 \multicolumn{1}{c|}{Min} &
 \multicolumn{1}{c|}{Max} &
 \multicolumn{1}{c|}{Dev.} &
 \multicolumn{1}{c|}{Skew.} &
 \multicolumn{1}{c|}{E-Ku.} \\
 \hline 
 \endfirsthead

 \multicolumn{10}{c}%
 {\tablename\ \thetable{} -- continued from previous page} \\
 \hline
 \multicolumn{1}{|c|}{Ticker} &
 \multicolumn{1}{|c|}{Interval} &
 \multicolumn{1}{c|}{Sess.} &
 \multicolumn{1}{c|}{Size} &
 \multicolumn{1}{c|}{Mean} &
 \multicolumn{1}{c|}{Min} &
 \multicolumn{1}{c|}{Max} &
 \multicolumn{1}{c|}{Dev.} &
 \multicolumn{1}{c|}{Skew.} &
 \multicolumn{1}{c|}{E-Ku.} \\
 \hline 
 \endhead

 \hline \multicolumn{10}{|r|}{{Continued on next page}} \\ \hline
 \endfoot

 \hline
 \endlastfoot
ZCZ15 & 2015/08/07, 2015/12/14 & 88 & 891395 & 7.7e-5 & -27 & 26 & 0.48 & -0.040 & -3.0\\
ZCH16 & 2015/08/07, 2016/03/14 & 144 & 922345 & -3.3e-6 & -23 & 26 & 0.46 & -0.084 & -3.0\\
ZCK16 & 2015/08/07, 2016/05/13 & 189 & 844353 & 5.6e-5 & -37 & 39 & 0.47 & -0.11 & -1.9\\
ZCN16 & 2015/08/07, 2016/07/14 & 242 & 1111165 & -7.1e-5 & -46 & 38 & 0.47 & -0.51 & -2.6\\
ZCU16 & 2015/08/07, 2016/09/14 & 271 & 645837 & -2.6e-4 & -27 & 36 & 0.56 & 0.34 & -2.9\\
ZCZ16 & 2015/08/07, 2016/12/14 & 346 & 1993327 & 3.3e-5 & -28 & 24 & 0.46 & 0.0071 & -3.0\\
ZCH17 & 2015/08/26, 2017/02/24 & 378 & 1153547 & 2.1e-4 & -21 & 49 & 0.49 & 0.91 & -2.7\\
ZCK17 & 2015/09/22, 2017/02/24 & 347 & 383320 & 3.1e-4 & -25 & 21 & 0.57 & -0.25 & -2.6\\
ALL & 2015/08/07, 2017/02/24 & 2005 & 7945289 & 3.6e-5 & -46 & 49 & 0.48 & 0.072 & -3.0\\
\end{longtable}
\end{center}
\begin{center}
\begin{longtable}{|r|r|r|r|r|r|r|r|}
\caption[Frequencies of b-Increments]{Empirical frequencies of b-increments corresponding to ALL in Table \ref{b-increments} with Mean 0.000036, StdDev 0.48, and $\sum n_k = 7945289$.} \label{fb-increments} \\

 \hline
 \multicolumn{1}{|c|}{$k$} &
 \multicolumn{1}{|c|}{$n_k$} &
 \multicolumn{1}{c|}{Frequency} &
 \multicolumn{1}{c|}{$\frac{k - \textrm{Mean}}{\textrm{StdDev}}$} &
 \multicolumn{1}{|c|}{$k$} &
 \multicolumn{1}{|c|}{$n_k$} &
 \multicolumn{1}{c|}{Frequency} &
 \multicolumn{1}{c|}{$\frac{k - \textrm{Mean}}{\textrm{StdDev}}$} \\
 \hline 
 \endfirsthead

 \multicolumn{8}{c}%
 {\tablename\ \thetable{} -- continued from previous page} \\
 \hline
 \multicolumn{1}{|c|}{$k$} &
 \multicolumn{1}{|c|}{$n_k$} &
 \multicolumn{1}{c|}{Frequency} &
 \multicolumn{1}{c|}{$\frac{k - \textrm{Mean}}{\textrm{StdDev}}$} &
 \multicolumn{1}{|c|}{$k$} &
 \multicolumn{1}{|c|}{$n_k$} &
 \multicolumn{1}{c|}{Frequency} &
 \multicolumn{1}{c|}{$\frac{k - \textrm{Mean}}{\textrm{StdDev}}$} \\
 \hline 
 \endhead

 \hline \multicolumn{8}{|r|}{{Continued on next page}} \\ \hline
 \endfoot

 \hline
 \endlastfoot
-46 & 1 & 1.26E-07 & -95.8 & 0 & 6383586 & 0.803 & 0.0\\
-37 & 1 & 1.26E-07 & -77.1 & 1 & 760044 & 0.0957 & 2.1\\
-33 & 1 & 1.26E-07 & -68.8 & 2 & 16107 & 0.00203 & 4.2\\
-30 & 1 & 1.26E-07 & -62.5 & 3 & 2467 & 0.00031 & 6.2\\
-29 & 2 & 2.52E-07 & -60.4 & 4 & 883 & 0.000111 & 8.3\\
-28 & 1 & 1.26E-07 & -58.3 & 5 & 392 & 4.93E-05 & 10.4\\
-27 & 2 & 2.52E-07 & -56.3 & 6 & 240 & 3.02E-05 & 12.5\\
-25 & 2 & 2.52E-07 & -52.1 & 7 & 126 & 1.59E-05 & 14.6\\
-24 & 1 & 1.26E-07 & -50.0 & 8 & 82 & 1.03E-05 & 16.7\\
-23 & 3 & 3.78E-07 & -47.9 & 9 & 43 & 5.41E-06 & 18.7\\
-22 & 1 & 1.26E-07 & -45.8 & 10 & 36 & 4.53E-06 & 20.8\\
-21 & 2 & 2.52E-07 & -43.8 & 11 & 27 & 3.40E-06 & 22.9\\
-20 & 1 & 1.26E-07 & -41.7 & 12 & 11 & 1.38E-06 & 25.0\\
-19 & 7 & 8.81E-07 & -39.6 & 13 & 24 & 3.02E-06 & 27.1\\
-18 & 6 & 7.55E-07 & -37.5 & 14 & 10 & 1.26E-06 & 29.2\\
-17 & 5 & 6.29E-07 & -35.4 & 15 & 5 & 6.29E-07 & 31.2\\
-16 & 4 & 5.03E-07 & -33.3 & 16 & 7 & 8.81E-07 & 33.3\\
-15 & 5 & 6.29E-07 & -31.3 & 17 & 5 & 6.29E-07 & 35.4\\
-14 & 12 & 1.51E-06 & -29.2 & 18 & 8 & 1.01E-06 & 37.5\\
-13 & 16 & 2.01E-06 & -27.1 & 19 & 3 & 3.78E-07 & 39.6\\
-12 & 23 & 2.89E-06 & -25.0 & 20 & 4 & 5.03E-07 & 41.7\\
-11 & 27 & 3.40E-06 & -22.9 & 21 & 2 & 2.52E-07 & 43.7\\
-10 & 40 & 5.03E-06 & -20.8 & 22 & 1 & 1.26E-07 & 45.8\\
-9 & 46 & 5.79E-06 & -18.8 & 23 & 1 & 1.26E-07 & 47.9\\
-8 & 82 & 1.03E-05 & -16.7 & 24 & 2 & 2.52E-07 & 50.0\\
-7 & 111 & 1.40E-05 & -14.6 & 26 & 3 & 3.78E-07 & 54.2\\
-6 & 209 & 2.63E-05 & -12.5 & 32 & 1 & 1.26E-07 & 66.7\\
-5 & 395 & 4.97E-05 & -10.4 & 35 & 1 & 1.26E-07 & 72.9\\
-4 & 927 & 0.000117 & -8.3 & 36 & 1 & 1.26E-07 & 75.0\\
-3 & 2422 & 0.000305 & -6.3 & 38 & 1 & 1.26E-07 & 79.2\\
-2 & 15278 & 0.00192 & -4.2 & 39 & 1 & 1.26E-07 & 81.2\\
-1 & 761530 & 0.0958 & -2.1 & 49 & 1 & 1.26E-07 & 102.1\\
\end{longtable}
\end{center}
\begin{center}
\begin{longtable}{|r|r|r|r|r|r|r|r|}
\caption[Frequencies of absolute b-Increments]{Empirical frequencies of absolute b-increments corresponding to ALL in Table \ref{b-increments} with Mean 0.000036, StdDev 0.48, and $\sum n_k = 7945289$.} \label{fabsb-increments} \\

 \hline
 \multicolumn{1}{|c|}{$|k|$} &
 \multicolumn{1}{|c|}{$n_k$} &
 \multicolumn{1}{c|}{Frequency} &
 \multicolumn{1}{c|}{|$\frac{|k| - \textrm{Mean}}{\textrm{StdDev}}|$} &
 \multicolumn{1}{|c|}{$|k|$} &
 \multicolumn{1}{|c|}{$n_k$} &
 \multicolumn{1}{c|}{Frequency} &
 \multicolumn{1}{c|}{|$\frac{|k| - \textrm{Mean}}{\textrm{StdDev}}|$} \\
 \hline 
 \endfirsthead

 \multicolumn{8}{c}%
 {\tablename\ \thetable{} -- continued from previous page} \\
 \hline
 \multicolumn{1}{|c|}{$|k|$} &
 \multicolumn{1}{|c|}{$n_k$} &
 \multicolumn{1}{c|}{Frequency} &
 \multicolumn{1}{c|}{|$\frac{|k| - \textrm{Mean}}{\textrm{StdDev}}|$} &
 \multicolumn{1}{|c|}{$|k|$} &
 \multicolumn{1}{|c|}{$n_k$} &
 \multicolumn{1}{c|}{Frequency} &
 \multicolumn{1}{c|}{|$\frac{|k| - \textrm{Mean}}{\textrm{StdDev}}|$} \\
 \hline 
 \endhead

 \hline \multicolumn{8}{|r|}{{Continued on next page}} \\ \hline
 \endfoot

 \hline
 \endlastfoot
0 & 6383586 & 0.803 & 0.0 & 20 & 5 & 6.29E-07 & 41.7\\
1 & 1521574 & 0.192 & 2.1 & 21 & 4 & 5.03E-07 & 43.7\\
2 & 31385 & 0.00395 & 4.2 & 22 & 2 & 2.52E-07 & 45.8\\
3 & 4889 & 0.000615 & 6.2 & 23 & 4 & 5.03E-07 & 47.9\\
4 & 1810 & 0.000228 & 8.3 & 24 & 3 & 3.78E-07 & 50.0\\
5 & 787 & 9.91E-05 & 10.4 & 25 & 2 & 2.52E-07 & 52.1\\
6 & 449 & 5.65E-05 & 12.5 & 26 & 3 & 3.78E-07 & 54.2\\
7 & 237 & 2.98E-05 & 14.6 & 27 & 2 & 2.52E-07 & 56.2\\
8 & 164 & 2.06E-05 & 16.7 & 28 & 1 & 1.26E-07 & 58.3\\
9 & 89 & 1.12E-05 & 18.7 & 29 & 2 & 2.52E-07 & 60.4\\
10 & 76 & 9.57E-06 & 20.8 & 30 & 1 & 1.26E-07 & 62.5\\
11 & 54 & 6.80E-06 & 22.9 & 32 & 1 & 1.26E-07 & 66.7\\
12 & 34 & 4.28E-06 & 25.0 & 33 & 1 & 1.26E-07 & 68.7\\
13 & 40 & 5.03E-06 & 27.1 & 35 & 1 & 1.26E-07 & 72.9\\
14 & 22 & 2.77E-06 & 29.2 & 36 & 1 & 1.26E-07 & 75.0\\
15 & 10 & 1.26E-06 & 31.2 & 37 & 1 & 1.26E-07 & 77.1\\
16 & 11 & 1.38E-06 & 33.3 & 38 & 1 & 1.26E-07 & 79.2\\
17 & 10 & 1.26E-06 & 35.4 & 39 & 1 & 1.26E-07 & 81.2\\
18 & 14 & 1.76E-06 & 37.5 & 46 & 1 & 1.26E-07 & 95.8\\
19 & 10 & 1.26E-06 & 39.6 & 49 & 1 & 1.26E-07 & 102.1\\
\end{longtable}
\end{center}

\begin{figure}[!h]
  \centering
  \includegraphics[width=120mm]{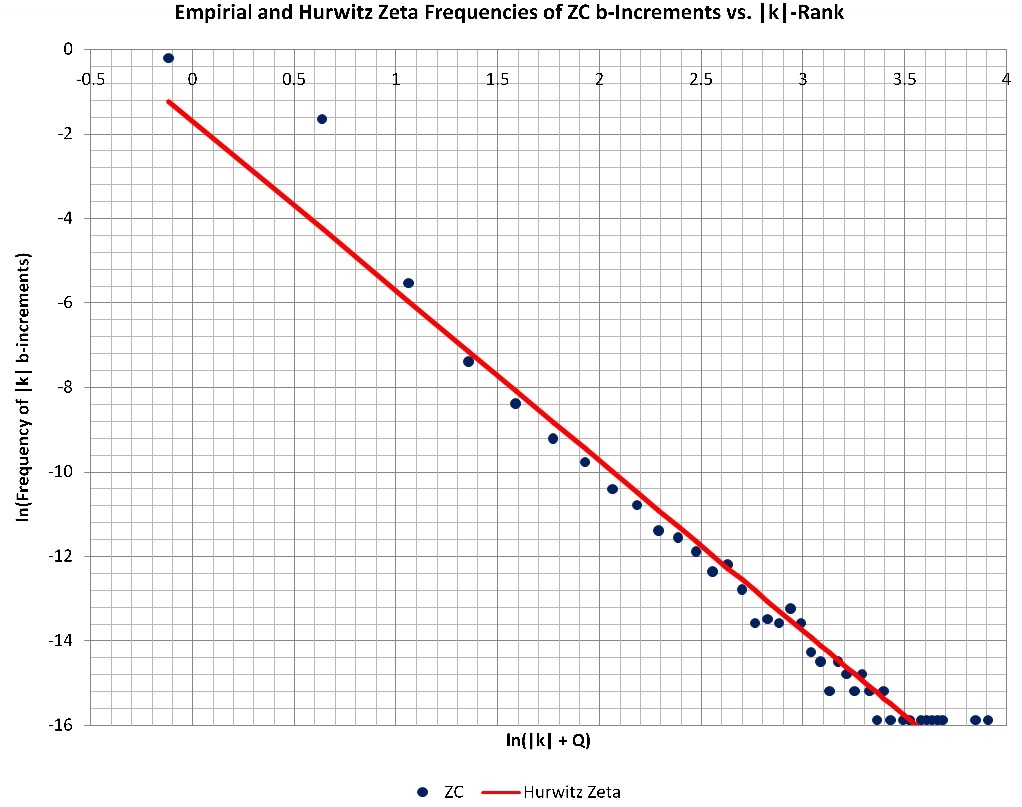}
  \caption[FigBIncrementsFrequency]
   {Empirical frequencies of absolute b-increments combined in one sample: ZCZ15, ZCH16, ZCK16, ZCN16, ZCU16, ZCZ16, ZCH17, ZCK17; 2015/08/07 - 2017/02/24; 08:30:00 - 13:15:00; 2005 sessions; Table \ref{fabsb-increments}, row ALL in Table \ref{b-increments}. Unweighted Hurwitz Zeta $S=4.024$, $Q=0.8908$, slope = -4.024, intercept = -1.691, sum of squares of deviations = 17.16.} 
  \label{FigBIncrementsFrequency}
\end{figure}

Empirical frequencies of the absolute b-increments $|0\delta_{ZC}|$, $|1\delta_{ZC}|$, $|2\delta_{ZC}|$, ..., $|k\delta_{ZC}|$, ... in Table \ref{fabsb-increments}, create for all investigated ZC contracts a universal ZC-dependence on the rank $|k|$ in bi-logarithmic coordinates, Figure \ref{FigBIncrementsFrequency}. The work \cite[Figures 20, 21; pp. 46 - 47]{salov2013} presents similar plots and proposes approximating such dependencies by discrete \textit{Zipf-Mandelbrot} and \textit{Hurwitz Zeta distributions}. It describes an algorithm for evaluation of the required \textit{Hurwitz Zeta function}  \cite[pp. 45 - 49]{salov2013}.
\begin{figure}[!h]
  \centering
  \includegraphics[width=120mm]{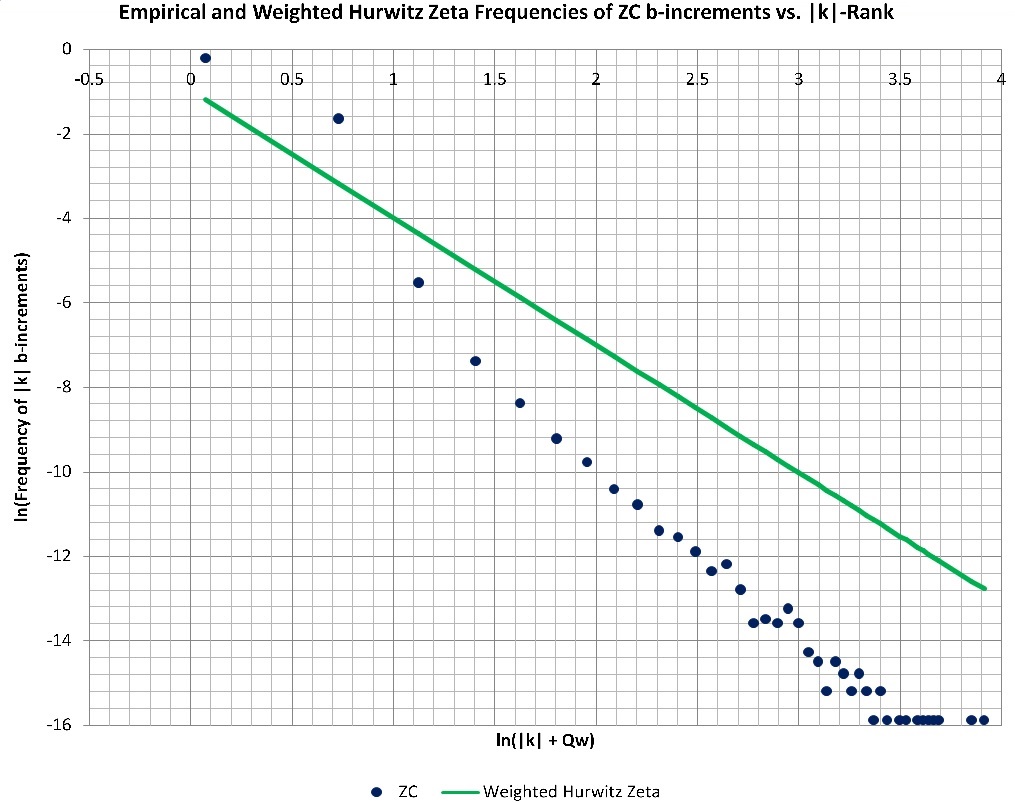}
  \caption[FigBIncrementsFrequencyW]
   {Empirical frequencies of absolute b-increments combined in one sample: ZCZ15, ZCH16, ZCK16, ZCN16, ZCU16, ZCZ16, ZCH17, ZCK17; 2015/08/07 - 2017/02/24; 08:30:00 - 13:15:00; 2005 sessions; Table \ref{fabsb-increments}, row ALL in Table \ref{b-increments}. Weighted Hurwitz Zeta $S=3.015$, $Q=1.078$, slope = -3.015, intercept = -0.9788, sum of the weighted squares of deviations = 1.242.} 
  \label{FigBIncrementsFrequencyW}
\end{figure}

A multinomial distribution assumes fixed $K > 2$ events with probabilities $p_1$, $p_2$, ..., $p_K$, where $\sum_{i=1}^{i=K} p_i=1$. This extends the \textit{Bernoulli distribution} for $K=2$ with probabilities $p$ and $1-p$. While the $p_1$, $p_2$, ..., $p_K$ could be arbitrary and experiment would be the only way to estimate them using empirical frequencies of the $K$ event, such laws and distributions as \textit{power}, \textit{Zipf}, Zipf-Mandelbrot, \textit{Riemann Zeta}, and Hurwitz Zeta imply additional relationships for frequencies. Usually, the reasons of the power laws remain unclear but studying asymptotic behavior and logarithmic corrections can give theoretical explanations \cite{arnold2004}. Such empirical relationships can simplify models.

The Zipf-Mandelbrot law probability mass function is for a finite number of events $N$
\begin{equation}
\label{EqZipfMandelbrot}
\begin{split}
PMF_{ZM}(k) = \frac{(k+Q)^{-S}}{\sum_{i=1}^{i=N}(i+Q)^{-S}}, Q > 0, S > 0,\\
\ln(PMF_{ZM}(k)) = -S\ln(k+Q) - \ln(\sum_{i=1}^{i=N}(i+Q)^{-S}).
\end{split}
\end{equation}
It is more flexible than the Zipf law with $Q=0$. For ZC contracts traded under price limits the number of ranks is finite. The range of the lattice distribution $k_{min} = -\frac{L_{ZC}}{\delta_{ZC}}, ..., -1, 0, 1, ..., k_{max} = \frac{L_{ZC}}{\delta_{ZC}}$ is symmetrical around the previous settlement price $P_S$. If the previous closing price $P_C \ne P_S$, then with respect to $P_C$ the ranks are asymmetric and either $|k_{min}|$ or $|k_{max}|$ exceeds $\frac{L_{ZC}}{\delta_{ZC}} = \frac{\$0.25}{\$0.0025}=100$. Always, $L_{ZC} < P_S$. The observed extremes -46 and 49 are far from the maximums. If the price drops to the down limit and then jumps to the up limit, the b-increment is 200. For any price $P_S - L_{ZC} < P < P_S + L_{ZC}$ in a session,
\begin{equation}
\label{EqMinMaxRanks}
\begin{split}
k_{min}(P, P_S) = \frac{P_S-L_{ZC}-P}{\delta_{ZC}} \le 0,\\
k_{max}(P, P_S) = \frac{P_S+L_{ZC}-P}{\delta_{ZC}} \ge 0.
\end{split}
\end{equation}
The author is unfamiliar with cases, where a corn future price move down to the limit was followed by an opposite move up to the limit or vice versa. Opening a market up or down to the limit after news arriving during off hours are known. Electronic markets and extended trading hours make such events less frequent. To simulate asymmetric ranks, fit separately the Zipf-Mandelbrot law to non-negative and non-positive branches of the distribution connected at rank zero. The branches may get different optimal $Q$ and $S$.

For contracts traded without price limits such as corn futures traded in the closing months, the $0 \le k_{max}$ is unlimited. Since price $P > 0$, $-\frac{P}{\delta_{ZC}} < k_{min}$. The infinite branch can be modeled with the Hurwitz Zeta distribution
\begin{equation}
\label{EqHirwitzZetaDistribution}
\begin{split}
PMF_{H}(k) = \frac{(k+Q)^{-S}}{\sum_{i=0}^{i=\infty}(i+Q)^{-S}}=\frac{(k+Q)^{-S}}{\zeta(Q,S)}, Q > 0, S > 1,\\
\ln(PMF_{H}(k)) = -S\ln(k+Q) - \ln(\zeta(Q,S)),
\end{split}
\end{equation}
where $\zeta(Q,S)$ is the Hurwitz Zeta function generalizing the \textit{Riemann Zeta function} \cite[p. 45]{salov2013}. It is enough to compute $\zeta(Q,S)$ for \textit{real} arguments.

For a fixed size of sample, obeying a power law, the greater by absolute value rank is, the less accurate its frequency estimate is. In bi-logarithmic plots such values appear as horizontal chains of dots on the right, Figure \ref{FigBIncrementsFrequency}. Since these dots correspond to the lowest frequencies, accurate plotting requires enormous sample increase. In fitting, such dots should get lower weights.

The so-called \textit{King effect} is observed as the highest frequency rank outlier \cite[p. 49]{salov2013}. The most accurate estimates of frequencies of b-increments equal to 0 and $|\delta_{ZC}|$, $|k|=0, 1$, are outliers on Figures \ref{FigBIncrementsFrequency} , \ref{FigBIncrementsFrequencyW}. Minimizing unweighted sum of squares of deviations of points from the line $\ln(PMF_{H}(|k|))=\mathrm{slope} \times \ln(|k|+Q) + \mathrm{intercept}$ fits the line giving preference to the majority of points, Figure \ref{FigBIncrementsFrequency}. Applying frequencies as weights of the squares fits the line to the three points with $|k|=0, 1, 2$, Figure \ref{FigBIncrementsFrequencyW}. Since abscissas of experimental points depend on the Hurwitz Zeta distribution parameter $Q$, the same points are plotted differently on both charts. \textit{Instead of using one fitting line, the author sees sense in applying two or three greatest most accurately estimated frequencies "as is" and approximating remaining ranks by the discrete Hurwitz Zeta distribution to estimate on the tails.}

\paragraph{Distribution and characteristic functions. Infinite divisibility.} For a random variable $\xi$, there is \textit{one-to-one correspondence} between its distribution $F_{\xi}(x)=P(\xi < x)$ \cite[p. 19]{gnedenko1949} and characteristic $f_{\xi}(t)=\int e^{itx}dF_{\xi}(x)$, $i=\sqrt{-1}$, \cite[p. 50]{gnedenko1949} function, c.f., \cite[pp. 52 - 53]{jessen1935}, \cite[Theorem 7, p. 10]{khinchin1938}, \cite[6. A uniqueness theorem, pp. 22 - 26]{esseen1945}, \cite[Theorems 1, 2, pp. 54 - 55]{gnedenko1949}, \cite[Theorem 3.1.1, p. 29]{lukacs1970}. Without a proof, Gnedenko and Kolmogorov \cite[Example 4, p. 78 - 79]{gnedenko1949} present the \textit{probability density function} $p(x)=0$ for $x \le 0$, $p(x) = \frac{\beta^{\alpha}}{\Gamma(\alpha)}x^{\alpha - 1}e^{-\beta x}$ for $x >0$ and corresponding c.f. $f(t)=(1-\frac{it}{\beta})^{-\alpha}$. For illustration: $f(t)=\int_{-\infty}^{\infty}e^{itx}p(x)dx=\int_0^{\infty}e^{itx}\frac{\beta^{\alpha}}{\Gamma(\alpha)}x^{\alpha - 1}e^{-\beta x}dx=\frac{\beta^{\alpha}}{\Gamma(\alpha)}\int_0^{\infty}x^{\alpha - 1}e^{(it -\beta)x}dx = \{-y=(it-\beta)x\}=\frac{\beta^{\alpha}}{\Gamma(\alpha)(\beta - it)^{\alpha}}\int_0^{\infty}y^{\alpha - 1}e^{-y}dy=\frac{\beta^{\alpha}}{\Gamma(\alpha)(\beta - it)^{\alpha}}\Gamma(\alpha)=(1-\frac{it}{\beta})^{-\alpha}$. Taking the root of degree $n$ yields one of $n$ values $(1-\frac{it}{\beta})^{-\frac{\alpha}{n}}$, the same type of c. f. with $p(x)=\frac{\beta^{\frac{\alpha}{n}}}{\Gamma(\frac{\alpha}{n})}x^{\frac{\alpha}{n} - 1}e^{-\beta x}$. This c.f. is \textit{infinitely divisible} because $n$ is arbitrary: $f(t)=[f_n(t)]^n$ with $n$ identical factors. This property is important: the c.f. of the sum of $n$ i.i.d random variables is the $nth$ power of c.f. of the variable. We recollect these details to avoid a confusion with the names of zeta distributions.

\paragraph{"Riemann Zeta", "Zeta", "Hurwitz Zeta" distributions.}  The Riemann Zeta function $\zeta(z=\sigma + it)=\sum_{n \; \mathrm{natural}} n^{-z}=\prod_{p \; \mathrm{prime}}(1-p^{-z})^{-1}$, where $\sigma, t \in R$, $\sigma > 1$, and the sum and product indexes run via \textit{all} natural and prime numbers. The former do not include zero. The latter do not include one. Khinchin \cite[Example 3, p. 35]{khinchin1938} proves that $\frac{\zeta(\sigma+it)}{\zeta(\sigma)}$ is a \textit{characteristic function} corresponding to an \textit{infinitely divisible} distribution. Gnedenko and Kolmogorov \cite[Example 6, p. 82]{gnedenko1949} use this as an illustration slightly "expanding" the proof. All three, "not having Internet access in 1938 and 1949", could consider the proofs related to $\frac{\zeta(\sigma+it)}{\zeta(\sigma)}$ too simple for themselves and avoid visiting a library using the results just as an interesting illustration. Such a visit might be needed, if priority would be important for them. Allan Gut in the charming topic notes \cite{gut2005} among other things cites \cite{jessen1935}. The latter is not referenced explicitly in \cite{lin2001}, \cite{hu2006} but implicitly is contained, \textit{of course}, in Lukacs \cite[p. 399]{lukacs1970}. Aoyama and Nakamura include it \cite{aoyama2013}. Gnedenko and Kolmogorov cite Esseen \cite{esseen1945}, \textit{of course}, containing on page 125 the reference to Jessen and Wintner. A brief look at \cite[Theorem 19, pp. 70 - 72]{jessen1935} helps to estimate their contribution to infinite divisibility of the discussed function. Khinchin's result is "direct" but Gnedenko and Kolmogorov hardly could name it "Khinchin's characteristic function" being familiar with Esseen's essay and contributions to the discovery published in 1935 prior 1938. The author emphasizes the historical scientific tradition connecting Esseen and Gut - members of Uppsala University (created in 1477).

Lin and Hu \cite{lin2001} focus on the \textit{distribution} of discrete $\xi=-\ln(n)$ with probabilities $\frac{n^{-\sigma}}{\zeta(\sigma)}$. They prove the interesting condition of the infinite divisibility of Dirichlet type c.f. - completely multiplicative coefficients $c(mn)=c(m)c(n)$, where the Riemann Zeta function, $c(nm)=1=c(n)c(m)=1 \times 1$ is a trivial particular case and Remark 2 presents a nontrivial example. We read \cite[p. 817]{lin2001}: \textit{"For convenience the corresponding distribution $F_{\sigma}$, of $f_{\sigma}$, will be called the Riemann zeta distribution with parameter $\sigma$."} This naming convention was extended by \cite{hu2006} to the Hurwitz Zeta function and picked up by \cite{gut2005}, \cite{aoyama2013}, \cite{nakamura2015}. Following this proposal the answer on the question "Is the Riemann Zeta distribution infinitely divisible?" is "Yes".

The probabilities $\frac{n^{-\sigma}}{\zeta(\sigma)}$ are similar to those in Zipf and Zipf-Mandelbrot laws associating them with the random variable $\eta=n$, the \textit{rank}. Prior \cite{lin2001}, Mandelbrot names it \textit{Zeta distribution} \cite[pp. 201 - 202]{mandelbrot1997}. This is not infinitely divisible. Unless Bernhard Riemann name should be repeated in each function, theorem, distribution involving his famous, extremely important, and influential function and following the references, the names looking logical for the author could be: Khinchin characteristic function for $f(\sigma+it)=\frac{\zeta(\sigma+it)}{\zeta(\sigma)}$ and Khinchin distribution for corresponding $F_{\xi}(-\ln(n))=P(\xi \le -\ln(n))$, Zeta distribution (Mandelbrot) for $F_{\eta}(n)=P(\eta \le n)=\frac{\sum_{k=1}^{k=n}k^{\sigma}}{\zeta(\sigma)}$, and Hurwitz Zeta distribution for $F_{\eta}(n)=P(\eta \le n)=\frac{\sum_{k=0}^{k=n}(k+Q)^{-S}}{\sum_{k=0}^{k=\infty}(k+Q)^{-S}}$. In this sense, the "Zeta distribution" and "Riemann Zeta distribution" are interchangeably used in \cite{salov2013} and this paper together with the "Hurwitz Zeta distribution". To avoid confusions, the Equations \ref{EqHirwitzZetaDistribution} are presented.

\section{Non-Gaussian Relative b-Increments}

Sample statistics were computed for $\ln(\frac{P_i}{P_{i-1}})=\ln(\frac{m_i\delta_{ZC}}{m_{i-1}\delta_{ZC}})=\ln(m_i)-\ln(m_{i-1})$: $N=7945289$, $\textrm{mean} = 1.9793 \times 10^{-8}$, $\textrm{standard deviation} = 0.00032513$, $\textrm{skewness} = 0.0918$, $\textrm{excess kurtosis} = -3$, $\textrm{minimum} = -0.02920843$, $\textrm{maximum} = 0.02960047$. For small $|\Delta m_i|=|m_i-m_{i-1}|=|k_i|$ and big $m_i$, $\ln(m_i)-\ln(m_{i-1}) \approx \frac{k_i}{m_i}$, a rational number expressing the relative price change. Table \ref{b-log-increments} presents boundaries of sample distribution bins, counts, and frequencies.
\begin{center}
\begin{longtable}{|r|r|r|r|}
\caption[Sample distribution of  logarithms of prices of b-increments]{Sample distribution of $\ln(\frac{P_i}{P_{i-1}})$ for b-increments: ZCZ15, ZCH16, ZCK16, ZCN16, ZCU16, ZCZ16, ZCH17, ZCK17; 2015/08/07 - 2017/02/24; 08:30:00 - 13:15:00 CST.} \label{b-log-increments} \\

 \hline
 \multicolumn{1}{|c|}{Bin low} &
 \multicolumn{1}{|c|}{Bin high} &
 \multicolumn{1}{c|}{Count $n$} &
 \multicolumn{1}{c|}{Frequency} \\
 \hline 
 \endfirsthead

 \multicolumn{4}{c}%
 {\tablename\ \thetable{} -- continued from previous page} \\
 \hline
 \multicolumn{1}{|c|}{Bin low} &
 \multicolumn{1}{|c|}{Bin high} &
 \multicolumn{1}{c|}{Count $n$} &
 \multicolumn{1}{c|}{Frequency} \\
 \hline 
 \endhead

 \hline \multicolumn{4}{|r|}{{Continued on next page}} \\ \hline
 \endfoot

 \hline
 \endlastfoot
-0.029327383 & -0.028689831 & 1 & 1.26E-07\\
-0.024226968 & -0.023589417 & 1 & 1.26E-07\\
-0.020401658 & -0.019764106 & 2 & 2.52E-07\\
-0.018489002 & -0.01785145 & 2 & 2.52E-07\\
-0.01785145 & -0.017213899 & 3 & 3.78E-07\\
-0.016576347 & -0.015938795 & 1 & 1.26E-07\\
-0.015938795 & -0.015301243 & 1 & 1.26E-07\\
-0.015301243 & -0.014663691 & 4 & 5.03E-07\\
-0.014663691 & -0.01402614 & 2 & 2.52E-07\\
-0.01402614 & -0.013388588 & 1 & 1.26E-07\\
-0.012751036 & -0.012113484 & 4 & 5.03E-07\\
-0.012113484 & -0.011475932 & 8 & 1.01E-06\\
-0.011475932 & -0.010838381 & 6 & 7.55E-07\\
-0.010838381 & -0.010200829 & 7 & 8.81E-07\\
-0.010200829 & -0.009563277 & 3 & 3.78E-07\\
-0.009563277 & -0.008925725 & 12 & 1.51E-06\\
-0.008925725 & -0.008288173 & 21 & 2.64E-06\\
-0.008288173 & -0.007650622 & 20 & 2.52E-06\\
-0.007650622 & -0.00701307 & 15 & 1.89E-06\\
-0.00701307 & -0.006375518 & 42 & 5.29E-06\\
-0.006375518 & -0.005737966 & 41 & 5.16E-06\\
-0.005737966 & -0.005100414 & 60 & 7.55E-06\\
-0.005100414 & -0.004462863 & 101 & 1.27E-05\\
-0.004462863 & -0.003825311 & 165 & 2.08E-05\\
-0.003825311 & -0.003187759 & 301 & 3.79E-05\\
-0.003187759 & -0.002550207 & 676 & 8.51E-05\\
-0.002550207 & -0.001912655 & 1741 & 0.000219124\\
-0.001912655 & -0.001275104 & 10376 & 0.001305931\\
-0.001275104 & -0.000637552 & 622385 & 0.07833384\\
-0.000637552 & 0 & 6528748 & 0.821713093\\
0 & 0.000637552 & 144854 & 0.018231432\\
0.000637552 & 0.001275104 & 621729 & 0.078251276\\
0.001275104 & 0.001912655 & 10743 & 0.001352122\\
0.001912655 & 0.002550207 & 1699 & 0.000213837\\
0.002550207 & 0.003187759 & 639 & 8.04E-05\\
0.003187759 & 0.003825311 & 322 & 4.05E-05\\
0.003825311 & 0.004462863 & 197 & 2.48E-05\\
0.004462863 & 0.005100414 & 97 & 1.22E-05\\
0.005100414 & 0.005737966 & 71 & 8.94E-06\\
0.005737966 & 0.006375518 & 40 & 5.03E-06\\
0.006375518 & 0.00701307 & 32 & 4.03E-06\\
0.00701307 & 0.007650622 & 26 & 3.27E-06\\
0.007650622 & 0.008288173 & 10 & 1.26E-06\\
0.008288173 & 0.008925725 & 22 & 2.77E-06\\
0.008925725 & 0.009563277 & 9 & 1.13E-06\\
0.009563277 & 0.010200829 & 10 & 1.26E-06\\
0.010200829 & 0.010838381 & 7 & 8.81E-07\\
0.010838381 & 0.011475932 & 5 & 6.29E-07\\
0.011475932 & 0.012113484 & 5 & 6.29E-07\\
0.012113484 & 0.012751036 & 3 & 3.78E-07\\
0.012751036 & 0.013388588 & 2 & 2.52E-07\\
0.013388588 & 0.01402614 & 2 & 2.52E-07\\
0.01402614 & 0.014663691 & 3 & 3.78E-07\\
0.014663691 & 0.015301243 & 2 & 2.52E-07\\
0.016576347 & 0.017213899 & 3 & 3.78E-07\\
0.017213899 & 0.01785145 & 1 & 1.26E-07\\
0.019126554 & 0.019764106 & 1 & 1.26E-07\\
0.022314313 & 0.022951865 & 1 & 1.26E-07\\
0.022951865 & 0.023589417 & 1 & 1.26E-07\\
0.026139624 & 0.026777176 & 1 & 1.26E-07\\
0.027414727 & 0.028052279 & 1 & 1.26E-07\\
0.029327383 & 0.029964935 & 1 & 1.26E-07\\
\end{longtable}
\end{center}
Counts from Table \ref{b-log-increments} are grouped in 37 classes containing five or more points. Gaussian $(\textrm{mean}=1.9793 \times 10^{-8}, \textrm{standard deviation}=0.00032513)$ probabilities $p_j$ are computed for each class together with Pearson's $\chi^2$-quantities $\frac{(n_j-Np_j)^2}{Np_j}$, Table \ref{b-log-increments-classes}.
\begin{center}
\begin{longtable}{|r|r|r|r|r|r|r|}
\caption[Classes of  logarithms of prices of b-increments]{Classes of $\ln(\frac{P_i}{P_{i-1}})$ for b-increments: ZCZ15, ZCH16, ZCK16, ZCN16, ZCU16, ZCZ16, ZCH17, ZCK17; 2015/08/07 - 2017/02/24; 08:30:00 - 13:15:00 CST. $N=7945289.$} \label{b-log-increments-classes} \\

 \hline
 \multicolumn{1}{|c|}{Class $j$} &
 \multicolumn{1}{|c|}{Class low} &
 \multicolumn{1}{|c|}{Class high} &
 \multicolumn{1}{c|}{Gaussian $p_j$} &
 \multicolumn{1}{c|}{$n_j$} &
 \multicolumn{1}{c|}{$Np_j$} &
 \multicolumn{1}{c|}{$\chi^2$} \\
 \hline 
 \endfirsthead

 \multicolumn{7}{c}%
 {\tablename\ \thetable{} -- continued from previous page} \\
 \hline
 \multicolumn{1}{|c|}{Class $j$} &
 \multicolumn{1}{|c|}{Class low} &
 \multicolumn{1}{|c|}{Class high} &
 \multicolumn{1}{c|}{Gaussian $p_j$} &
 \multicolumn{1}{c|}{$n_j$} &
 \multicolumn{1}{c|}{$Np_j$} &
 \multicolumn{1}{c|}{$\chi^2$} \\
 \hline 
 \endhead

 \hline \multicolumn{7}{|r|}{{Continued on next page}} \\ \hline
 \endfoot

 \hline
 \endlastfoot
1 & $-\infty$ & -0.011475932 & 3.32e-273 & 30 & 2.64e-266 & 3.41e+268\\
2 & -0.011475932 & -0.010838381 & 5.89e-244 & 6 & 4.68e-237 & 7.69e+237\\
3 & -0.010838381 & -0.009563277 & 1.83e-190 & 10 & 1.45e-183 & 6.87e+184\\
4 & -0.009563277 & -0.008925725 & 3.21e-166 & 12 & 2.55e-159 & 5.64e+160\\
5 & -0.008925725 & -0.008288173 & 1.21e-143 & 21 & 9.62e-137 & 4.58e+138\\
6 & -0.008288173 & -0.007650622 & 9.82e-123 & 20 & 7.8e-116 & 5.13e+117\\
7 & -0.007650622 & -0.00701307 & 1.71e-103 & 15 & 1.36e-96 & 1.65e+98\\
8 & -0.00701307 & -0.006375518 & 6.45e-86 & 42 & 5.13e-79 & 3.44e+81\\
9 & -0.006375518 & -0.005737966 & 5.24e-70 & 41 & 4.17e-63 & 4.03e+65\\
10 & -0.005737966 & -0.005100414 & 9.23e-56 & 60 & 7.33e-49 & 4.91e+51\\
11 & -0.005100414 & -0.004462863 & 3.52e-43 & 101 & 2.8e-36 & 3.64e+39\\
12 & -0.004462863 & -0.003825311 & 2.94e-32 & 165 & 2.33e-25 & 1.17e+29\\
13 & -0.003825311 & -0.003187759 & 5.38e-23 & 301 & 4.27e-16 & 2.12e+20\\
14 & -0.003187759 & -0.002550207 & 2.19e-15 & 676 & 1.74e-08 & 2.63e+13\\
15 & -0.002550207 & -0.001912655 & 2.02e-09 & 1741 & 0.016 & 1.89e+08\\
16 & -0.001912655 & -0.001275104 & 4.39e-05 & 10376 & 349 & 2.88e+05\\
17 & -0.001275104 & -0.000637552 & 0.0249 & 622385 & 1.98e+05 & 9.11e+05\\
18 & -0.000637552 & 0 & 0.475 & 6528748 & 3.77e+06 & 2.01e+06\\
19 & 0 & 0.000637552 & 0.475 & 144854 & 3.77e+06 & 3.49e+06\\
20 & 0.000637552 & 0.001275104 & 0.0249 & 621729 & 1.98e+05 & 9.08e+05\\
21 & 0.001275104 & 0.001912655 & 4.39e-05 & 10743 & 349 & 3.09e+05\\
22 & 0.001912655 & 0.002550207 & 2.02e-09 & 1699 & 0.016 & 1.8e+08\\
23 & 0.002550207 & 0.003187759 & 2.19e-15 & 639 & 1.74e-08 & 2.35e+13\\
24 & 0.003187759 & 0.003825311 & 5.38e-23 & 322 & 4.27e-16 & 2.43e+20\\
25 & 0.003825311 & 0.004462863 & 2.94e-32 & 197 & 2.33e-25 & 1.66e+29\\
26 & 0.004462863 & 0.005100414 & 3.52e-43 & 97 & 2.8e-36 & 3.36e+39\\
27 & 0.005100414 & 0.005737966 & 9.23e-56 & 71 & 7.33e-49 & 6.88e+51\\
28 & 0.005737966 & 0.006375518 & 5.24e-70 & 40 & 4.17e-63 & 3.84e+65\\
29 & 0.006375518 & 0.00701307 & 6.45e-86 & 32 & 5.13e-79 & 2e+81\\
30 & 0.00701307 & 0.007650622 & 1.71e-103 & 26 & 1.36e-96 & 4.96e+98\\
31 & 0.007650622 & 0.008288173 & 9.82e-123 & 10 & 7.8e-116 & 1.28e+117\\
32 & 0.008288173 & 0.008925725 & 1.21e-143 & 22 & 9.62e-137 & 5.03e+138\\
33 & 0.008925725 & 0.009563277 & 3.21e-166 & 9 & 2.55e-159 & 3.17e+160\\
34 & 0.009563277 & 0.010200829 & 1.83e-190 & 10 & 1.45e-183 & 6.87e+184\\
35 & 0.010200829 & 0.010838381 & 2.24e-216 & 7 & 1.78e-209 & 2.75e+210\\
36 & 0.010838381 & 0.011475932 & 5.89e-244 & 5 & 4.68e-237 & 5.34e+237\\
37 & 0.011475932 & $\infty$ & 3.32e-273 & 27 & 2.64e-266 & 2.76e+268\\
$\sum$ &  &  & 1 & 7945289 & 7945289 & 6.17e+268\\
\end{longtable}
\end{center}
The Microsoft Excel $\textrm{NORMDIST}(x, \textrm{mean}, \textrm{standard deviation}, \textrm{TRUE})$, where TRUE means cumulative distribution function, is applied to compute $p_j$. Minor asymmetry of counts around zero is caused by tiny positive mean and skewness, and rounding bin lows and highs to six meaningful decimal digits. \textit{The tail $\chi^2$ values are gigantic making the middle not principal for the conclusion that the Gaussian hypothesis is unsound for relative b-increments}.

\section{Price Increments vs. Waiting Times}

Brownian motion $B(t)$ is self-similar. Zooming in or out displays a process with the variance of $\Delta B(t)$ proportional to the time interval $\Delta t$. The a-b-c process is not self-similar already because of finite $0 < \delta_{ZC}$ and discreteness of b- and c-increments. The \cite[pp. 62 - 64, 71 - 75]{salov2013} checks dependences between b- and a-increments. The difficulties are 1) a-increments, waiting times, are random, 2) the reported accuracy of a-increments is one second but the number of ticks for liquid contracts is greater than the number of seconds in a corresponding time interval, 3) the concepts of statistical and non-statistical dependence and the methods of it experimental detection are not well developed.

\begin{figure}[!h]
  \centering
  \includegraphics[width=120mm]{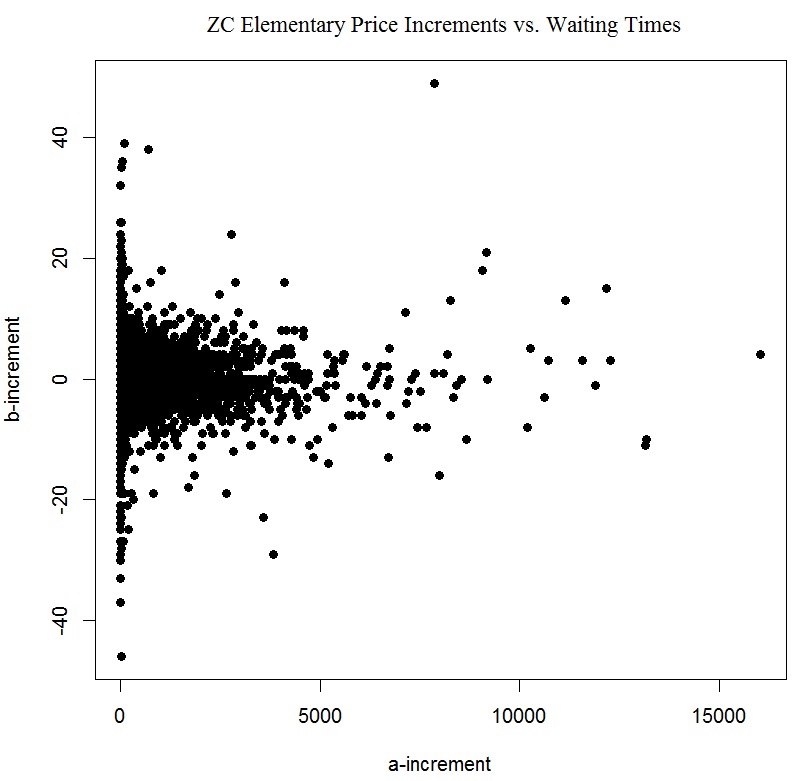}
  \caption[FigABIncrements]
   {b-Increments vs. a-increments in seconds: ZCZ15, ZCH16, ZCK16, ZCN16, ZCU16, ZCZ16, ZCH17, ZCK17; 2015/08/07 - 2017/02/24; 08:30:00 - 13:15:00; 2005 sessions; 7945289 dots. Plotted with R version 3.0.2, the R Foundation for Statistical Computing Platform.} 
  \label{FigABIncrements}
\end{figure}

Neighboring ticks $0 \le i-1$, $i$ form dots $(\Delta t_{i}, \Delta P_{i})$, Figure \ref{FigABIncrements}. Ranges of b-increments at lower waiting times are wider than on the right. However, sample variance is determined not only by extreme values but the number of intermediate dots millions of which are hidden due to $\delta_{ZC}$ discreteness of b-increments and rounding a-increments to seconds.
The first look is misleading.

\begin{figure}[!h]
  \centering
  \includegraphics[width=120mm]{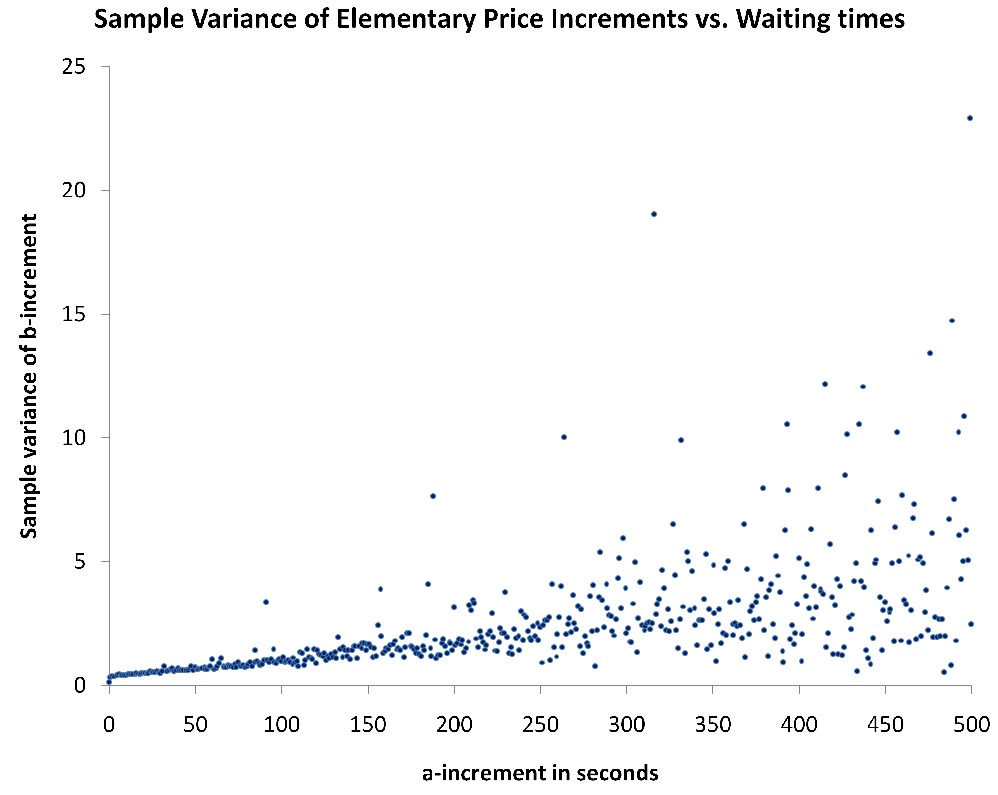}
  \caption[FigBVarianceAIncrement500]
   {Sample variance of b-increments vs. a-increments: ZCZ15, ZCH16, ZCK16, ZCN16, ZCU16, ZCZ16, ZCH17, ZCK17; 2005 sessions; 2015/08/07 - 2017/02/24; 08:30:00 - 13:15:00; 7942115 increments, 501 dots.} 
  \label{FigBVarianceAIncrement500}
\end{figure}

\begin{figure}[!h]
  \centering
  \includegraphics[width=100mm]{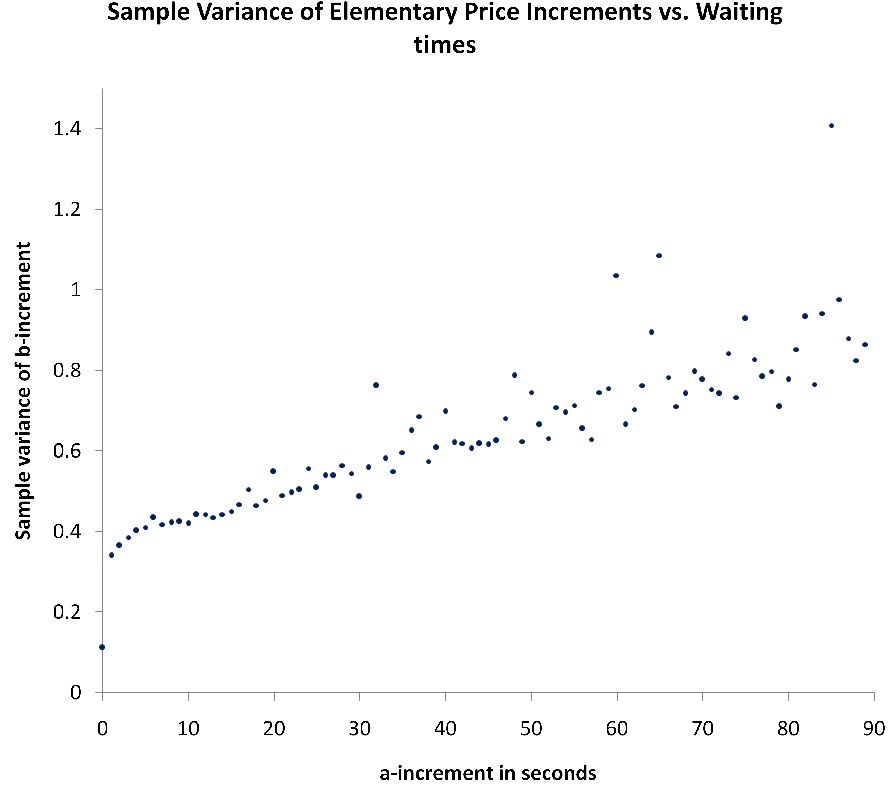}
  \caption[FigBVarianceAIncrement100]
   {Zoom-in to Figure \ref{FigBVarianceAIncrement500}: 7903192 increments, 91 dots.} 
  \label{FigBVarianceAIncrement100}
\end{figure}

Let us divide b-increments on samples, corresponding to a-increments $0, 1, ..., n$ seconds, and for each sample evaluate empirical distribution. Samples are time slices on Figure \ref{FigABIncrements}. Rounding times to seconds and randomness of a-increments can mis-distribute b-increments affecting sample statistics. Figures \ref{FigBVarianceAIncrement500} and \ref{FigBVarianceAIncrement100} show if sample variances are proportional to seconds. Errors of these estimates increase on the right side, where samples are getting smaller. There is no line crossing $(0, 0)$ but increasing dependence with dots concentrating around a line, except the initial point, where errors for the interval $[0, 1]$ seconds are significant due to rounding. Table \ref{b-increments-a-increments} presents 101 initial sample statistics.

\begin{center}
\begin{longtable}{|r|r|r|r|r|r|r|r|r|r|}
\caption[Sample Statistics of b-Increments per a-increment]{Sample Statistics of b-Increments in $\delta_{ZC}$ corresponding to a-increments in seconds: ZCZ15, ZCH16, ZCK16, ZCN16, ZCU16, ZCZ16, ZCH17, ZCK17; 2015/08/07 - 2017/02/24; 08:30:00 - 13:15:00 CST.} \label{b-increments-a-increments} \\
 \hline
 \multicolumn{1}{|c|}{Seconds} &
 \multicolumn{1}{|c|}{Size} &
 \multicolumn{1}{c|}{Mean} &
 \multicolumn{1}{c|}{Min} &
 \multicolumn{1}{c|}{$n_{min}$} &
 \multicolumn{1}{c|}{Max} &
 \multicolumn{1}{c|}{$n_{max}$} &
 \multicolumn{1}{c|}{StdDev} &
 \multicolumn{1}{c|}{Skew.} &
 \multicolumn{1}{c|}{E-Kurt.} \\
 \hline 
 \endfirsthead
 \multicolumn{10}{c}%
 {\tablename\ \thetable{} -- continued from previous page} \\
 \hline
 \multicolumn{1}{|c|}{Seconds} &
 \multicolumn{1}{|c|}{Size} &
 \multicolumn{1}{c|}{Mean} &
 \multicolumn{1}{c|}{Min} &
 \multicolumn{1}{c|}{$n_{min}$} &
 \multicolumn{1}{c|}{Max} &
 \multicolumn{1}{c|}{$n_{max}$} &
 \multicolumn{1}{c|}{StdDev} &
 \multicolumn{1}{c|}{Skew.} &
 \multicolumn{1}{c|}{E-Kurt.} \\
 \hline 
 \endhead
 \hline \multicolumn{10}{|r|}{{Continued on next page}} \\ \hline
 \endfoot
 \hline
 \endlastfoot
0 & 4923314 & -0.00085 & -37 & 1 & 32 & 1 & 0.33 & -0.92 & -2.97\\
1 & 826494 & 0.0016 & -28 & 1 & 23 & 1 & 0.58 & 0.19 & -2.99\\
2 & 443250 & -0.00017 & -10 & 1 & 13 & 1 & 0.6 & 0.051 & -2.91\\
3 & 300638 & 0.003 & -10 & 1 & 26 & 1 & 0.62 & 0.28 & -2.97\\
4 & 217470 & 0.0048 & -19 & 1 & 14 & 1 & 0.63 & 0.0029 & -2.99\\
5 & 168587 & 0.0015 & -19 & 1 & 13 & 1 & 0.64 & -0.11 & -1.97\\
6 & 134079 & -0.0012 & -46 & 1 & 19 & 1 & 0.66 & -2.5 & 5.45\\
7 & 105755 & 0.00042 & -14 & 1 & 8 & 2 & 0.65 & -0.13 & -1.33\\
8 & 87872 & 0.0068 & -12 & 1 & 11 & 1 & 0.65 & -0.038 & 0.0993\\
9 & 73606 & 0.0035 & -8 & 1 & 13 & 1 & 0.65 & 0.11 & -1.53\\
10 & 63554 & 0.0025 & -12 & 1 & 9 & 1 & 0.65 & -0.11 & 4.4\\
11 & 52528 & 0.002 & -10 & 1 & 13 & 1 & 0.66 & 0.13 & 7.01\\
12 & 44920 & 0.003 & -10 & 2 & 17 & 1 & 0.67 & 0.43 & 17\\
13 & 38895 & 0.0031 & -8 & 1 & 5 & 1 & 0.66 & -0.054 & 1.06\\
14 & 34486 & -0.0038 & -6 & 2 & 10 & 1 & 0.67 & 0.2 & 3.36\\
15 & 30273 & 0.011 & -6 & 1 & 8 & 1 & 0.67 & 0.053 & 1.97\\
16 & 26346 & -7.6e-05 & -5 & 2 & 10 & 2 & 0.68 & 0.25 & 4.29\\
17 & 23622 & 0.0061 & -13 & 1 & 26 & 1 & 0.71 & 1.4 & 86.1\\
18 & 21094 & -0.0033 & -8 & 1 & 6 & 1 & 0.68 & -0.042 & 1.78\\
19 & 18786 & 0.0099 & -6 & 1 & 7 & 1 & 0.69 & 0.1 & 2.18\\
20 & 17418 & 0.005 & -7 & 1 & 35 & 1 & 0.74 & 6 & 287\\
21 & 15593 & 0.0064 & -8 & 1 & 5 & 1 & 0.7 & -0.33 & 3.95\\
22 & 14300 & -0.0018 & -8 & 1 & 9 & 1 & 0.7 & 0.09 & 4.74\\
23 & 13084 & -0.0092 & -6 & 1 & 11 & 1 & 0.71 & 0.21 & 5.95\\
24 & 11823 & -0.0028 & -23 & 1 & 8 & 1 & 0.75 & -2.5 & 78.5\\
25 & 10822 & 0.0098 & -6 & 1 & 4 & 3 & 0.71 & -0.036 & 1.35\\
26 & 10178 & 0.0013 & -12 & 1 & 7 & 1 & 0.73 & -0.34 & 9.01\\
27 & 9279 & -0.0061 & -12 & 1 & 8 & 1 & 0.73 & -0.42 & 10.4\\
28 & 8643 & -0.0071 & -6 & 1 & 20 & 1 & 0.75 & 2.2 & 58.8\\
29 & 8158 & 0.00061 & -4 & 5 & 8 & 1 & 0.74 & 0.13 & 2.59\\
30 & 8794 & -0.0031 & -4 & 3 & 7 & 1 & 0.7 & 0.16 & 3.03\\
31 & 7161 & -0.0067 & -7 & 1 & 6 & 2 & 0.75 & 0.083 & 3.95\\
32 & 6594 & 0.011 & -11 & 1 & 36 & 1 & 0.87 & 10 & 441\\
33 & 6122 & -0.014 & -8 & 1 & 6 & 1 & 0.76 & -0.18 & 4.29\\
34 & 5816 & -0.0067 & -4 & 3 & 4 & 1 & 0.74 & -0.045 & 1.01\\
35 & 5571 & -0.012 & -4 & 5 & 9 & 2 & 0.77 & 0.44 & 7.35\\
36 & 5258 & -0.018 & -14 & 1 & 12 & 1 & 0.81 & -0.21 & 28.6\\
37 & 4799 & -0.013 & -8 & 1 & 13 & 1 & 0.83 & 1.1 & 23.2\\
38 & 4625 & -0.013 & -7 & 2 & 3 & 6 & 0.76 & -0.34 & 3.35\\
39 & 4323 & -0.0023 & -5 & 2 & 11 & 1 & 0.78 & 0.4 & 10.6\\
40 & 4091 & -0.0064 & -12 & 1 & 19 & 1 & 0.84 & 2.2 & 75.4\\
41 & 3868 & 0.023 & -5 & 2 & 7 & 1 & 0.79 & 0.21 & 3.17\\
42 & 3771 & 0.024 & -5 & 2 & 4 & 1 & 0.79 & -0.13 & 1.77\\
43 & 3454 & -0.015 & -5 & 1 & 4 & 1 & 0.78 & -0.14 & 1.41\\
44 & 3372 & -0.02 & -5 & 1 & 6 & 1 & 0.79 & 0.025 & 1.65\\
45 & 3340 & -0.0063 & -7 & 1 & 5 & 1 & 0.79 & -0.32 & 5.17\\
46 & 3049 & -0.014 & -3 & 4 & 6 & 1 & 0.79 & 0.29 & 2.1\\
47 & 2847 & 0.013 & -9 & 1 & 8 & 1 & 0.83 & -0.031 & 8.79\\
48 & 2943 & 0.017 & -14 & 1 & 13 & 1 & 0.89 & 0.14 & 38.6\\
49 & 2693 & 0.039 & -4 & 1 & 4 & 2 & 0.79 & 0.03 & 1.17\\
50 & 2535 & 0.015 & -5 & 1 & 6 & 2 & 0.86 & 0.26 & 3.25\\
51 & 2461 & 0.012 & -6 & 1 & 7 & 1 & 0.82 & -0.013 & 4.34\\
52 & 2476 & 0.01 & -3 & 10 & 4 & 2 & 0.8 & 0.069 & 1.25\\
53 & 2272 & 0.0048 & -5 & 1 & 3 & 8 & 0.84 & -0.2 & 1.44\\
54 & 2243 & -0.025 & -10 & 1 & 4 & 1 & 0.83 & -0.8 & 10.3\\
55 & 2024 & 0.003 & -4 & 2 & 3 & 7 & 0.84 & -0.1 & 0.99\\
56 & 2053 & 0.013 & -3 & 1 & 3 & 10 & 0.81 & 0.28 & 0.639\\
57 & 1943 & -0.013 & -4 & 2 & 4 & 1 & 0.79 & 0.13 & 1.35\\
58 & 1904 & -0.036 & -8 & 1 & 4 & 2 & 0.86 & -0.47 & 4.86\\
59 & 1835 & -0.0071 & -4 & 1 & 6 & 1 & 0.87 & 0.16 & 2.03\\
60 & 1891 & -0.038 & -27 & 1 & 4 & 1 & 1 & -9.9 & 260\\
61 & 1674 & -0.011 & -3 & 6 & 5 & 1 & 0.82 & 0.24 & 2.02\\
62 & 1638 & 0.0061 & -5 & 1 & 4 & 1 & 0.84 & -0.18 & 1.79\\
63 & 1507 & 0.013 & -3 & 6 & 4 & 2 & 0.87 & 0.15 & 1.25\\
64 & 1531 & -0.024 & -6 & 1 & 14 & 1 & 0.95 & 2 & 33\\
65 & 1449 & -0.0028 & -4 & 2 & 17 & 1 & 1 & 4.7 & 70.9\\
66 & 1405 & 0.00071 & -4 & 2 & 4 & 2 & 0.89 & 0.13 & 1.78\\
67 & 1385 & -0.015 & -3 & 6 & 3 & 8 & 0.84 & 0.05 & 0.964\\
68 & 1297 & -0.047 & -4 & 3 & 3 & 6 & 0.86 & -0.2 & 1.59\\
69 & 1300 & 0.019 & -6 & 1 & 6 & 1 & 0.89 & 0.073 & 3.99\\
70 & 1233 & -0.0041 & -4 & 2 & 3 & 3 & 0.88 & -0.33 & 1.21\\
71 & 1183 & 0.0059 & -4 & 2 & 4 & 1 & 0.87 & -0.011 & 1.84\\
72 & 1219 & -0.014 & -4 & 2 & 4 & 2 & 0.86 & 0.073 & 1.57\\
73 & 1184 & -0.011 & -4 & 3 & 4 & 2 & 0.92 & 0.16 & 2.02\\
74 & 1119 & -0.0098 & -3 & 3 & 4 & 1 & 0.86 & 0.25 & 0.906\\
75 & 1066 & -0.023 & -4 & 1 & 11 & 1 & 0.96 & 1.6 & 17.5\\
76 & 1062 & -0.032 & -3 & 2 & 5 & 2 & 0.91 & 0.52 & 2.29\\
77 & 981 & -0.082 & -4 & 3 & 3 & 3 & 0.89 & -0.24 & 1.35\\
78 & 959 & 0.0094 & -4 & 1 & 4 & 1 & 0.89 & -0.045 & 1.19\\
79 & 955 & -0.01 & -5 & 1 & 3 & 3 & 0.84 & -0.12 & 1.72\\
80 & 869 & -0.028 & -4 & 1 & 3 & 3 & 0.88 & -0.17 & 0.87\\
81 & 888 & 0.098 & -4 & 1 & 7 & 1 & 0.92 & 0.61 & 4.46\\
82 & 870 & -0.045 & -5 & 1 & 4 & 2 & 0.97 & -0.12 & 2.29\\
83 & 887 & -0.021 & -4 & 1 & 4 & 1 & 0.88 & 0.11 & 1.16\\
84 & 806 & 0.0012 & -5 & 1 & 5 & 1 & 0.97 & 0.23 & 2.8\\
85 & 745 & 0.015 & -19 & 1 & 6 & 1 & 1.2 & -5.6 & 88.9\\
86 & 788 & -0.016 & -6 & 2 & 8 & 1 & 0.99 & 0.45 & 11.5\\
87 & 747 & 0.023 & -6 & 1 & 4 & 3 & 0.94 & -0.045 & 3.51\\
88 & 743 & -0.015 & -4 & 1 & 4 & 1 & 0.91 & 0.029 & 1.02\\
89 & 707 & -0.065 & -3 & 8 & 4 & 1 & 0.93 & -0.04 & 1.22\\
90 & 697 & -0.057 & -6 & 1 & 5 & 1 & 1 & -0.52 & 4.3\\
91 & 686 & 0.029 & -9 & 1 & 39 & 1 & 1.8 & 14 & 299\\
92 & 665 & 0.065 & -6 & 1 & 5 & 1 & 1 & 0.069 & 3.69\\
93 & 644 & -0.034 & -5 & 2 & 5 & 2 & 0.96 & -0.27 & 4.42\\
94 & 632 & -0.054 & -7 & 1 & 5 & 1 & 1 & -0.73 & 5.26\\
95 & 662 & 0.023 & -13 & 1 & 9 & 1 & 1.2 & -1.3 & 27.3\\
96 & 641 & 0.031 & -8 & 1 & 4 & 1 & 0.97 & -0.44 & 8.03\\
97 & 542 & -0.017 & -4 & 1 & 4 & 2 & 0.95 & -0.019 & 1.83\\
98 & 595 & -0.034 & -5 & 1 & 4 & 1 & 1 & -0.17 & 2.17\\
99 & 535 & 0.1 & -4 & 1 & 5 & 1 & 1 & 0.33 & 2.11\\
100 & 533 & -0.013 & -4 & 1 & 7 & 1 & 0.99 & 0.68 & 5.24\\
\end{longtable}
\end{center}
\paragraph{Are sample variances stochastic?} The sample variance of b-increments $\mu_{2(s,r)}^{\Delta P}$ obtained from Equation \ref{EqSampleStatistics} by replacing $t$ with $P$ is an unbiased estimator of the variance and random itself. This is so, if the sample is from a generalized population. If the stochastic properties of the population change, then this "natural" estimator should not be applied "mechanically". Here we discuss a different stochasticity. In contrast with other \textit{constant parameters} of the famous Black-Scholes-Merton \textit{European option} value formula, its volatility $\sigma$ is not observed. The last three decades have been "fixing" the \textit{implied} $\sigma$ dependent on the option expiration time and strike price. The advanced models represent $\sigma$ as a stochastic process - parameter of the underlying price stochastic process. Their goal is fitting option premiums by equations. After calibration, the options with characteristics different than those used for fitting can be evaluated. As long as the fitting is good, the model and interpolated or extrapolated "unknown" options values are accepted. Ironically, these models formulated for underlying prices and rates are rarely applied to simulate the later. They do not deal with discrete prices, rounded random waiting times, daily price limits important for the high leveraged futures profits and losses. While fitting option prices, they are useless for trading futures sensitive to the details described in this paper and \cite{salov2013}.

\textit{Why} do these theories fit option prices and not futures prices, b-increments, and waiting times? Not \textit{why} \textit{classical mechanics} works in macro and not micro world giving up to \textit{quantum mechanics}. But \textit{what} are the worlds differences? \textit{High frequency trading is where scaling down fails. The elementary indecomposable further a- and b-increments studied here and \cite{salov2013} serve as building blocks, algebraic summands, for all increments.}

Instead of fitting a deterministic curve to points on Figures \ref{FigBVarianceAIncrement500} and \ref{FigBVarianceAIncrement100}, we switch the interpretation: each value associated with time $t=0+\textrm{a-increment}$ is a single observation of a random variable at $t$. A collection of random variables marked by time is a representation of a stochastic process \cite{korn1968}. There is visible drift and increasing dispersion. A non-negative function is on a top of a random variable. This interpretation makes the property a stochastic process starting at $t=0$. \textit{It remains to recollect that the property interpreted as a stochastic process on the figures is the sample variance of b-increments.}

7903192 pairs of a- and b-increments grouped by equal a-increments allowing to estimate sample properties of b-increments for each group were treated as following. The 871 samples with three and more b-increments were extracted. The a-increment = 0 has the sample variance of b-increments 0.33, Table \ref{b-increments-a-increments}. 10 sequential a-increments were used to compute \textit{sample moments of the sample variance of b-increments} for $\frac{871-1}{10}=87$ intervals of a-increments. It is assumed that the sample variance does not change much within each interval, Table \ref{sample_variance_b_increments}.

\begin{figure}[!h]
  \centering
  \includegraphics[width=100mm]{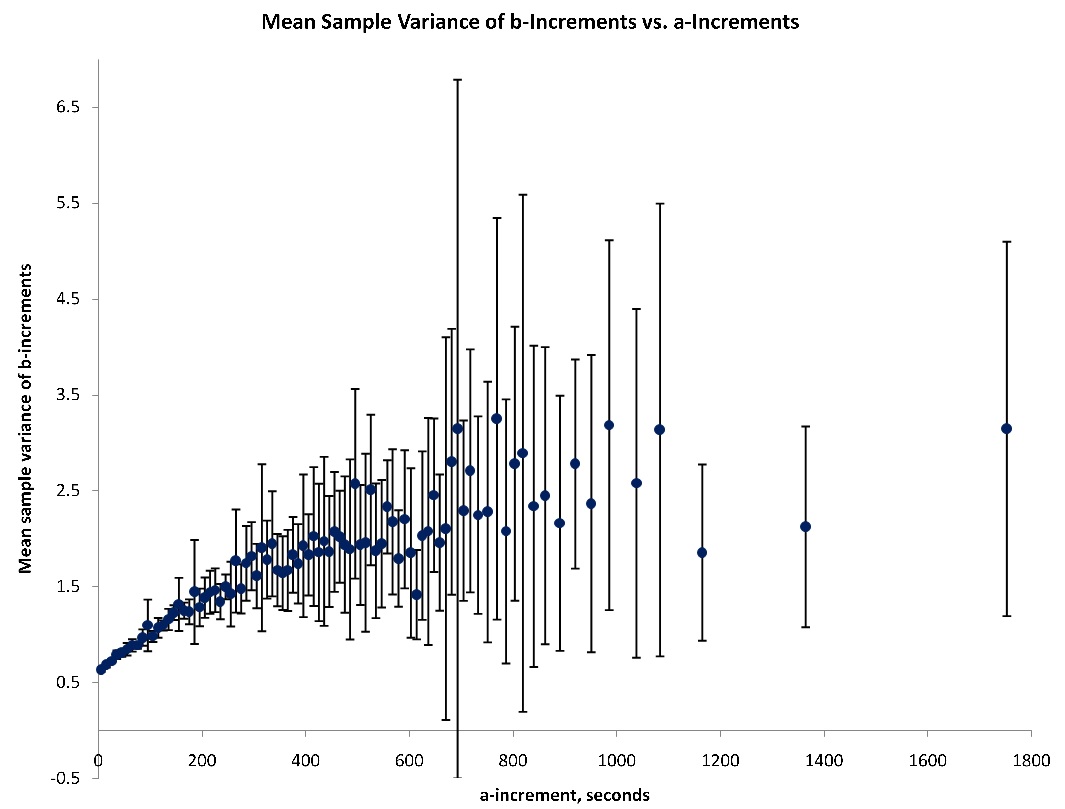}
  \caption[FigSampleVarianceOfBVarianceAIncrement]
   {Mean sample variance of b-increments vs. a-increments in seconds, 87 dots from Table \ref{sample_variance_b_increments}.} 
  \label{FigSampleVarianceOfBVarianceAIncrement}
\end{figure}

\begin{figure}[!h]
  \centering
  \includegraphics[width=100mm]{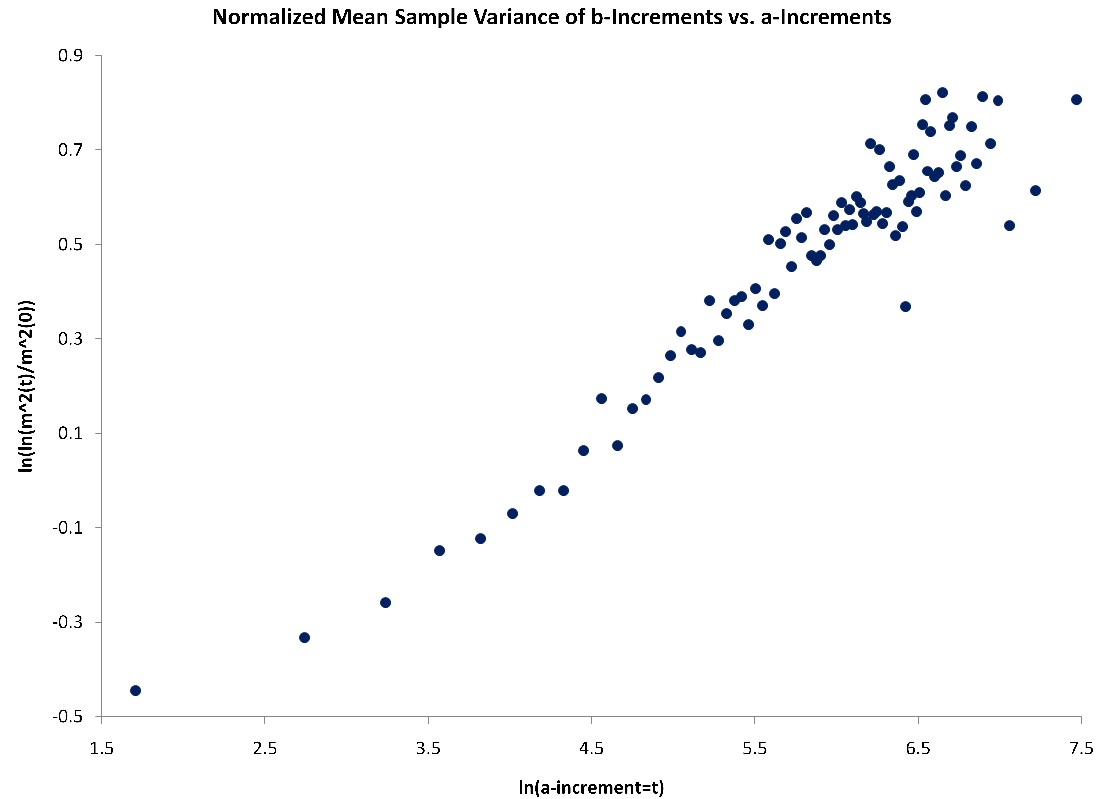}
  \caption[FigNormalizedSampleVarianceOfBVarianceAIncrement]
   {Normalized mean sample variance of b-increments vs. a-increments in seconds, 87 dots from Table \ref{sample_variance_b_increments}: $\ln(\ln(\frac{\mu_{2}^{\Delta P}(\textrm{a-increment})}{\mu_{2}^{\Delta P}(0)=0.33377})) \approx (-1.01 \pm 0.09) + (0.25 \pm 0.02) \times \ln(\textrm{a-increment)}$, $R \approx 0.96$.} 
  \label{FigNormalizedSampleVarianceOfBVarianceAIncrement}
\end{figure}

\begin{center}
\begin{longtable}{|r|r|r|r|r|r|r|r|r|r|}
\caption[Sample Statistics of Sample Variances of b-Increments]{Sample statistics of sample variances of b-increments in $\delta_{ZC}$ corresponding to a-increments in seconds: ZCZ15, ZCH16, ZCK16, ZCN16, ZCU16, ZCZ16, ZCH17, ZCK17; 2015/08/07 - 2017/02/24; 08:30:00 - 13:15:00 CST. Each interval of a-increments $[a_{left}, a_{right}]$ contains 10 sample variances estimated by at least three points each.} \label{sample_variance_b_increments} \\

 \hline
 \multicolumn{1}{|c|}{$a_{left}$} &
 \multicolumn{1}{|c|}{$a_{right}$} &
 \multicolumn{1}{c|}{Mean} &
 \multicolumn{1}{c|}{Min} &
 \multicolumn{1}{c|}{$n_{min}$} &
 \multicolumn{1}{c|}{Max} &
 \multicolumn{1}{c|}{$n_{max}$} &
 \multicolumn{1}{c|}{StdDev} &
 \multicolumn{1}{c|}{Skew.} &
 \multicolumn{1}{c|}{E-Kurt.} \\
 \hline 
 \endfirsthead

 \multicolumn{10}{c}%
 {\tablename\ \thetable{} -- continued from previous page} \\
 \hline
 \multicolumn{1}{|c|}{$a_{left}$} &
 \multicolumn{1}{|c|}{$a_{right}$} &
 \multicolumn{1}{c|}{Mean} &
 \multicolumn{1}{c|}{Min} &
 \multicolumn{1}{c|}{$n_{min}$} &
 \multicolumn{1}{c|}{Max} &
 \multicolumn{1}{c|}{$n_{max}$} &
 \multicolumn{1}{c|}{StdDev} &
 \multicolumn{1}{c|}{Skew.} &
 \multicolumn{1}{c|}{E-Kurt.} \\
 \hline 
 \endhead

 \hline \multicolumn{10}{|r|}{{Continued on next page}} \\ \hline
 \endfoot

 \hline
 \endlastfoot
1 & 10 & 0.634 & 0.584 & 1 & 0.659 & 1 & 0.024 & -1.27 & 0.112\\
11 & 20 & 0.684 & 0.661 & 1 & 0.741 & 1 & 0.0253 & 1.47 & 0.864\\
21 & 30 & 0.723 & 0.698 & 1 & 0.752 & 1 & 0.02 & 0.111 & -1.86\\
31 & 40 & 0.791 & 0.742 & 1 & 0.873 & 1 & 0.0436 & 0.761 & -0.898\\
41 & 50 & 0.808 & 0.778 & 1 & 0.888 & 1 & 0.0382 & 1.49 & 0.153\\
51 & 60 & 0.849 & 0.794 & 1 & 1.02 & 1 & 0.0651 & 2.25 & 3.94\\
61 & 70 & 0.888 & 0.816 & 1 & 1.04 & 1 & 0.0648 & 1.67 & 1.96\\
71 & 80 & 0.888 & 0.843 & 1 & 0.964 & 1 & 0.0354 & 0.986 & 0.307\\
81 & 90 & 0.97 & 0.875 & 1 & 1.19 & 1 & 0.0859 & 2 & 3.29\\
91 & 100 & 1.1 & 0.947 & 1 & 1.83 & 1 & 0.269 & 2.77 & 5.44\\
101 & 110 & 0.981 & 0.883 & 1 & 1.07 & 1 & 0.0574 & 0.132 & -0.528\\
111 & 120 & 1.07 & 0.909 & 1 & 1.21 & 1 & 0.103 & -0.0151 & -1.39\\
121 & 130 & 1.1 & 1 & 1 & 1.2 & 1 & 0.0535 & 0.156 & 0.113\\
131 & 140 & 1.16 & 1.02 & 1 & 1.39 & 1 & 0.111 & 0.772 & -0.0433\\
141 & 150 & 1.23 & 1.05 & 1 & 1.3 & 1 & 0.0768 & -1.62 & 1.86\\
151 & 160 & 1.32 & 1.06 & 1 & 1.97 & 1 & 0.277 & 1.71 & 1.77\\
161 & 170 & 1.25 & 1.09 & 1 & 1.39 & 1 & 0.0844 & 0.0328 & -0.116\\
171 & 180 & 1.24 & 1.06 & 1 & 1.45 & 1 & 0.13 & 0.728 & -0.519\\
181 & 190 & 1.45 & 1.04 & 1 & 2.76 & 1 & 0.542 & 1.99 & 2.27\\
191 & 200 & 1.28 & 1.09 & 1 & 1.77 & 1 & 0.197 & 1.81 & 2.72\\
201 & 210 & 1.39 & 1.16 & 1 & 1.8 & 1 & 0.211 & 1.42 & 0.155\\
211 & 220 & 1.44 & 1.21 & 1 & 1.85 & 1 & 0.224 & 1.19 & -0.301\\
221 & 230 & 1.46 & 1.17 & 1 & 1.94 & 1 & 0.228 & 0.896 & 0.297\\
231 & 240 & 1.34 & 1.13 & 1 & 1.73 & 1 & 0.184 & 0.875 & 0.225\\
241 & 250 & 1.5 & 1.34 & 1 & 1.7 & 1 & 0.129 & 0.251 & -1.55\\
251 & 260 & 1.42 & 0.976 & 1 & 2.02 & 1 & 0.338 & 0.137 & -1.1\\
261 & 270 & 1.77 & 1.24 & 1 & 3.16 & 1 & 0.538 & 2.24 & 3.86\\
271 & 280 & 1.48 & 1.14 & 1 & 1.89 & 1 & 0.256 & 0.469 & -1.4\\
281 & 290 & 1.74 & 0.884 & 1 & 2.32 & 1 & 0.39 & -0.998 & 0.999\\
291 & 300 & 1.82 & 1.43 & 1 & 2.43 & 1 & 0.355 & 0.576 & -1.28\\
301 & 310 & 1.61 & 1.15 & 1 & 2.23 & 1 & 0.336 & 0.665 & -0.692\\
311 & 320 & 1.91 & 1.5 & 1 & 4.36 & 1 & 0.871 & 3.05 & 6.6\\
321 & 330 & 1.78 & 1.22 & 1 & 2.55 & 1 & 0.407 & 0.598 & -0.78\\
331 & 340 & 1.95 & 1.14 & 1 & 3.15 & 1 & 0.548 & 0.979 & 0.851\\
341 & 350 & 1.67 & 1.2 & 1 & 2.3 & 1 & 0.377 & 0.421 & -1.05\\
351 & 360 & 1.64 & 0.978 & 1 & 2.23 & 1 & 0.385 & 0.045 & -0.69\\
361 & 370 & 1.67 & 1.07 & 1 & 2.55 & 1 & 0.423 & 1.02 & 0.345\\
371 & 380 & 1.83 & 1.44 & 1 & 2.82 & 1 & 0.395 & 1.93 & 2.92\\
381 & 390 & 1.74 & 1.08 & 1 & 2.28 & 1 & 0.414 & -0.528 & -1.4\\
391 & 400 & 1.93 & 0.974 & 1 & 3.25 & 1 & 0.744 & 0.596 & -1.16\\
401 & 410 & 1.83 & 0.987 & 1 & 2.51 & 1 & 0.423 & -0.528 & 0.135\\
411 & 420 & 2.02 & 1.12 & 1 & 3.49 & 1 & 0.724 & 0.841 & -0.104\\
421 & 430 & 1.86 & 1.1 & 1 & 3.18 & 1 & 0.716 & 0.902 & -0.604\\
431 & 440 & 1.97 & 0.756 & 1 & 3.48 & 1 & 0.881 & 0.488 & -0.789\\
441 & 450 & 1.87 & 0.926 & 1 & 2.72 & 1 & 0.579 & -0.188 & -1.19\\
451 & 460 & 2.07 & 1.33 & 1 & 3.2 & 1 & 0.625 & 0.523 & -1.09\\
461 & 470 & 2.02 & 1.32 & 1 & 2.7 & 1 & 0.479 & -0.134 & -1.41\\
471 & 480 & 1.94 & 1.39 & 1 & 3.66 & 1 & 0.711 & 1.82 & 2.14\\
481 & 490 & 1.89 & 0.738 & 1 & 3.84 & 1 & 0.939 & 0.951 & 0.0249\\
491 & 500 & 2.57 & 1.36 & 1 & 4.79 & 1 & 0.988 & 1.22 & 0.97\\
501 & 510 & 1.94 & 0.983 & 1 & 2.81 & 1 & 0.626 & -0.0839 & -1.58\\
511 & 520 & 1.96 & 0.817 & 1 & 2.94 & 1 & 0.928 & -0.303 & -2.12\\
521 & 530 & 2.51 & 1.61 & 1 & 3.85 & 1 & 0.785 & 0.556 & -1.38\\
531 & 540 & 1.87 & 0.976 & 1 & 3.07 & 1 & 0.702 & 0.819 & -0.538\\
541 & 552 & 1.95 & 0.957 & 1 & 2.81 & 1 & 0.665 & -0.221 & -1.78\\
553 & 562 & 2.33 & 1.5 & 1 & 2.9 & 1 & 0.486 & -0.41 & -1.37\\
563 & 572 & 2.18 & 0.957 & 1 & 3.11 & 1 & 0.757 & -0.509 & -1.09\\
574 & 584 & 1.79 & 1 & 1 & 2.87 & 1 & 0.503 & 0.719 & 0.797\\
585 & 597 & 2.2 & 1.24 & 1 & 3.21 & 1 & 0.72 & 0.31 & -1.73\\
598 & 608 & 1.85 & 0.817 & 1 & 3.81 & 1 & 0.883 & 1.25 & 0.809\\
609 & 619 & 1.42 & 0.548 & 1 & 2.22 & 1 & 0.466 & -0.216 & -0.15\\
620 & 630 & 2.03 & 0.957 & 1 & 4.04 & 1 & 0.879 & 1.21 & 1.28\\
631 & 642 & 2.08 & 0.577 & 1 & 3.91 & 1 & 1.18 & 0.177 & -1.72\\
643 & 652 & 2.45 & 0.837 & 1 & 3.61 & 1 & 0.801 & -0.813 & 0.033\\
653 & 664 & 1.96 & 1 & 2 & 2.98 & 1 & 0.71 & -0.0263 & -1.6\\
665 & 676 & 2.11 & 0.577 & 1 & 7.37 & 1 & 2 & 2.38 & 4.3\\
677 & 686 & 2.8 & 1.27 & 1 & 5.72 & 1 & 1.39 & 0.908 & 0.0138\\
687 & 699 & 3.14 & 1 & 2 & 13.2 & 1 & 3.64 & 2.87 & 5.97\\
700 & 709 & 2.29 & 0.577 & 1 & 3.74 & 1 & 0.941 & -0.179 & -0.572\\
711 & 724 & 2.71 & 0.894 & 1 & 5.1 & 1 & 1.27 & 0.787 & -0.22\\
725 & 740 & 2.25 & 0.957 & 1 & 3.79 & 1 & 1.03 & 0.368 & -1.67\\
741 & 761 & 2.28 & 0.817 & 1 & 4.93 & 1 & 1.36 & 0.741 & -0.72\\
762 & 776 & 3.25 & 1.26 & 2 & 8.5 & 1 & 2.09 & 1.9 & 3.04\\
777 & 796 & 2.08 & 0.577 & 1 & 5.48 & 1 & 1.38 & 1.84 & 2.63\\
797 & 810 & 2.78 & 0.535 & 1 & 5.8 & 1 & 1.43 & 0.718 & 0.613\\
812 & 826 & 2.89 & 0 & 1 & 9.35 & 1 & 2.7 & 1.61 & 1.98\\
830 & 850 & 2.34 & 0.577 & 2 & 6.11 & 1 & 1.67 & 1.24 & 0.951\\
853 & 871 & 2.45 & 0.577 & 1 & 4.8 & 1 & 1.55 & 0.34 & -1.77\\
877 & 904 & 2.16 & 0.577 & 1 & 5.51 & 1 & 1.33 & 1.91 & 3.17\\
905 & 935 & 2.78 & 0.577 & 1 & 4.19 & 1 & 1.09 & -0.59 & -0.0698\\
938 & 964 & 2.37 & 0.837 & 1 & 5.77 & 1 & 1.55 & 1.32 & 0.503\\
967 & 1004 & 3.18 & 0.577 & 1 & 6.98 & 1 & 1.93 & 0.743 & -0.419\\
1007 & 1069 & 2.58 & 0.577 & 3 & 5.13 & 1 & 1.82 & 0.221 & -1.89\\
1073 & 1094 & 3.13 & 0 & 1 & 6.66 & 1 & 2.36 & 0.26 & -1.57\\
1111 & 1219 & 1.86 & 0.577 & 2 & 3.21 & 1 & 0.919 & -0.0682 & -1.39\\
1248 & 1481 & 2.12 & 0.577 & 1 & 4 & 1 & 1.05 & 0.306 & -0.893\\
1484 & 2021 & 3.14 & 1 & 1 & 8.02 & 1 & 1.95 & 1.88 & 2.95\\
\end{longtable}
\end{center}
The plot of mean sample variance against the center of interval $[a_{left}, a_{right}]$ with $\pm$ StdDev from Table \ref{sample_variance_b_increments} is on Figure \ref{FigSampleVarianceOfBVarianceAIncrement}. This error can be multiplied by the t-Student distribution coefficient 2.296 for $10-1=9$ degrees of freedom and two-sided 95\% confident interval. Errors in sample higher moments estimates are significant. Double logarithm of $\frac{\mu_{2}^{\Delta P}(\textrm{a-increment})}{\mu_{2}^{\Delta P}(0)=0.33377}$ vs. $\ln(\textrm{a-increment})$ forms linear regression, Figure \ref{FigNormalizedSampleVarianceOfBVarianceAIncrement}, with Microsoft Excel Data, Data Analysis, Regression estimates: coefficient of linear correlation $R \approx 0.96$, intercept $-1.01 \pm 0.09$, slope $0.25 \pm 0.02$, where confidence intervals correspond to two-sided 95\% with 85 degrees of freedom of residuals.

\paragraph{Dependence statistics $L_1\textrm{ distance}$, and Log-likelihood.} The $n=7945289$ pairs of a- and b-increments have $m_n^A=1866$ different a-increments, $m_n^B=64$ different b-increments, and $m_n^{AB}=6773$ different combinations of a- and b-increments. For each value, class $i$, the empirical frequency $\nu_{A_i}$, $\nu_{B_i}$, $\nu_{A_iB_i}$ is evaluated. For independent events and their probabilities $\nu_{A_iB_i}=\nu_{A_i} \nu_{B_i}$ \cite[\$23.2 Kolmogorov's advice, pp. 72 - 73]{salov2013}. Empirical frequencies are only estimates of probabilities for which the following statistics represent interest \cite{gretton2008}, \cite[pp. 73 - 75]{salov2013}
\begin{displaymath}
L_n(\nu_{AB}, \nu_A\nu_B)=\sum_{A \in A_n}\sum_{B \in B_n}|\nu_{AB} - \nu_A\nu_B|,
\end{displaymath}
\begin{displaymath}
I_n(\nu_{AB}, \nu_A\nu_B)=2\sum_{A \in A_n}\sum_{B \in B_n}\nu_{AB}\log\frac{\nu_{AB}}{\nu_A \nu_B},
\end{displaymath}
\begin{displaymath}
\chi_n^2(\nu_{AB}, \nu_A\nu_B)=\sum_{A \in A_n}\sum_{B \in B_n}\frac{(\nu_{AB}-\nu_A \nu_B)^2}{\nu_A \nu_B}.
\end{displaymath}
Gretton and Gy\"{o}rfi prove that \textit{almost surely} $L_n(\nu_{AB}, \nu_A\nu_B) > \sqrt{2\ln2}\sqrt{\frac{m_n^A m_n^B}{n}}=\epsilon_{L_n}$, if $\lim_{n \rightarrow \infty}\frac{m_n^A m_n^B}{n} = 0 \approx \frac{1866 \times 64}{7945289}=0.015, \; \lim_{n \rightarrow \infty}\frac{m_n^A}{\ln(n)} = \infty \approx \frac{1866}{\ln(7945289)}=117.4, \; \lim_{n \rightarrow \infty}\frac{m_n^B}{\ln(n)} = \infty \approx \frac{64}{\ln(7945289)} = 4.0$, rejects the hypothesis of independence of a- and b-increments. They also suggest to reject independence, if $I_n(\nu_{AB}, \nu_A\nu_B) > \frac{m_n^A m_n^B (2\ln(n + m_n^A m_n^B) + 1)}{n}=\epsilon_{I_n}$. Finally, they derive $\xi_{\chi_n^2}=\frac{n \chi_n^2(\nu_{AB}, \nu_A\nu_B) - m_n^A m_n^B}{\sqrt{2m_n^A m_n^B}} \rightarrow \textrm{Gaussian}(\alpha_1 = 0, \mu_2 = 1)$ meaning convergence on distribution. We got $L_{7945289}=0.26 > \epsilon_{L_{7945289}} = 0.14$, $I_{7945289}=0.13 < \epsilon_{I_{7945289}} = 0.49$. The former rejects and the latter does not reject the hypothesis of independence. The author cannot interpret the values $\chi_{7945289}^2=7.5$ and $\xi_{\chi_{7945289}^2}=1.2 \times 10^5$. \textit{These results are controversial}.

\section{Jumps. Chain Reactions}

\begin{figure}[!h]
  \centering
  \includegraphics[width=110mm]{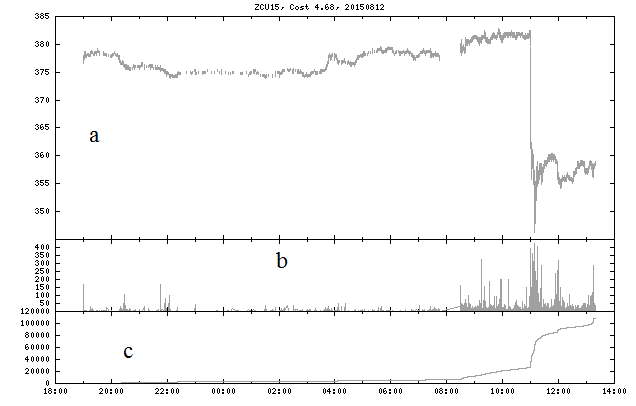}
  \caption[FigZCU15_20150812_ALL]
   {ZCU15, Wednesday August 12, 2015: a - prices, b - volume, c - cumulative volume.} 
  \label{FigZCU15_20150812_ALL}
\end{figure}

\begin{figure}[!h]
  \centering
  \includegraphics[width=110mm]{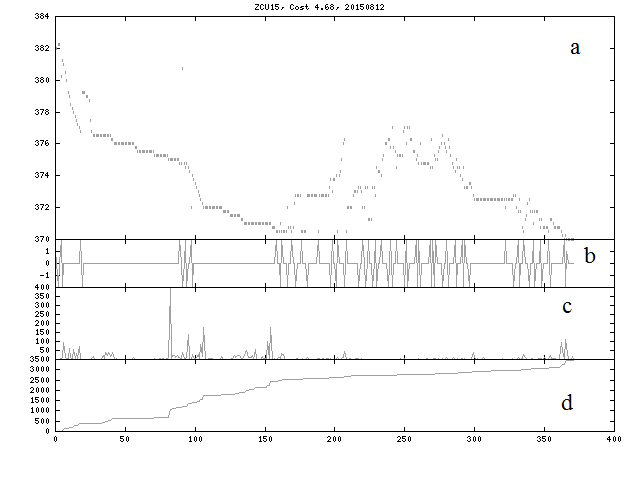}
  \caption[FigZCU15_20150812_ALL]
   {ZCU15, Wednesday August 12, 2015, 10 seconds [11:00:00 - 11:00:10]: a - prices, b - MPS0 (transaction cost \$4.68), c - volume, d - cumulative volume.} 
  \label{FigZCU15_20150812_ZOOM}
\end{figure}

Mathematics and numerical methods of stochastic integration of jump-diffusions are described by Floyd Hanson \cite{hanson2007}. For "fast price moves", rounding to seconds places prices marked by one date and time stamp to a vertical line on a chart, Figure \ref{FigZCU15_20150812_ALL}, creating an illusion of a jump from one diffusion level to another. Plotting prices vs. arriving time indexes zooms in: the dots are "time equidistant" and uncover details, Figure \ref{FigZCU15_20150812_ZOOM}. The United States Department of Agriculture, USDA, news announced at 11:00:00 CST are responsible for this drama. This is not boiling water but an \textit{explosion}. It is not a Poisson jump from a Gaussian diffusion \cite[pp. 101 - 105, Figures 1, 14 - 16, 18, 19, 40 - 45]{salov2013}. \textit{Less of all it resembles an equilibrium}.

Economics borrows "equilibrium" from other branches of science. In \cite{salov2013}, it is suggested that economics will follow thermodynamics, where evolutionary studying equilibria has been replaced with revolutionary understanding of non-equilibrium systems \cite{prigogine1977}, \cite{prigogine1993}. Figures \ref{FigZCU15_20150812_ALL}, \ref{FigZCU15_20150812_ZOOM} strengthen the author's conviction. The MPS is a measure of the frequency and magnitude of market opportunities attracting to trading and serving as essential condition of market's existence and deviation from equilibrium - disequilibrium.

Chemical kinetics explains explosions by chain branching reactions \cite{semenov1967}, \cite{semiokhin1995}. With the reference to Lewis, Semenov describes \cite[pp. 15 - 16]{semenov1967} 1) initiation of combustion of hydrogen in oxygen $\textrm{H}_2+\textrm{O}_2 \rightarrow 2\textrm{OH}$, 2) branching $\textrm{H}+\textrm{O}_2 \rightarrow \textrm{OH} + \textrm{O}$, and inhibition on "walls". Branching increases the number of intermediate active atoms and radicals, yielding an avalanche.

In trading, China Yuan Devaluation, USDA, Brexit, Election news (N) create a specific state in minds of traders and input of computer programs - trading robots (M) - initiation. They send orders to a Trading Book (B). The state of (B) interacts with arriving information creating Time \& Sales output (TS) returned to M. TS being interpreted as "confirmation" activates more parts of M - branching  - until M is exhausted by taken positions, limited capital, losses and profits targets: $\textrm{N} \rightarrow (\textrm{M} \rightarrow \textrm{B} \rightarrow \textrm{TS} \rightarrow \textrm{M})_{\textrm{iterations}}$. Notation for iterated functions and iterals are proposed in \cite{salov2012b}, \cite{salov2012c}. This sketch could be improved by studying the stages $\textrm{M} \rightarrow \textrm{B}$ and $\textrm{B} \rightarrow \textrm{TS}$, where times at all gates to and from B are accurately registered: \textit{rounding to a second is a bottleneck}.

\section{Extreme b-Increments}

\begin{figure}[!h]
  \centering
  \includegraphics[width=130mm]{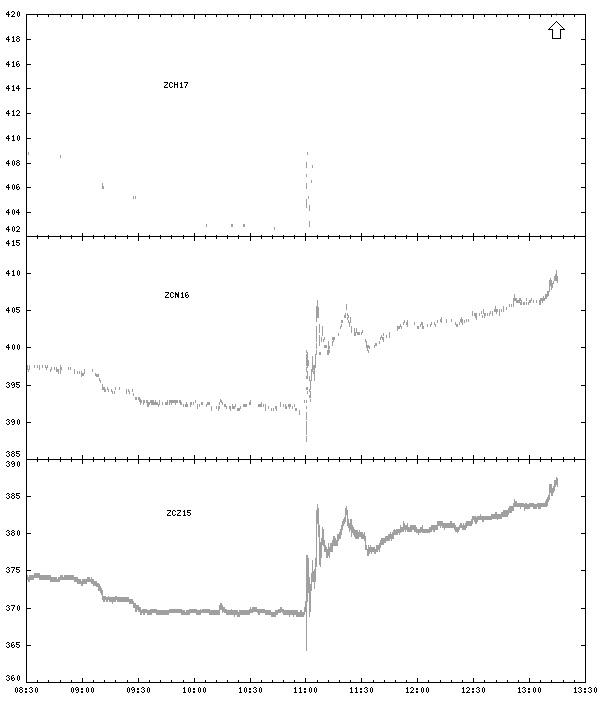}
  \caption[FigZCH17_ZCN16_ZCZ15_20150911]
   {ZCH17, ZCN16, ZCZ15 tick prices, Friday September 11, 2015 [08:30:00 - 13:15:00].} 
  \label{FigZCH17_ZCN16_ZCZ15_20150911}
\end{figure}

The market is always right: the extreme b-increments are not errors. Ignoring outliers, spoiling a theory, can be costly. They ruin accounts and create fortunes. In studied sessions, the greatest b-increments -46 and 49 are not data or program errors: Figure \ref{FigZCH17_ZCN16_ZCZ15_20150911}, Table \ref{TBL_Extreme_b}. Both are from illiquid futures.

\begin{center}
\begin{longtable}{|r|r|r|r|r|r|r|}
\caption[Extreme_b]{Ticks brining the greatest b-increments and their neighbors; 2015/08/07 - 2017/02/24; 08:30:00 - 13:15:00 CST.} \label{TBL_Extreme_b} \\

\hline
\multicolumn{1}{|c|}{Ticker} &
\multicolumn{1}{|c|}{Date} &
\multicolumn{1}{c|}{Time} &
\multicolumn{1}{c|}{Price $P$} &
\multicolumn{1}{c|}{$\Delta P$} &
\multicolumn{1}{c|}{$k$} &
\multicolumn{1}{c|}{Size} \\
\hline 
\endfirsthead

\multicolumn{7}{c}%
{\tablename\ \thetable{} -- continued from previous page} \\
\hline
\multicolumn{1}{|c|}{Ticker} &
\multicolumn{1}{|c|}{Date} &
\multicolumn{1}{c|}{Time} &
\multicolumn{1}{c|}{Price $P$} &
\multicolumn{1}{c|}{$\Delta P$} &
\multicolumn{1}{c|}{$k$} &
\multicolumn{1}{c|}{Size} \\
\hline 
\endhead

\hline \multicolumn{7}{|r|}{{Continued on next page}} \\ \hline
\endfoot

\hline
\endlastfoot
ZCN16 & 2015-09-11 & 11:00:04 & 399.50 &  1.75 &  7 & 1\\
ZCN16 & 2015-09-11 & 11:00:10 & 388.00 & -11.50 & -46 & 8\\
ZCN16 & 2015-09-11 & 11:00:10 & 388.00 &   0.00 &  0 & 1\\
 & & & & & &\\
ZCH17 & 2015-09-11 & 11:03:22 & 407.75 &  1.25 &  5 & 1\\
ZCH17 & 2015-09-11 & 13:14:22 & 420.00 & 12.25 & 49 & 2\\
ZCH17 & 2015-09-11 & 13:14:23 & 420.00 &  0.00 &  0 & 1\\
\end{longtable}
\end{center}
On Friday September 11, 2015, the most liquid ZCZ15 contract moved far by shorter steps. Avoiding arbitrage, the distant ZCN16, ZCH17 contract prices in parallel sessions must adapt. But the transactions for ZCN16, and, especially, ZCH17 are rare. Thus - greater steps. For a trader of the ZCN16, ZCH17 this is additional risk unfortunately (or fortunately) obeying the plot on Figure \ref{FigBIncrementsFrequency}. There is plenty of not the greatest but great b-increments.

\begin{figure}[!h]
  \centering
  \includegraphics[width=120mm]{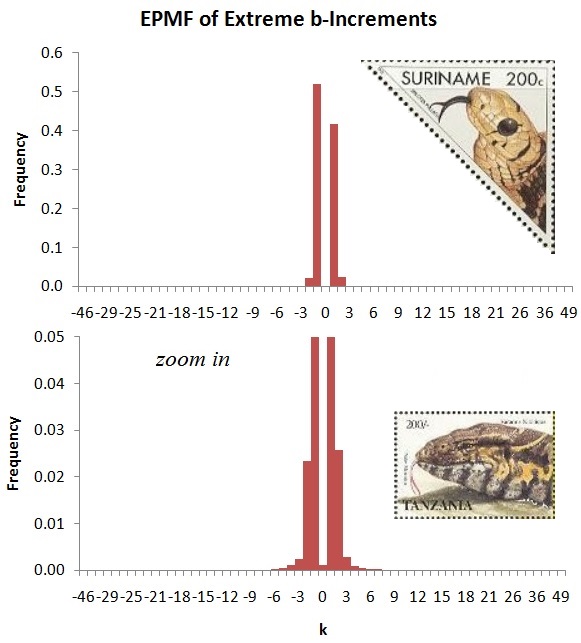}
  \caption[FigFrequenciesExtremeB]
   {EPMF of extreme b-increments: ZCZ15, ZCH16, ZCK16, ZCN16, ZCU16, ZCZ16, ZCH17, ZCK17; 1953 sessions; 2015/08/07 - 2017/02/24; 08:30:00 - 13:15:00. The postage stamps images for collage are borrowed from \url{https://www.freestampcatalogue.com/stamps/nature/snakes}.} 
  \label{FigFrequenciesExtremeB}
\end{figure}

\begin{figure}[!h]
  \centering
  \includegraphics[width=120mm]{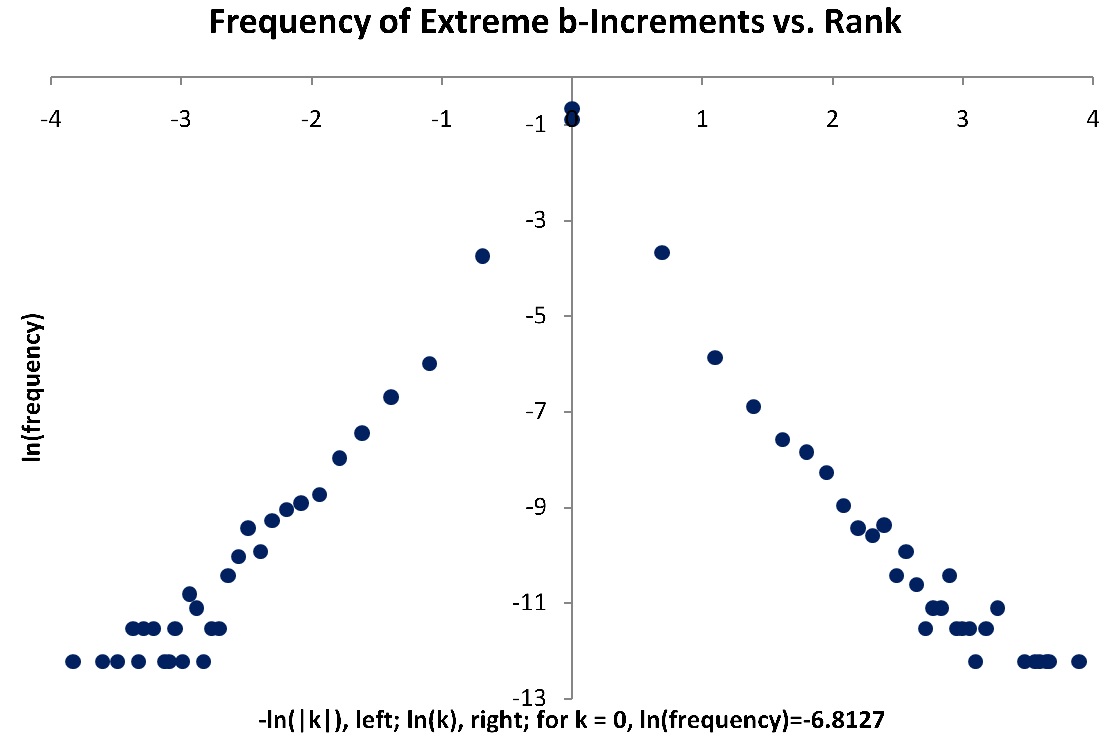}
  \caption[FigFrequenciesExtremeBvsRank]
   {Frequencies of extreme b-increments vs. rank in bi-logarithmic coordinates: ZCZ15, ZCH16, ZCK16, ZCN16, ZCU16, ZCZ16, ZCH17, ZCK17; 1953 sessions; 2015/08/07 - 2017/02/24; 08:30:00 - 13:15:00. This is the same data as on Figure \ref{FigFrequenciesExtremeB}. Both symmetric "wings" are candidates for application of the Hurwitz Zeta distribution and can be combined using absolute values $|k|$, with the exception of the point $(k=0\delta, \; frequency \approx 0.0011)$. The zero rank frequency is not zero for illiquid contracts where price has not changed. By the same reasons a maximum rank in a session can be negative and minimum rank can be positive.}
  \label{FigFrequenciesExtremeBvsRank}
\end{figure}

The C++, Python, AWK, and Bourn shell programs, written by the author, among other quantities report the extreme values of b-increments and the numbers of their occurrences in a trading range and session. The extreme value theory has been applied to them in \cite[pp. 50 - 56]{salov2013}, reviewing contributions of Frechet, Fisher, Tippet, von Mises, Gnedenko, Gumbel, Haan. The discrete version of the II type of the Fisher-Tippet-Gnedenko continuous $PDF_{FTG}^{\textrm{II}}(x)$ has been suggested \cite[p. 54]{salov2013}:
\begin{displaymath}
PMF^{\textrm{II}}(n) = \frac{\frac{kb}{(bn+a)^{k+1}}e^{-(bn+a)^{-k}}}{\sum_{i=1}^{\infty} \frac{kb}{(bi+a)^{k+1}}e^{-(bi+a)^{-k}}  }, \; n \in \textrm{N}, \; k > 0, \; b > 0, \; a \ge 0,
\end{displaymath}
where $PMF^{\textrm{II}}(0)=0$. The denominator passes the Maclaurin-Cauchy integral test of convergence of series allowing to adopt the Euler-Maclaurin formula for the distribution and come to a robust algorithm \cite[pp. 54 - 56]{salov2013}. Naturally, the most frequent $0\delta$ b-increment rarely becomes extreme, even, for the futures contracts with low liquidity. Accordingly, an EPMF of the extreme b-increments resembles the end of a forked tongue of some snakes and lizards, Figures \ref{FigFrequenciesExtremeB}, \ref{FigFrequenciesExtremeBvsRank}. These reptiles benefit from the tongue splitting due to \textit{discretional smelling} supporting a \textit{stereo effect}. Another common feature: both can be dangerous for human beings - travelers to exotic places, and traders - travelers to the most exotic and exciting place ...

Redistributing continuous probability density between discrete mass points, $PMF^{\textrm{II}}(n)$ better handles fat tails, where theoretical behavior is not observed maybe due to small samples sizes and slow convergence to the asymptotic limit. Starting from the $PDF_{FTG}^{\textrm{II}}(x)$, this approach embeds discreteness, computing the original function at integer arguments, and yields $PMF^{\textrm{II}}(n)$ after proper normalization making sure that the total summed probability mass is one. It enjoys the improved fitting without \textit{theoretical} justification. In a similar manner many continuous distributions can be converted to discrete \cite[pp. 56 - 58]{salov2013}. If one would argue that $PDF_{FTG}^{\textrm{II}}(x)$ is one of the only three possible asymptotic limits and "must be used", then the author will say that the original i.i.d. assumption leading to this asymptotic is violated by statistical non-stationarity of the market. We should not forget \cite[\$22.2 Einstein's suspended particle, pp. 65 - 69; \$22.3 Solution which Einstein did and Black did not know]{salov2013} that behind "Einstein's Brownian motion" the number of participating particles is of the order of the Avogadro number $N_A=6.02 \times 10^{23} \; \textrm{mol}^{-1}$. The remarkable experimental artifact - the number of ZC ticks $n=7945289$ - supporting (or disproving) "Bachelier's Brownian motion" is more than modest. Prokhorov's estimates \cite[pp. 13 - 14]{salov2015} indicate that convergence can be slow. The numbers of ticks generated by modern high frequency trading are far from $N_A$. Not only the "most exotic place" works under conditions, chaotically and randomly deviating from an equilibrium, and violates the rules of stationarity but it is likely far from the convergence requirements of the limit theorems. Plus, discreteness, rounding, price limits ... Under such conditions, simple empirical dependencies on Figures \ref{FigVN}, \ref{EqCumulativeVC123}, \ref{FigExKurtosisSkewness}, \ref{FigStdDevMean}, \ref{FigBIncrementsFrequency}, \ref{FigFrequenciesExtremeBvsRank} are useful. The laws of nature fold letters of complexity into envelopes of simplicity on different hierarchy levels of a system. If this would not be so, then as Goldenfeld and Kadanoff emphasize \cite{goldenfeld1999}: \textit{"In order to model a bulldozer, we would need to be careful to model its constituting quarks!"}

Another complication: every futures time series is unique and non-repeatable. Hmm..., Bernoulli trials \cite{salov2014}!? This limits the application of the probability theory requiring either \textit{multiple} process realizations or \textit{several} values for each random variable in time. The role of \textit{individual random objects} will increase \cite[\$24 A Comment on Randomness, pp. 75 - 81]{salov2013}.

\paragraph{Things we have no time for.} Figure \ref{FigNewCorn} is the Beauty portrait. Focusing on the elementary market time and price moves of the wandering corn, we leave untouched \textit{Fundamental Analysis}. Figures \ref{FigESH16_20160122} and \ref{FigEmptyField} prove: corn prices fluctuate, even, if there is no corn on fields. Corn is in granaries, factories, restaurants, South America fields ... The weather in northern Argentina in January, when the early corn approaching maturity, can boost prices on the CBOT. Taking a bicycle ride in June, the author was surprised to find soybean instead of corn, Figure \ref{FigSoybeanField}. They do not switch every year. One field does not make a difference: intentions to plant soybean or corn influence on the crop predictions statistically. USDA during the growing season weekly reports on planting, crop development, and harvesting progress. Special raids to the fields uncover real situation. A beneficial rain in Illinois, Ohio, Idaho decreasing soybean and corn prices but continuing and flooding, Figure \ref{FigSoybeanField}, soars them. Multifactor linear and non-linear models attempt to uncover the next crop yields. Nicely looking plants Figures \ref{FigCornField}, \ref{FigNewCornField} do not mean that corn is harvested. In a private conversation, a farmer has "opened the author's eyes": corn is stable enough and can remain on the fields during autumn. This may create a shortage of corn after the high plants quality has been announced. The majority incorrectly anticipates events, Figure \ref{FigZCU15_20150812_ALL}. An amateur trader having a position is seeking a confirmation in every thing in the world. Can a too big long position bring a bullish taker to a restaurant in order to eat as much corn as possible to "increase demand" and "hit prices"? Before going, take a look at Figures \ref{FigESH16_20160122}, \ref{FigEmptyField}. A minority with independent mind wins. Quick recognizing \textit{always own} errors is a must. The best teacher and advisor is \textit{the always right market}.

And the best traders known to the author are MPS0, MPS1, and MPS2 trading on ... a hindsight. Still, it makes sense to study them. With transaction costs and a limited budget, MPS2 takes every market offer reinvesting gains as soon as it becomes profitable. However, MPS2 cannot \textit{beat the market} in the sense that taking everything offered is only a draw.

\bigskip

\noindent\textbf{Valerii Salov} received his M.S. from the Moscow State University, Department of Chemistry in 1982 and his Ph.D. from the Academy of Sciences of the USSR, Vernadski Institute of Geochemistry and Analytical Chemistry in 1987.  He is the author of the articles on analytical, computational, and physical chemistry, the book Modeling Maximum Trading Profits with C++, \textit{John Wiley and Sons, Inc., Hoboken, New Jersey}, 2007, and papers in \textit{Futures Magazine} and \textit{ArXiv}.

\noindent\textit{v7f5a7@comcast.net}

\end{document}